\pdfoutput=1
\documentclass[aip,11pt]{revtex4-1}
\usepackage{epsfig}
\usepackage{amssymb}
\usepackage{amsmath,amsfonts}
\usepackage{bm}
\usepackage{color,soul} 
\usepackage{booktabs}
\usepackage[colorlinks,
            linkcolor=blue,
            anchorcolor=green,   citecolor=blue]{hyperref}

\usepackage{nomencl}
\usepackage{subcaption}
\usepackage[percent]{overpic}
\usepackage{dcolumn}% Align table columns on decimal point
\usepackage[export]{adjustbox} % 控制图像位置
\usepackage{caption}
\usepackage[T1]{fontenc}
\usepackage{lipsum}
\usepackage{dcolumn}% Align table columns on decimal point
\makenomenclature
\draft

\begin{document}

\preprint{Preprint submitted to Physics of Fluids}

\title{Residual U-Net for accurate and efficient prediction of hemodynamics in two-dimensional asymmetric stenosis} %Title of paper



\author{Xintong Zou}
\affiliation{Department of Mechanics and Aerospace Engineering, Southern University of Science and
Technology, Shenzhen 518055, People's Republic of China}

\author{Suiyang Tong}
\affiliation{Hubei University of Medicine, 
442000, People's Republic of China}
\affiliation{Department of Cardiology, Suizhou Hospital, 441300, People's Republic of China}

\author{Wenhui Peng}
\affiliation{Hubei University of Medicine, 
442000, People's Republic of China}

\author{Qiuxiang Huang}
\email[Corresponding author: ]{qxhuang2013@gmail.com}
\affiliation{School of Mechanical, Medical and Process Engineering, Queensland University of Technology, Brisbane, QLD 4000, Australia}
\affiliation{Centre for Biomedical Technologies, Queensland University of Technology, Brisbane, QLD 4000, Australia}

\author{Jianchun Wang}
\email[Corresponding author: ]{wangjc@sustech.edu.cn}
\affiliation{Department of Mechanics and Aerospace Engineering, Southern University of Science and Technology, Shenzhen 518055, People's Republic of China}


%\date{\today}

\begin{abstract}
This study presents residual U-Net (U-ResNet), a deep learning surrogate model for predicting steady hemodynamic fields in two-dimensional asymmetric stenotic channels at Reynolds numbers ranging from 200 to 800. By integrating residual connections with multi-scale feature extraction, U-ResNet achieves exceptional accuracy while significantly reducing computational costs compared to computational fluid dynamics (CFD) approaches. Comprehensive evaluation against U-Net, Fourier Neural Operator (FNO), and U-Net enhanced Fourier Neural Operator (UFNO) demonstrates U-ResNet's superior performance in capturing sharp hemodynamic gradients and complex flow features. For pressure prediction, U-ResNet achieves a normalized mean absolute error ($NMAE$) of 1.10\%, approximately 6.2× lower than FNO (6.79\%). Similarly, U-ResNet's performance for wall shear stress ($NMAE$: 0.56\%, $7.1\times$ improvement), velocity ($NMAE$: 1.06\%, $6.1\times$ improvement), and vorticity ($NMAE$: 0.69\%, $13\times$ improvement) consistently surpasses alternative architectures. Notably, U-ResNet demonstrates robust generalization to interpolated Reynolds numbers without retraining - a capability rarely achieved in existing models. From a computational perspective, U-ResNet delivers a 180-fold acceleration over CFD, reducing simulation time from approximately 30 minutes to 10 seconds per case. The model's non-dimensional formulation ensures scalability across vessel sizes and anatomical locations, enhancing its applicability to diverse clinical scenarios. Statistical analysis of prediction errors reveals that U-ResNet maintains substantially narrower error distributions compared to FNO and UFNO, confirming its reliability. These advances position U-ResNet as a promising auxiliary tool to complement CFD simulations for real-time clinical decision support, treatment planning, and medical device optimization. Future work will focus on extending the framework to three-dimensional geometries and integrating it with patient-specific data.
\end{abstract}

% \pacs{}% insert suggested PACS numbers in braces on next line

\maketitle %\maketitle must follow title, authors, abstract and \pacs

%======================================================
\section{Introduction}
Hemodynamics, the study of blood flow dynamics and the physical laws governing vascular circulation, plays a pivotal role in elucidating the pathogenesis of cardiovascular diseases  \citep{ku1997blood,davies2009hemodynamic,cecchi2011role,tarbell2014fluid}. The complex interplay between hemodynamic forces, endothelial function, and vascular remodeling significantly influences the initiation, progression, and spatial distribution of atherosclerotic lesions and other vasculopathies  \citep{malek1999hemodynamic,cecchi2011role}. Computational fluid dynamics (CFD) - as a methodology for numerically reconstructing hemodynamic environments - has become an essential tool for improving our understanding of intricate blood flow phenomena  \citep{gijsen1999influence,thomas2003reproducibility}. CFD facilitates the simulation of blood flow and prediction of critical hemodynamic parameters, thus supporting comprehensive assessment of vascular diseases, patient-specific diagnosis, and medical device evaluation. In contrast to \emph{in vivo} or experimental methods - which face high costs (e.g., imaging, surgery, and animal models), ethical risks (e.g., invasive procedures), and limitations in spatial resolution and temporal measurement capabilities - CFD offers cost-effective, high-resolution, risk-free, and real-time measurement alternatives for hemodynamic analysis  \citep{taylor1998finite, cebral2005characterization,li2021prediction,candreva2022current}. 

Despite these advantages, traditional CFD workflows face two primary constraints hindering clinical translation. Firstly, high-fidelity simulations of complex vascular geometries under unsteady flow conditions require access to high-performance computing clusters, with typical runtime durations ranging from hours to days even for idealized models  \citep{taylor2009patient,steinman2019patient,li2021prediction}. Additionally, the iterative nature of mesh generation, solver convergence, and postprocessing renders conventional CFD impractical for real-time applications such as intraoperative decision support  \citep{kissas2020machine,candreva2022current}. Given these challenges, deep learning has emerged as a transformative adjunct to traditional computational methods, enabling rapid hemodynamic predictions with accuracy comparable to high-fidelity simulations while dramatically reducing computational costs. Neural networks trained on CFD datasets achieve inference times under 1 second for three-dimensional (3-D) velocity field reconstruction - a $10^3 - 10^4$ acceleration factor compared to traditional CFD solvers  \citep{li2021prediction,lino2023current}. This computational efficiency stems from deep learning’s ability to approximate nonlinear mappings between vascular morphology and flow characteristics without explicitly solving the Navier-Stokes equations.

Deep learning has emerged as a transformative paradigm for flow field prediction, providing accurate surrogate models that complement computationally intensive Navier-Stokes solvers by learning mappings between geometric inputs and hydrodynamic outputs \citep{raissi2020hidden,li2023long,wang2024artificial}. Operator learning frameworks, such as the deep operator network (DeepONet) \citep{lu2021learning}, leverage the universal approximation theorem to parameterize nonlinear operators, while physics-informed neural networks (PINNs) \citep{raissi2017part1,raissi2017part2,mao2020physics} embed partial differential equation (PDE) constraints directly into loss functions.  Recent extensions of PINNs address geometric variability \citep{burbulla2023physics} and scattered pressure field prediction \citep{nair2024physics}, yet struggle with hemodynamic applications requiring simultaneous resolution of velocity, pressure, and wall shear stress across varying Reynolds numbers. The Fourier neural operator (FNO) \citep{liFourierNeuralOperator2021} revolutionized turbulence modeling through spectral parameterization of integral kernels, enabling super-resolution predictions of 3-D flows \citep{liFourierNeuralOperator2022a}. Subsequent variants - including implicit FNO \citep{you2022learning}, factorized FNO \citep{tran2023ffno}, and geometry-adapted FNO \citep{li2023gino, li2023fourier} - improve efficiency and adapt for complex geometries. Transformer architectures, originally developed for natural language processing \citep{vaswani2023attention}, now drive advances in PDE solving through factorized attention mechanisms \citep{li2023scalable} and implicit representations \citep{yang2024implicit}. Models like Transolver \citep{wu2024tran} achieve state-of-the-art industrial flow predictions but require structured geometric inputs, limiting applicability to patient-specific vascular anatomies with irregular stenoses (the pathological narrowing of blood vessels).

Deep learning has gained significant traction in fluid dynamics, enabling advancements in hemodynamics prediction \citep{raissi2020hidden,li2021prediction,yevtushenko2021deep,arzani2022machine,kadem2022hemodynamic}. In the context of cardiovascular flows, Kissas \emph{et al.} \citep{kissas2020machine} introduced a PINN framework that integrates four-dimensional flow magnetic resonance imaging data with mass and momentum conservation laws to non-invasively predict arterial blood pressure. This approach eliminates the need for traditional CFD solvers while preserving physical consistency in velocity, pressure, and wall displacement predictions. Similarly, Liang \emph{et al.} \citep{liang2020feasibility} developed a deep neural network capable of directly estimating steady-state pressure and velocity distributions in the thoracic aorta, significantly improving computational efficiency. Building on these efforts, Du \emph{et al.} \citep{du2022deep} proposed a 3-D graph neural network surrogate model that predicts patient-specific blood flow fields from vascular geometries with $10^4\times$ faster inference speeds than conventional CFD methods while maintaining <3\% mean absolute error in velocity and pressure predictions. Sarabian \emph{et al.} \citep{sarabian2022physics} highlighted the limitations of traditional physics-based models in brain hemodynamics due to uncertainties in outlet boundary conditions and modeling deficiencies, underscoring the need for data-driven approaches. Li \emph{et al.} \citep{li2024novel} introduced a hybrid machine learning framework combining convolutional neural networks for geometric feature extraction and deep neural networks trained on CFD simulations to predict blood flow dynamics in stenotic arteries. This approach bypasses conventional CFD workflows, enabling efficient analysis of complex stenoses while achieving accuracy comparable to high-fidelity simulations. Furthermore, Versnjak \emph{et al.} \citep{versnjak2024deep} conducted a comparative study of artificial neural networks trained on datasets of varying dimensionalities, demonstrating successful prediction of hemodynamic parameters in patients with aortic coarctation. 

Despite these advancements, existing studies often focus on predicting single hemodynamic quantities such as pressure or velocity and lack generalization across varying vessel scales and Reynolds numbers. Moreover, critical metrics like wall shear stress (WSS) and vorticity - essential for assessing vascular health and disease progression - are rarely considered comprehensively. These limitations hinder the applicability of current approaches in clinical applications that require timely feedback for further therapeutic assessment and treatment planning. To overcome these challenges, we introduce U-ResNet, a residual-based neural network designed as a surrogate model for accurately predicting pressure, WSS, velocity fields, and vorticity fields in two-dimensional (2-D) stenotic channels across a wide range of Reynolds numbers. U-ResNet leverages residual learning and multi-scale feature extraction to capture sharp hemodynamic gradients induced by stenosis while maintaining strong generalization across vessel sizes and flow regimes. Unlike traditional CFD methods, which require iterative meshing and solver convergence over hours or days, U-ResNet achieves sub-second inference speeds with predictive accuracy comparable to high-fidelity simulations. All simulations and model evaluations in this work are based on steady-state flow conditions.

The novelty of this research lies in U-ResNet’s ability to simultaneously predict multiple hemodynamic quantities with high accuracy while ensuring scalability across varying anatomical geometries and Reynolds numbers - a capability rarely achieved by existing models. Additionally, we systematically evaluate U-ResNet against state-of-the-art neural network architectures to demonstrate its advantages in predictive accuracy, computational efficiency, and generalizability. The results indicate that U-ResNet functions as an efficient surrogate model that complements traditional CFD methods by providing rapid and accurate approximations for real-time or large-scale hemodynamic assessments, while not replacing the need for high-fidelity simulations in novel or complex scenarios.

The remainder of this paper is organized as follows: Section~\ref{model} provides a detailed description of the physical and mathematical models underlying the problem. Section~\ref{numerical} outlines the numerical methods employed for data generation and deep learning model training. Numerical results are presented and discussed in Section~\ref{results}. Discussions are drawn in Section~\ref{discussion}. Finally, Section~\ref{conclusions} concludes with key findings and their implications for future research and clinical applications.

%======================================================
\section{Model description}\label{model}
In this study, blood flow within the lumen is modeled as incompressible, with the vessel walls assumed to be rigid. Blood is treated as a viscous Newtonian fluid. Under these conditions, the governing equations for the fluid flow are derived from the fundamental principles of conservation of mass and momentum \cite{pope2000,sagaut2006}: 
\begin{equation}
\bigtriangledown \cdot \textbf{\textit{u}} =0,
\end{equation}
\begin{equation}
\rho \frac{\partial \textbf{\textit{u}}}{\partial \mathit{t} } + \rho \textbf{\textit{u}}\cdot \bigtriangledown \textbf{\textit{u}}=-\bigtriangledown\mathit{p} +\mu\bigtriangledown ^{2}\textbf{\textit{u}}+\textbf{\textit{f}},
\end{equation}
where $\textbf{\textit{u}}$ is the fluid velocity, $\rho$ is the constant fluid density, $p$ is the pressure, $\mu$ is the dynamic viscosity, and $\textbf{\textit{f}}$ is the body force. 

\begin{figure}\centering
	\begin{subfigure}{\textwidth}
        \begin{overpic}[width=\linewidth]{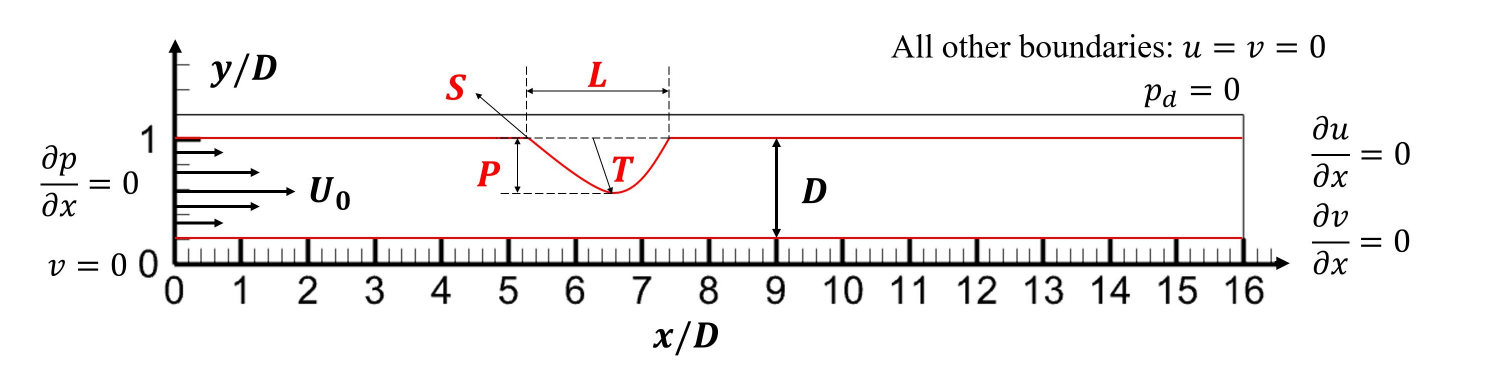}
            \put(3,20){\small (a)}  
        \end{overpic}
	\end{subfigure}%
    
	\begin{subfigure}{\textwidth}
        \begin{overpic}[width=0.9\linewidth]{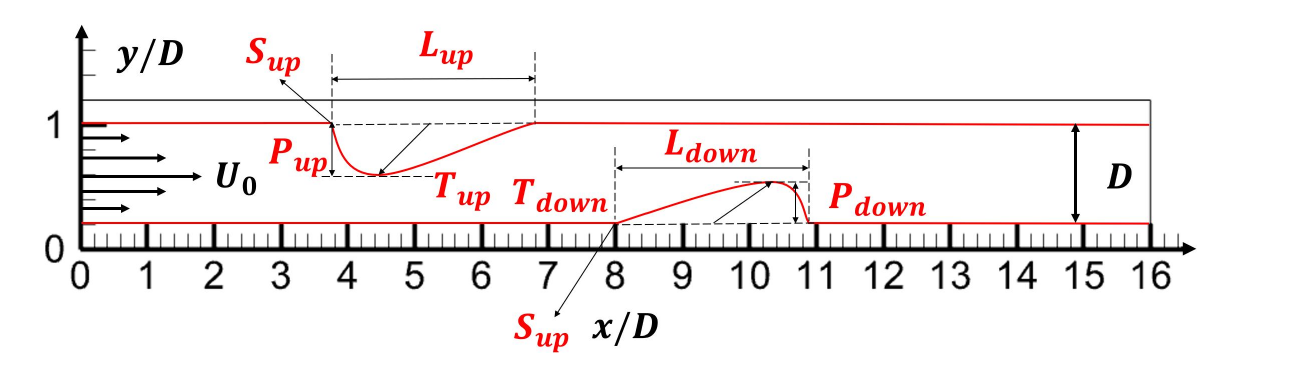}
            \put(-2,23){\small (b)}  
        \end{overpic}
	\end{subfigure}%
	
	\caption{Schematic diagram of fluid flow through 2-D asymmetric stenosis: (a) unilateral stenosis, (b) bilateral stenosis. Red lines represent the upper and lower channel walls. $D$ is the diameter of the non-stenosed channel.}\label{fig:1}
\end{figure}

Stenosis - the pathological narrowing of blood vessels - is a critical contributor to cardiovascular morbidity, altering hemodynamic forces such as WSS and pressure gradients that drive atherosclerosis progression and plaque rupture \citep{ku1997blood,tarbell2014fluid}. This study focuses on 2-D incompressible and steady-state flow through asymmetric stenotic channels (as illustrated in Fig.~\ref{fig:1}), chosen for their ability to capture essential flow physics while enabling systematic exploration of geometric variability. To model the stochastic nature of vascular narrowing, we investigate both unilateral and bilateral stenosis configurations, representing localized and diffuse arterial disease, respectively. The unilateral stenosis geometry (Fig.~\ref{fig:1} (a)) is mathematically defined by the wall profile:
\begin{equation}
y_{up}=-P\cdot \sin [ \frac{\pi }{L} \cdot (x-S)+T\cdot \sin(\frac{\pi }{L}\cdot (x-S))] +Y_{up},
\label{eq:yup1}
\end{equation}
\begin{equation}
y_{down}=Y_{down},
\label{eq:ydw1}
\end{equation}
where $P \in [0.15D, 0.55D]$ is the peak height of stenosis at the upper wall, $L \in [0.5\pi D, 2\pi D]$ is the stenosis length, $T \in [-D, D]$ represents the stenosis tilt, $S \in [2D, 8D]$ denotes the starting position of stenosis, $D$ is the diameter of the non-stenosed channel, $Y_{up}$ and $Y_{down}$ are the y-coordinates of the upper and lower walls, respectively. The stenosis geometries employed in this study were designed to represent physiologically relevant vascular narrowing based on clinical observations and established computational models. The range of diameter reduction (15-55\%) corresponds to mild-to-moderate stenoses commonly encountered in clinical practice \citep{lee2019identification}. The length-to-diameter ratios ($L \in [0.5\pi D, 2\pi D]$) were selected based on clinical imaging studies that typically report coronary plaque lengths of 1-3 vessel diameters, or approximately 8-15 mm in standard coronary arteries \citep{lee2019identification}. The inclusion of tilt parameters ($T \in [-D, D]$) enables representation of both concentric and eccentric lesions, reflecting the morphological heterogeneity of atherosclerotic disease commonly observed in patients \citep{lee2019identification,chen2024hemodynamic}. This comprehensive parameterization approach ensures systematic coverage of clinically relevant stenosis profiles while enabling controlled analysis of hemodynamic effects across geometric variations.

For bilateral stenosis (Fig.~\ref{fig:1} (b)), the upper and lower wall geometries are defined as:
\begin{equation}
y_{up}=-P_{up}\cdot \sin  [ \frac{\pi }{L_{up}} \cdot (x-S_{up})+T_{up}\cdot \sin(\frac{\pi }{L_{up}}\cdot (x-{S_{up}}))] +Y_{up},
\label{eq:yup2}
\end{equation}
\begin{equation}
y_{down}=P_{down}\cdot \sin [ \frac{\pi }{L_{down}} \cdot (x-S_{down})+T_{down}\cdot \sin(\frac{\pi }{L_{down}}\cdot (x-{S_{down}}))] +Y_{down},
\label{eq:ydw2}
\end{equation}
where $P_{up}$ and $P_{down}$ $\in [0.05D, 0.30D]$ are peak stenosis heights (clinically relevant for 10–60\% diameter reduction \citep{sarabian2022physics}) for the upper and lower channel wall, respectively. $L_{up}$ and $L_{down}$ $\in [0.5\pi D, 2\pi D]$ are stenosis axial lengths, $T_{up}$ and $T_{down}$ $\in [-D, D]$ are tilt parameters inducing flow asymmetry, and $S_{up}$ and $S_{down}$ $\in [2D, 8D]$ denotes stenosis starting positions. A random seed to generate random numbers within the above ranges. The downstream-oriented stenosis geometry (e.g., the shape of the stenosis on the upper wall in Fig.~\ref{fig:1} (b)) was intentionally designed to amplify adverse pressure gradients and flow separation, providing a rigorous test case for neural networks to resolve sharp hemodynamic transitions. While physiological stenoses often exhibit upstream asymmetry, this idealised configuration enables systematic evaluation of model performance under controlled gradient-rich conditions. This methodology is consistent with foundational hemodynamic studies that employ idealised geometries to probe flow physics \citep{ahmed1983velocity,ahmed1984pulsatile,westein2013atherosclerotic}.

  %======================================================
\section{Numerical methods}\label{numerical}
The hemodynamics of steady stenotic flows in this study are simulated using an in-house and high-fidelity immersed boundary-lattice Boltzmann method (IB-LBM) solver \citep{huang2021transition, huang2022streamline}, which combines the computational efficiency of LBM with the geometric flexibility of IBM to handle complex vascular geometries. The 2-D nine discrete velocity (D2Q9) LBM with multi-relaxation-time (MRT) model is adopted to simulate the fluid dynamics. The IBM is employed to enforce the no-slip and no-penetration boundary conditions at the fluid-structure interface. 

\subsection{Lattice Boltzmann method for fluid dynamics}
In the LBM, the macroscopic fluid dynamics emerge from the statistical behaviour of particles, which are described by the distribution function $g_i\left(\boldsymbol{x},t\right)$. The evolution of the distribution function $g_i(\boldsymbol{x},t)$, representing particles moving with discrete velocity $\boldsymbol{e}_i$, follows \citep{Lallemand2000,Luo2011}, 
\begin{equation}
g_i\left(\boldsymbol{x}+\boldsymbol{e}_i\Delta t, t+\Delta t\right)-g_i\left(\boldsymbol{x},t\right)=\Omega_i\left(\boldsymbol{x},t\right)+ \Delta t G_i, \label{LBE}
\end{equation}
where $\boldsymbol{x}$ is the particle position, $t$ is the time, $\Delta t$ is the time increment, $\Omega_i(\boldsymbol{x},t)$ is the collision operator, and $G_i$ is the forcing term accounting for the body force $\boldsymbol{f}$. The D2Q9 model \cite{Lallemand2000} is employed, utilizing nine discrete velocity vectors on a square grid to resolve 2-D flow dynamics. Compared with the single-relaxation-time collision model, the MRT model has been shown to provide enhanced numerical stability \citep{Lallemand2000,Xu2018}. Therefore, the MRT collision model is adopted, which is given by \cite{Lallemand2000}:
\begin{eqnarray}
\Omega_i=-(\boldsymbol{M}^{-1}\boldsymbol{S}\boldsymbol{M})_{ij}[g_i(\boldsymbol{x},t)-g_i^{eq}(\boldsymbol{x},t)],\label{omega}
\end{eqnarray}
where $\boldsymbol{M}$ is a $9\times 9$ transform matrix, and $\boldsymbol{S}=diag(\tau_0,\tau_1,\ldots,\tau_{8})^{-1}$ is a non-negative diagonal $9\times 9$ relaxation matrix. The determination of $\boldsymbol{S}$ in 2-D model can be found in  \citet{Luo2011}. The equilibrium distribution function $g_i^{eq}$ is defined as
\begin{eqnarray}
g^{eq}_i=\rho\omega_i\Big[1+\frac{\boldsymbol{e_i}\cdot\boldsymbol{u}}{c^2_s}+
\frac{\boldsymbol{uu}:(\boldsymbol{e}_i\boldsymbol{e}_i-c^2_s\boldsymbol{I})}{2c^4_s}\Big],
\label{LBM1}
\end{eqnarray}
where $c_s=\Delta x/(\sqrt{3}\Delta t)$ is the speed of sound, $\Delta x$ is the lattice spacing, $\boldsymbol{I}$ is the unit tensor, and the weighting factors $\omega_i$ are given by $\omega_0=4/9$,
$\omega_{1-4}=1/9$ and $\omega_{5-8}=1/36$. The velocity $\boldsymbol{u}$, mass density $\rho$ and pressure $p$ can be obtained according to
\begin{eqnarray}
\rho=\sum_{i}g_i,\quad p=\rho c_s^2, \quad \boldsymbol{u}=\left(\sum_{i}\textbf{e}_i g_i+\frac{1}{2}\boldsymbol{f}\Delta t\right)\Bigg/\rho, \label{LBM2}
%\label{LBM3}
\end{eqnarray}
The force scheme proposed in  \citet{Guo2002} is adopted to determine $G_i$
\begin{eqnarray}
G_i=[\boldsymbol{M}^{-1}(\boldsymbol{I}-\boldsymbol{S}/2)\boldsymbol{M}]_{ij}F_i, \label{LBM4}\\
F_i=\Big(1-\frac{1}{2\tau}\Big)\omega_i\Big[\frac{\boldsymbol{e}_i-\boldsymbol{u}}{c^2_s}+
\frac{\boldsymbol{e}\cdot\boldsymbol{u}}{c^4_s}\boldsymbol{e}_i\Big]\cdot\boldsymbol{f},\label{LBM5}
\end{eqnarray}
where $\tau$ is the non-dimensional relaxation time. 

\subsection{Immersed boundary method for fluid-structure interface coupling}
A diffused-interface IBM \citep{GOLDSTEIN1993, HuangTian2019, huang2021transition} is employed to enforce no-slip/no-penetration conditions at fluid-structure interfaces while maintaining computational efficiency on Cartesian grids. This method introduces a body force $\boldsymbol{f}$ into the Navier-Stokes equations to mimic a boundary condition according to,
\begin{equation}
\boldsymbol{f}(\boldsymbol{x},t)=-\int \boldsymbol{F}_{ib}(s,t) \delta (\boldsymbol{x}-\boldsymbol{X}(s,t))ds,\label{ibm1}
\end{equation}
\begin{equation}
\boldsymbol{F}_{ib}(s,t)=\kappa \rho(\boldsymbol{x},t)(\boldsymbol{U}_{ib}(s,t)-\boldsymbol{U}(s,t)),\label{ibm2}
\end{equation}
\begin{equation}
\boldsymbol{U}_{ib}(s,t)=\int \boldsymbol{u}(x,t)\delta(\boldsymbol{x}-\boldsymbol{X}(s,t))d\boldsymbol{x},\label{ibm3}
\end{equation}
where $\boldsymbol{F}_{ib}(s,t)$ is the Lagrangian force density, $ds$ is the arc length of the immersed boundary, $\delta (\boldsymbol{x}-\boldsymbol{X}(s,t))$ is Dirac's delta function, $\boldsymbol{x}$ is the coordinate of the fluid lattice nodes, $\boldsymbol{X}$ is the coordinate of the structure (i.e. the channel wall here). The feedback coefficient $\kappa=5.2$ is calibrated to minimize boundary penetration while maintaining numerical stability with improved computational efficiency \citep{huang2022streamline,huang2024self}. $\boldsymbol{U}_{ib}(s,t)$ is the immersed boundary velocity interpolated at the immersed boundary (i.e. the channel wall here), and $\boldsymbol{U}(s,t)$ is the velocity of the wall ($\boldsymbol{U}(s,t)=0$ for a static and rigid channel wall). The 4-point discrete delta function $\delta_h(\boldsymbol{x}-\boldsymbol{X}(s,t))$ is used to approximate the Dirac delta function $\delta(\boldsymbol{x}-\boldsymbol{X}(s,t))$, ensuring numerical stability and accuracy in resolving fluid-structure interactions \cite{Peskin2002,huang2022streamline}. 
\begin{equation}
        \delta_h( \boldsymbol{x}-\boldsymbol{X}(s,t))=\frac{1}{\Delta x \Delta y}\phi\left(\frac{x-X(s,t)}{\Delta x}\right) \phi\left(\frac{y-Y(s,t)}{\Delta y}\right),\label{delta}
\end{equation}

\begin{equation}\label{Eq:Delta_r}
\phi(\emph{r})=\left\{ 
\begin{aligned}
 &\frac{1}{8}\left(3-2|r|+\sqrt{1+4|r|-4r^2}\right),   &    0\leq |r|\leq 1,  \\
&\frac{1}{8}\left(5-2|r|-\sqrt{-7+12|r|-4r^2}\right), &    1\leq |r|\leq 2, \\
&                                              0,     &    |r|>2,
\end{aligned}
\right.
\end{equation} 
where $r=({x-X(s,t)})/{\Delta x}$ for $x$-component, and $({y-Y(s,t)})/{\Delta y}$ for $y$-component.

The computational domain spans $16D \times 1.4D$. At the inlet, a fully-developed Poiseuille velocity profile with mean velocity $U_0$ is prescribed, with $\partial p/\partial x = 0$ and $v=0$. The fully developed velocity profile assumption at the inlet represents a standard practice in computational hemodynamics that enables standardized model comparison and validation. This standardization facilitates reproducible research and allows isolation of geometric effects without confounding variables from inlet flow variability. While this assumption simplifies the upstream flow conditions, the stenosis itself acts as a flow disruptor that fundamentally alters flow dynamics regardless of inlet assumptions. The hemodynamic changes observed in this study are primarily attributable to stenosis-induced flow separation and recirculation rather than inlet flow development, making this assumption appropriate for the study's objectives of investigating geometry-flow interactions in controlled parametric conditions.

At the downstream outlet, zero pressure ($p_d=0$) with convective outflow conditions ($\partial u/\partial x = 0 = \partial v/\partial x = 0$). At the channel walls, no-slip condition enforced via the IBM. Grid independence studies determined optimal resolutions of $dx=0.01D$ for the fluid domain and $ds=0.005D$ (i.e., wall grid resolution $N_g=3201$) for the immersed boundary discretization \citep{huang2021transition,huang2022streamline}. The simulations were deemed converged when the normalized iterative velocity error fell below a threshold of $5.0\times10^{-4}$, calculated as:
\begin{equation}
    E_{u} =  \frac{\sum_{i=1}^N \left |\sqrt{\boldsymbol{u}^2_{n+1}(i)+\boldsymbol{v}^2_{n+1}(i)} - \sqrt{\boldsymbol{u}^2_{n}(i)+\boldsymbol{v}^2_{n}(i)} \right |}{\sum_{i=1}^N \sqrt{\boldsymbol{u}^2_{n+1}(i)+\boldsymbol{v}^2_{n+1}(i)}} \leq 5.0\times10^{-4},
\end{equation}
where $u_n(i)$ and $v_n(i)$ are the $x-$ and $y-$velocity components at grid point $i$ for iteration $n$, $N$ is the total number of grid points. This criterion ensures the velocity field stabilizes spatially, with relative changes in velocity magnitude below $0.05\%$ across successive iterations.

The IB-LBM framework employed in this study has undergone extensive validation across diverse flow regimes and geometric configurations. For internal flows, the method has been rigorously validated in both 2-D \citep{huang2021transition, huang2021low} and 3-D \citep{huang2022streamline, huang2024self} collapsible channel/tube configurations. These studies demonstrated the framework's capability to accurately capture complex flow physics including flow separation, vortex shedding, and fluid-structure interactions. The method's versatility extends to external flow applications, including hydrodynamic simulations of flapping foils \citep{liu2023partial} and aerodynamic modeling of flapping wings \citep{huang2023power}. Particularly relevant to hemodynamic applications, the framework has been validated for microscale flows involving blood cells in viscoelastic fluids \citep{ma2023effects}, accurately reproducing cell deformation and migration phenomena observed in microfluidic experiments. For high-Reynolds-number regimes, the algorithm has successfully simulated turbulent channel flows \citep{wang2024wall}, capturing near-wall turbulence statistics and coherent structures. These comprehensive validation studies across multiple flow regimes and length scales demonstrate the robustness and accuracy of the IB-LBM approach for the stenotic flow configurations investigated in the present work. While IBMs simplify grid generation for complex geometries, their diffuse interface formulation introduces systematic errors in near-wall gradient resolution. To mitigate this, a high-fidelity interpolation scheme is employed that reconstructs wall shear stress from resolved velocity fields \citep{huang2021transition, huang2022streamline}, achieving sub-2\% relative error compared to body-fitted mesh benchmarks \citep{huang2022streamline}. This approach retains IBM’s geometric flexibility while ensuring clinically relevant accuracy in WSS prediction.

\subsection{Deep learning models}
Four deep learning model architectures - U-Net, U-ResNet, FNO, and U-Net enhanced Fourier Neural Operator (UFNO) - are employed to predict pressure, WSS, velocity, and vorticity fields for asymmetric stenotic channels. The models take two inputs: (1) the channel geometry, represented as a binary mask encoding stenotic morphology, and (2) the Reynolds number ($Re\in[200,800]$), which governs flow regime transitions. The networks output spatially resolved fields across the computational domain, enabling simultaneous multi-physics prediction with sub-second inference times. All architectures are trained using a normalized $L_2$ loss to minimize relative errors between predicted ($y$) and ground-truth ($\hat{y}$) fields:
\begin{equation}
L_{2} (y,\hat{y})=\frac{\left \| y-\hat{y}  \right \| _{2}  }{\left \| \hat{y}  \right \| _{2} } =\frac{\sqrt{\sum_{N}^{i=1} (y_{i}-\hat{y_{i}} )^{2} } }{\sqrt{\sum_{N}^{i=1} \hat{y_{i}} ^{2} } }, 
\end{equation}
where $N$ is the total number of grid points.

\subsubsection{U-Net}

\begin{figure}[htbp]\centering
	\includegraphics[width=1\textwidth]{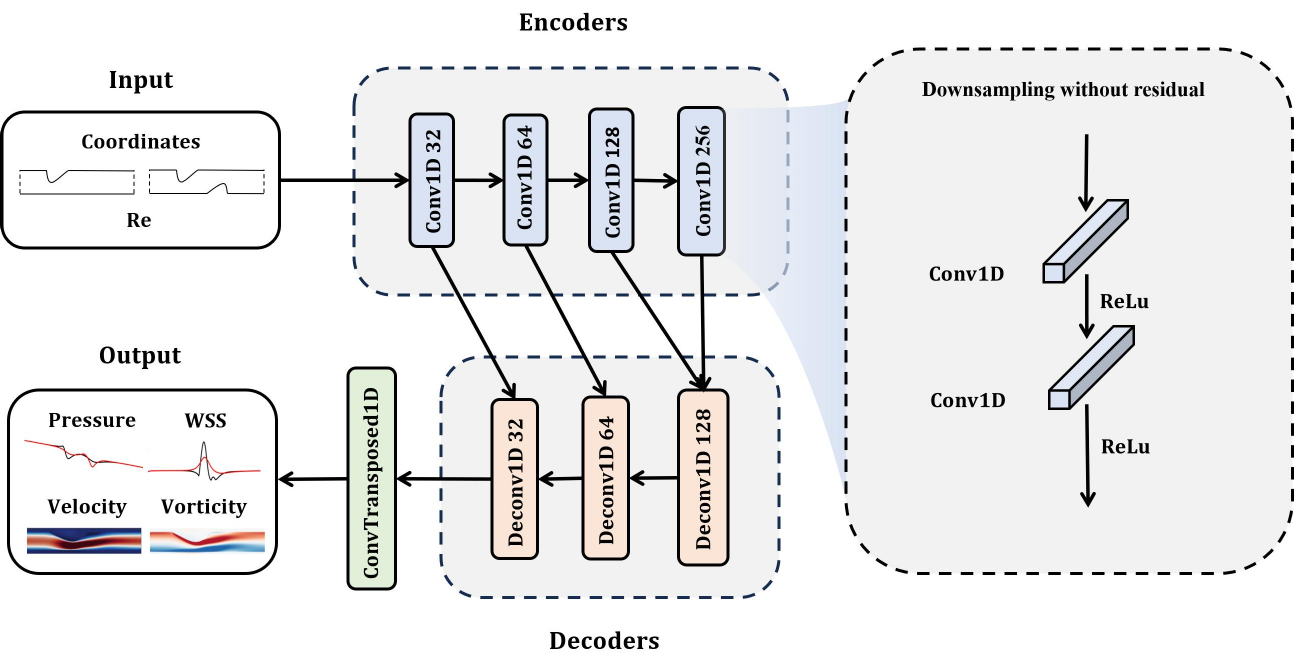}
	\caption{Architecture of the U-Net model, designed to predict physical quantities such as pressure, WSS, velocity, and vorticity fields based on input channel geometry and Reynolds number. Four separate U-Net models were trained independently for each physical quantity to optimize accuracy and computational efficiency.}\label{fig:2}
\end{figure}

The U-Net architecture \cite{ronnebergerUnetConvolutionalNetworks2015} is adapted for hemodynamic prediction via a one-dimensional (1-D) convolutional framework that maps input features to multi-physics outputs, as shown in Fig.~\ref{fig:2}. Four independent U-Net models are trained to predict pressure, WSS, velocity, and vorticity fields, ensuring specialized feature learning for each quantity:
\begin{equation}
\mathcal{N}_\theta : a(x) \mapsto z(x) \in \mathbb{R}^{d_u},
\end{equation}
where $\mathcal{N}_\theta$ represents the neural network with trainable parameters $\theta$, inputs $a(x)=[x, \mathrm{Re}] \in \mathbb{R}^2$ combine spatial coordinates $x \in D \subset \mathbb{R}$ and Reynolds number $Re$, outputs $z(x)$ represent pressure, WSS, velocity, or vorticity fields, and $d_u$ is the dimension of the output physical quantities.

The architecture consists of three principal components:

1) Encoder (contracting path):
The input is processed through $L$ downsampling layers, progressively extracting hierarchical feature representations:
\begin{equation}
f_0(x) = a(x), \quad
f_i(x) = \phi\left( \text{Conv}_{2,i}(\phi(\text{Conv}_{1,i}(f_{i-1}(x)))) \right), \quad i = 1,\dots,L,
\end{equation}
where $\text{Conv}_{1,i}$ and $\text{Conv}_{2,i}$ are 1-D convolution operations with kernel size 5 and stride 5, and $\phi$ is the ReLU (Rectified Linear Unit) activation function, defined as $\text{ReLU}(x) = \max(0, x)$ \cite{agarapDeepLearningUsing2019}.

2) Decoder (expanding path):  
Each decoder stage performs transposed convolution (upsampling), followed by two convolution layers, with skip connections from the encoder preserving fine-scale features:
\begin{equation}
g_j(x) = \phi\left( \text{Conv}_{2,j}\left( \phi\left( \text{Conv}_{1,j}\left([\text{Up}(g_{j-1}(x)), f_{L-j}(x)]\right) \right) \right) \right), \quad j = 1,\dots,L,
\end{equation}
where $\text{Up}(\cdot)$ denotes transposed convolution and $[\cdot,\cdot]$ represents channel-wise concatenation.

3) Output:
The final decoder output is projected to the target quantity space via a transposed convolution layer:
\begin{equation}
z(x) = \text{Conv}^{T}_{1\times1}(g_L(x)) \in \mathbb{R}^{d_u}.
\end{equation}

\subsubsection{U-ResNet}
\begin{figure}[htbp]\centering
	\includegraphics[width=1\textwidth]{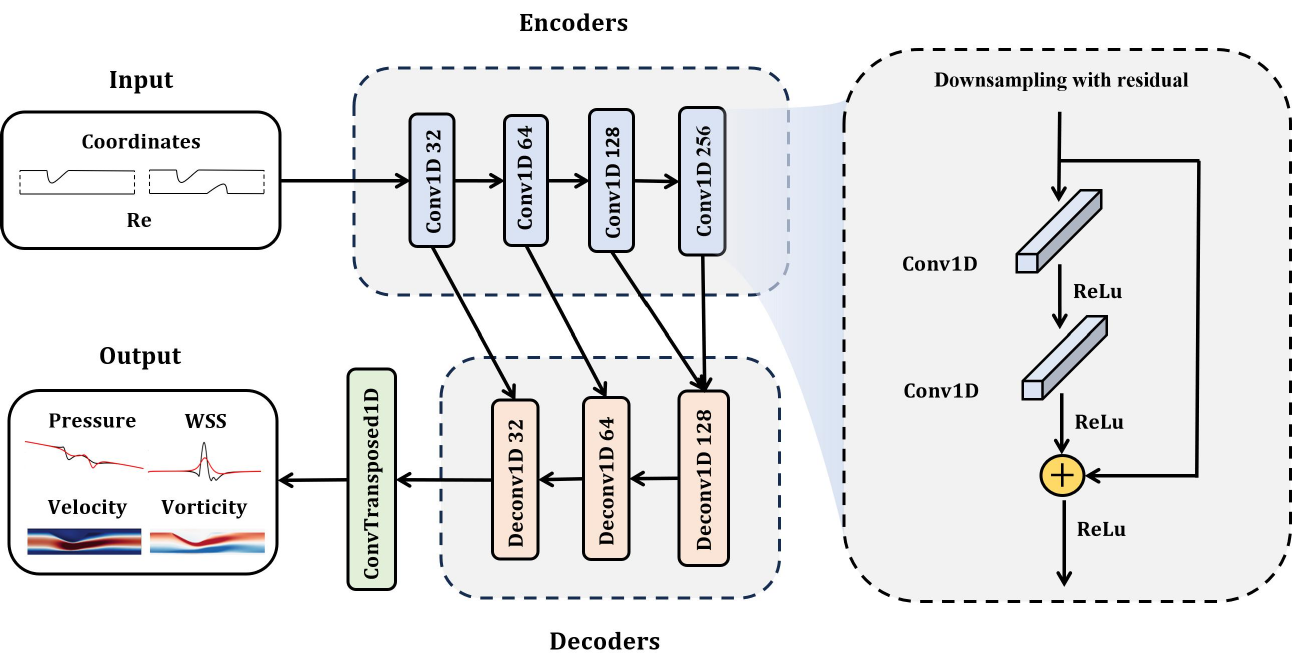}
	\caption{Architecture of the U-ResNet model for hemodynamic prediction, which maps input channel geometry and Reynolds number (Re) to spatially resolved physical fields. The network employs encoder-decoder blocks with residual connections to capture multi-scale flow features, enabling accurate reconstruction of hemodynamic forces.}\label{fig:3}
\end{figure}

To enhance gradient flow and feature representation capacity, we implement a 1-D U-Net architecture augmented with residual connections (U-ResNet) \cite{heDeepResidualLearning2015}. This architecture augments the standard U-Net framework with residual connections, enabling deeper network training while mitigating the vanishing gradient problem - particularly critical for capturing sharp hemodynamic gradients in stenotic regions. As illustrated in Fig.~\ref{fig:3}, U-ResNet learns mappings from spatially distributed inputs (geometry and Reynolds number) to target hemodynamic fields.

The U-ResNet architecture comprises the following components:

1) Encoder with residual block:  
The contracting path employs residual blocks to mitigate vanishing gradients while preserving high-frequency flow features (e.g., boundary layer velocity profiles):
\begin{equation}
f_0(x) = a(x), \quad
f_i(x) = \phi\left( \text{Conv}_{2,i}(\phi(\text{Conv}_{1,i}(f_{i-1}(x)))) + \text{Down}(f_{i-1}(x)) \right), \quad i = 1,\dots,L,
\end{equation}
where $\text{Down}(f_{i-1}(x))$ is a $1\times1$ convolution used when input and output channels differ. This residual path ensures better gradient propagation.

2) Decoder and output: 
The decoder and output projection layers are identical to the U-Net:
\begin{equation}
g_j(x) = \phi\left( \text{Conv}_{2,j}\left( \phi\left( \text{Conv}_{1,j}\left([\text{Up}(g_{j-1}(x)), f_{L-j}(x)]\right) \right) \right) \right), \quad j = 1,\dots,L,
\end{equation}
\begin{equation}
z(x) = \text{Conv}^{T}_{1\times1}(g_L(x)) \in \mathbb{R}^{d_u}.
\end{equation}

The residual connections in U-ResNet provide three key advantages for hemodynamic prediction. Firstly, residual pathways mitigate vanishing gradients during backpropagation, enabling more effective training on sharp WSS transitions that occur at stenosis peaks. Secondly, identity mappings allow direct propagation of early-layer features to deeper layers, preserving boundary information critical for accurate near-wall flow prediction. Finally, the architecture can learn more complex flow patterns without degradation, capturing recirculation zones and post-stenotic flow instabilities that emerge at higher Reynolds numbers.

% This architecture ensures both local feature extraction via convolutions and global context preservation through residual and skip connections. The padding and cropping operations guarantee dimension alignment between encoder and decoder branches \cite{ronnebergerUnetConvolutionalNetworks2015}. The addition of residual connections in the encoder allows U-ResNet to learn deeper and more expressive feature representations while maintaining architectural symmetry with the U-Net \cite{heDeepResidualLearning2015}.

\subsubsection{FNO}
\begin{figure}[htbp]\centering
	\includegraphics[width=1\textwidth]{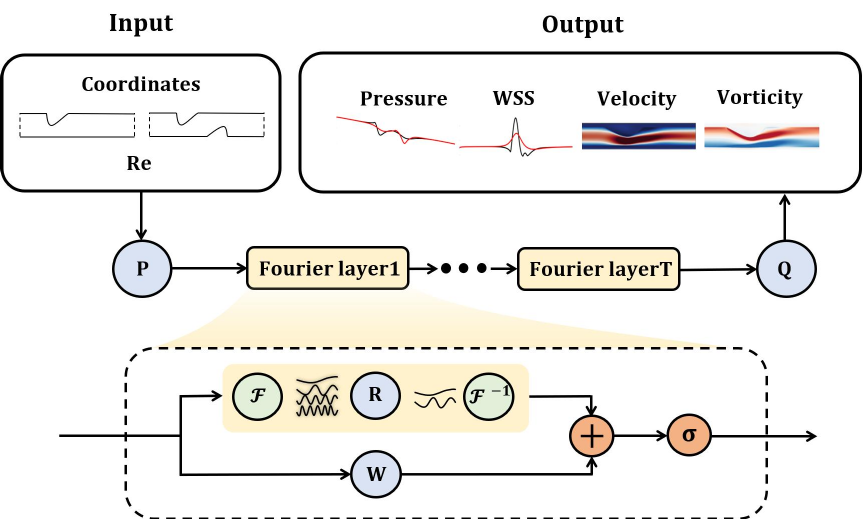}
	\caption{Architecture of the FNO model for hemodynamic prediction, mapping input channel geometry and Reynolds number (Re) to spatially resolved physical fields.}\label{fig:4}
\end{figure}

FNO seeks to learn mappings between infinite-dimensional function spaces based on a finite collection of input-output data pairs \cite{liFourierNeuralOperator2022a, liFourierNeuralOperator2021}. Unlike traditional neural networks that operate on fixed-dimensional vectors, FNO directly parameterizes operators between function spaces, enabling mesh-independent prediction of physical fields.

Consider a bounded open domain $D \subset \mathbb{R}^d$, with Banach spaces $\mathcal{A} = \mathcal{A}(D; \mathbb{R}^{d_a})$ and $\mathcal{U} = \mathcal{U}(D; \mathbb{R}^{d_u})$ representing input and output function spaces, respectively \cite{beauzamy2011introduction}. The FNO constructs a parameterized mapping $\theta \in \Theta$ that approximates the operator $\mathcal{A} \to \mathcal{U}$, with optimal parameters $\theta^{\dagger}$ obtained through empirical risk minimization on training data \cite{vapnik1999overview}. The underlying architecture adopts an iterative scheme of function updates $v_0 \mapsto v_1 \mapsto \dots \mapsto v_T$, where each $v_j \in \mathbb{R}^{d_v}$ for $j = 0, 1, \dots, T$ represents an intermediate functional representation which is set to 4 in our FNO structure. As illustrated in Fig.~\ref{fig:4}, the FNO architecture comprises three key computational stages.

1) Lifting operator: The input function $a \in \mathcal{A}$ is initially lifted to a higher-dimensional representation via a shallow fully connected network, yielding $v_0(x) = P(a(x))$ \cite{liFourierNeuralOperator2021}.

2) Iterative kernel integration: The lifted representation $v_0(x)$ is then iteratively updated as follows: 
\begin{equation}
v_{t+1} (x)= \sigma(W v_{t}(x)+(\mathcal{K}  (a;\phi )v_{t})(x)),\quad   \forall x\in D.
\label{eq:1}
\end{equation}
where $\mathcal{K}:\mathcal{A} \times \Theta _{\mathcal{K}}  \to \mathcal{L} (\mathcal{U}(D;\mathbb{R}^{d_{v}}),\mathcal{U}(D;\mathbb{R}^{d_{v}}))$ maps to  bounded linear operators on $\mathcal{U}(D;\mathbb{R}^{d_{v}}))$ and is parameterized by $\phi \in \Theta _{\mathcal{K} } ,\ W:\mathbb{R} ^{d_{v} } \to \mathbb{R} ^{d_{v} }$ is a linear transformation, and $\sigma : \mathbb{R} \to \mathbb{R}$ is a non-linear activation function, where ReLU is adopted in our model.

3) Projection operator: The final output $u \in \mathcal{U}$ is produced via a projection of the last iteration: $u(x) = Q(v_T(x))$, where $Q:\mathbb{R}^{d_v} \to \mathbb{R}^{d_u}$ is a fully connected neural network.

The key innovation of FNO lies in parameterizing the kernel integral operator $\mathcal{K}$ in the Fourier domain. Let $\mathcal{F}$ and $\mathcal{F}^{-1}$ denote the Fourier and inverse Fourier transforms of a function $f:D \to \mathbb{R}^{d_v}$, respectively. By substituting the kernel integral operator in Eq.~\eqref{eq:1} with a convolution operator in Fourier space, the operator can be reformulated as:
\begin{equation}
(\mathcal{K}(\phi)v_t)(x) = \mathcal{F}^{-1} \left(R_\phi \cdot \mathcal{F}(v_t)\right)(x), \quad \forall x \in D.
\end{equation}
here, $R_\phi$ is the Fourier transform of a periodic kernel $\mathcal{K}:\bar{D} \to \mathbb{R}^{d_v \times d_v}$, parameterized by $\phi \in \Theta_{\mathcal{K}}$. The frequency modes are indexed by $k \in \mathbb{Z}^d$. A truncated version of the Fourier series is used with cutoff index $k_{\max} = \left| Z_{k_{\max}} \right|$, where $Z_{k_{\max}} = \{k \in \mathbb{Z}^d : |k_j| \le k_{\max,j},\ j=1,\dots,d\}$.

When the domain $D$ is discretized into $n \in \mathbb{N}$ points, the function $v_t \in \mathbb{R}^{n \times d_v}$ can be transformed via the Fourier transform to $\mathcal{F}(v_t) \in \mathbb{C}^{n \times d_v}$, and after truncation, $\mathcal{F}(v_t) \in \mathbb{C}^{k_{\max} \times d_v}$. The kernel $R_\phi \in \mathbb{C}^{k_{\max} \times d_v \times d_v}$ is a complex-valued tensor containing learnable Fourier coefficients \cite{liFourierNeuralOperator2021}. The application of $R_\phi$ to the truncated transform is computed by:
\begin{equation}
(R_\phi \cdot \mathcal{F}(v_t))_{k,l} = \sum_{j=1}^{d_v} R_{\phi k,l,j} \mathcal{F}(v_t)_{k,j},\quad k = 1, \dots, k_{\max},\quad j = 1, \dots, d_v.
\end{equation}

\subsubsection{UFNO}
\begin{figure}[htbp]\centering
	\includegraphics[width=1\textwidth]{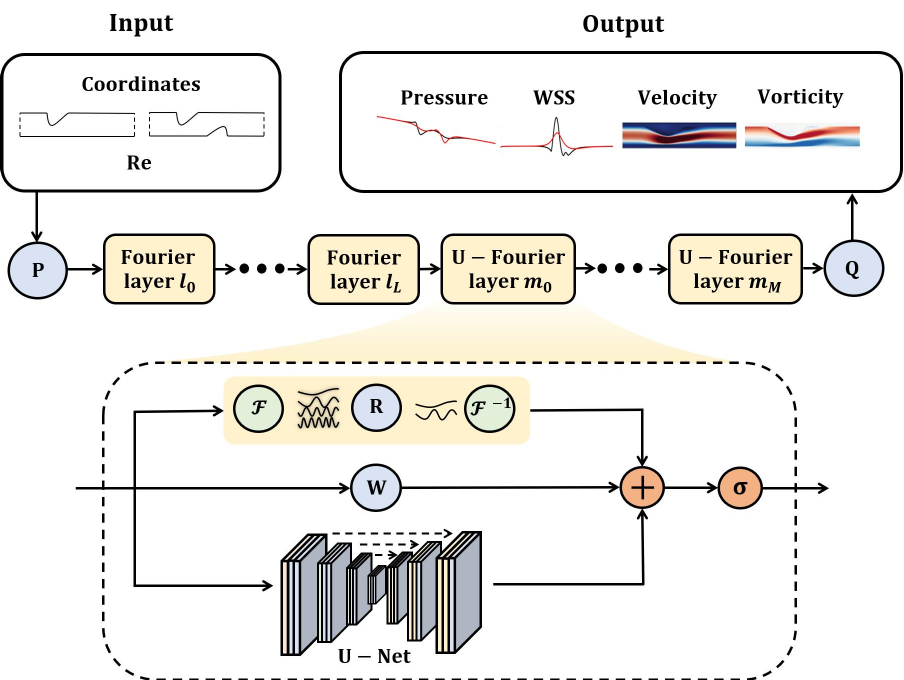}
	\caption{Architecture of the UFNO, designed to predict spatially resolved pressure, WSS, velocity, and vorticity fields in stenotic flows. The model integrates U-Net’s localized feature extraction with FNO spectral convolutions to resolve multi-scale hemodynamic patterns across Reynolds numbers.}\label{fig:5}
\end{figure}

Building upon the Fourier Neural Operator (FNO) framework, we implement a hybrid U-Net enhanced Fourier Neural Operator (UFNO) \citep{wenUFNOEnhancedFourier2022} that synergizes spectral convolutions with localized feature extraction. As shown in Fig.~\ref{fig:5}, the architecture employs four standard Fourier layers followed by four U-Fourier layers, leveraging both global spectral efficiency and local spatial precision. This hybrid architecture is designed to capture both global flow structures and fine-scale features, making it well-suited for complex fluid dynamics problems.

The UFNO comprises three principal computational stages:

1) Feature lifting: The input observation $a(x)$ is initially lifted into a higher-dimensional feature space via a fully connected neural network transformation $P$, yielding $v_{l_0}(x) = P(a(x))$.

2) Multi-scale feature processing: The lifted representation is then processed through a series of spectral layers, first involving standard Fourier layers, followed by U-Fourier layers, in the sequence $v_{l_0} \mapsto \dots \mapsto v_{l_L} \mapsto v_{m_0} \mapsto \dots \mapsto v_{m_M}$. Here, $v_{l_j}$ for $j = 0,1,\dots,L \ (L=4)$ and $v_{m_k}$ for $k = 0,1,\dots,M \ (M=4)$ denote intermediate representations at each layer, where each function maps to $\mathbb{R}^c$ with $c$ denoting the channel dimension \cite{wenUFNOEnhancedFourier2022}.

3) Output projection: The final output is obtained by projecting the last U-Fourier representation $v_{m_M}(x)$ back to the original function space using another fully connected neural network $Q$, resulting in $z(x) = Q(v_{m_M}(x))$.

Each U-Fourier layer is characterized by a spectral convolution defined in the Fourier domain as follows:
\begin{equation}
(R_\phi \cdot \mathcal{F}(v_t))_{k,l} = \sum_{j=1}^{d_v} R_{\phi k,l,j} \mathcal{F}(v_t)_{k,j},\quad k = 1, \dots, k_{\max},\quad j = 1, \dots, d_v,
\end{equation}
where $\mathcal{K}$ denotes the kernel integral operator introduced earlier, $\mathcal{U}$ represents a U-Net based convolutional neural network module, and $W$ is a learnable linear operator. The activation function $\sigma$ introduces nonlinearity, thereby enhancing the expressivity of each U-Fourier layer \cite{wenUFNOEnhancedFourier2022}. Here, ReLU is adopted as the activation function. All neural network models were trained and tested on an NVIDIA A100 GPU, with computations performed on a system equipped with an Intel(R) Xeon(R) Gold 6240 CPU @ 2.60GHz. A detail hyperparameter optimization for the four deep learning models are provided in Appendix \ref{appendix1}. 

Neural operators learn mappings between infinite-dimensional function spaces, whereas traditional neural networks learn mappings from finite-dimensional vectors to vectors. This fundamental difference enables neural operators to achieve discretization invariance and resolution-independent predictions. In this study, U-ResNet and U-Net are conventional neural networks that process structured numerical inputs (coordinate grids and Reynolds numbers) as finite-dimensional vectors. FNO and UFNO are neural operators designed to learn operators that map one function space to another. Although the actual inputs provided to FNO/UFNO consist of discrete numerical values (spatial coordinates and Reynolds number), these are interpreted as discretized evaluations of input functions defined over the spatial domain.

\subsection{Evaluation criteria for neural networks}
To rigorously assess the predictive performance of the neural networks, four error metrics are employed: mean absolute error ($MAE$), normalized mean absolute error ($NMAE$), root mean squared error ($RMSE$), and normalized root mean squared error ($NRMSE$). These metrics quantify deviations between network predictions and high-fidelity CFD results, providing insights into both absolute and scale-invariant errors.

1) $MAE$: Measures the average magnitude of pointwise errors across the domain
\begin{equation}
MAE=\frac{1}{N} \sum_{i=1}^{N} \left | P_{i}-\hat{P}_{i}   \right |.
\end{equation}

2) $NMAE$: Provides a dimensionless error measure normalized by the dynamic range of the CFD data:
\begin{equation}
NMAE=\frac{1}{N} \sum_{i=1}^{N}\frac{ \left | P_{i}-\hat{P}_{i}   \right |}{\max(P)-\min (P)  } \times 100\%.
\end{equation}

3) $RMSE$: Emphasizes larger errors due to quadratic weighting, critical for assessing peak-sensitive quantities like WSS:
\begin{equation}
RMSE=\sqrt{\frac{1}{N} \sum_{i=1}^{N}(P_{i} -\hat{P} _{i} )^{2} }.  
\end{equation}

4) $NRMSE$: Scales $RMSE$ by the dynamic range for cross-variable comparison:
\begin{equation}
NRMSE=\frac{\sqrt{\frac{1}{N} \sum_{i=1}^{N}(P_{i} -\hat{P} _{i} )^{2} }  }{\max(P)-\min(P) } \times 100\%,
\end{equation}
where $P$ is the physical quantity of interest (e.g., pressure, vorticity), which can be either a scalar or a vector, $P_{i}$ is the physical quantity at node-\textit{i} predicted by network, and $\hat{P}_{i}$ is the corresponding ground truth value calculated by CFD, $\max(P)$ represents the maximum value and $\min(P)$ represents the minimum value from CFD, and $N$ denotes the number of grid points.

%======================================================
\section{Results} \label{results}
This section evaluates the computational efficiency of four neural network architectures, U-Net, U-ResNet, FNO, and UFNO, when predicting pressure, WSS, velocity, and vorticity fields. Model accuracy was evaluated using four key hemodynamic parameters (pressure, WSS, velocity, vorticity) derived from ground truth CFD simulations. Synthetic and CFD-generated data with systematically varied stenosis geometries were utilized to ensure comprehensive coverage of training scenarios. Statistical validation employed MAE, NMAE, RMSE, and NRMSE metrics, while spatial analysis utilized quantitative line plots and box plot distributions to assess prediction fidelity across stenotic geometries. This comprehensive framework ensures robust performance evaluation consistent with computational modeling standards. Tables~\ref{tab:1}–\ref{tab:4} compare model complexity (parameter count), GPU memory usage, and training time per epoch, providing critical insights into their suitability for real-time hemodynamic applications.

For parameter efficiency, U-ResNet and U-Net utilize approximately $5$--$8\times$ fewer parameters compared to FNO and UFNO, as shown in Tables~\ref{tab:1} and~\ref{tab:2}. In the prediction of pressure and wall shear stress (WSS), U-ResNet employs $2.07$ million parameters, whereas FNO requires $9.97$ million and $16.37$ million, respectively. This disparity becomes even more pronounced for vector field predictions: U-ResNet still uses $2.07$ million parameters for both velocity and vorticity fields, while FNO requires $16.37$ million, as summarized in Tables~\ref{tab:3} and~\ref{tab:4}. Architectural parameter counts reflect inherent design philosophies: neural operators (FNO/UFNO) prioritize spectral efficiency, while conventional networks (U-ResNet/U-Net) emphasize spatial feature resolution. Each model was optimized within its paradigm to balance accuracy and computational feasibility, ensuring fair comparison of their hemodynamic prediction capabilities.

For GPU memory consumption, U-ResNet reduces GPU memory usage by 85–98\% compared to spectral methods. For velocity prediction, U-ResNet requires $182.32 MB$ vs FNO’s $8,201.17 MB$ (Table~\ref{tab:3}). UFNO exhibits the highest memory demand ($1,153.73 MB$ for scalar fields), limiting its applicability to resource-constrained environments.

In terms of training efficiency, U-ResNet trains approximately $1.5$--$7.5\times$ faster than FNO and UFNO. For instance, when predicting vorticity fields, U-ResNet completes one training epoch in $47.49$ GPU$\cdot$s compared to FNO’s $91.64$ GPU$\cdot$s, as shown in Table~\ref{tab:4}. The hybrid architecture of UFNO introduces additional computational overhead, resulting in significantly longer training times—$2.5\times$ and $5.0\times$ longer than FNO, for pressure ($116.58$ vs. $46.26$ GPU$\cdot$s/epoch) and WSS ($252.30$ vs. $50.17$ GPU$\cdot$s/epoch), respectively.

\begin{table}[tbp]
	\begin{center}
		\caption{Computational cost comparison of four neural network architectures (U-Net, U-ResNet, FNO, and UFNO) for predicting pressure on the wall. The evaluation includes the number of parameters, GPU memory usage, and training time per epoch.}\label{tab:1}
		\begin{tabular*}{1\textwidth}{@{\extracolsep{\fill}} lccc }
			\hline\hline
			\small    
			Model & Numbers of parameters & GPU memory usage(MB) & Training time per epoch(GPU$\cdot$s) \\ \hline
			U-Net & 2029632    & 152.04   & 33.60   \\
			U-ResNet & 2072672    & 155.36    & 41.50 \\
			FNO & 9969649    & 1017.05   & 46.26     \\            
			UFNO & 11947265     & 1153.73   & 116.58 \\ \hline\hline
		\end{tabular*}%
	\end{center}
\end{table}

\begin{table}[tbp]
	\begin{center}
		\caption{Computational cost comparison of four neural network architectures (U-Net, U-ResNet, FNO, and UFNO) for predicting WSS on the wall. The evaluation includes the number of parameters, GPU memory usage, and training time per epoch.}\label{tab:2}
		\begin{tabular*}{1\textwidth}{@{\extracolsep{\fill}} lccc }
			\hline\hline
			\small    
			Model & Numbers of parameters & GPU memory usage(MB) & Training time per epoch(GPU$\cdot$s) \\ \hline
			U-Net & 2029632    & 152.04   & 33.60   \\
			U-ResNet & 2072672    & 155.36    & 41.50 \\
			FNO & 16369649    & 1531.75   & 50.17     \\            
			UFNO & 11947265     & 1153.73   & 252.30 \\ \hline\hline
		\end{tabular*}%
	\end{center}
\end{table}

\begin{table}[tbp]
	\begin{center}
		\caption{Computational cost comparison of U-ResNet and FNO architectures for velocity field prediction, evaluating key metrics including the number of parameters, GPU memory usage, and training time per epoch.}\label{tab:3}
		\begin{tabular*}{1\textwidth}{@{\extracolsep{\fill}} lccc }
			\hline\hline
			\small    
			Model & Numbers of parameters & GPU memory usage(MB) & Training time per epoch(GPU$\cdot$s) \\ \hline
			U-ResNet & 2072832    & 182.32    & 47.50 \\        
			FNO & 16370874     & 8201.17   & 93.59 \\ \hline\hline
		\end{tabular*}%
	\end{center}
\end{table}

\begin{table}[tbp]
	\begin{center}
		\caption{Computational cost comparison of U-ResNet and FNO architectures for vorticity field prediction, evaluating model complexity, GPU memory usage, and training time per epoch. }\label{tab:4}
		\begin{tabular*}{1\textwidth}{@{\extracolsep{\fill}} lccc }
			\hline\hline
			\small    
			Model & Numbers of parameters & GPU memory usage (MB) & Training time per epoch (GPU$\cdot$s) \\ \hline
			U-ResNet & 2072672    & 157.79    & 47.49 \\           
			FNO & 16369849     & 8197.69   & 91.64 \\ \hline\hline
		\end{tabular*}%
	\end{center}
\end{table}

The stark efficiency gains of U-ResNet and U-Net make them preferable for clinical translation, where rapid inference and minimal hardware requirements are critical. While FNO/UFNO excel in spectral accuracy, their computational costs render them impractical for real-time applications. 

In this section, we present the predicted results and the generalization performance across different Reynolds numbers. The ground truth for the pressure, WSS, velocity, and vorticity fields is obtained from CFD simulation results, which are generated using the IB-LBM solver described earlier.

\subsection{Prediction results} \label{prediction results}
The baseline configuration employed a training dataset of $N_{s}=2000$ samples, a wall grid resolution of $N_{g}=3201$ points per wall (i.e., $ds=0.005D$), and a Reynolds number of $Re=200$. To systematically evaluate model robustness, we further varied training sample sizes ($N_{s}=$ $500$, $1000$, $3000$) and wall grid resolutions ($N_{g}=$ $201$, $401$, $801$, $1601$), corresponding to wall grid spacings of $ds = {0.08D, 0.04D, 0.02D, 0.01D}$ respectively. In all CFD simulations, the wall grid spacing was maintained at half the fluid grid spacing ($ds = dx/2$), ensuring adequate resolution of near-wall gradients while optimizing computational efficiency \citep{huang2021transition,huang2022streamline}. This systematic discretization approach enabled quantitative assessment of spatial convergence for both the reference CFD solutions and the neural network predictions. 

\begin{table}[tbp]
	\begin{center}
		\caption{Performance evaluation of predicted pressure on the wall for 20 randomly generated stenosis at $Re=200$ using four neural network architectures, comparing mean absolute error ($MAE$), normalized mean absolute error ($NMAE$), root mean squared error ($RMSE$), and normalized root mean squared error ($NRMSE$). Here, these errors correspond to interpolation results.\label{tab:5}}
		\begin{tabular*}{1\textwidth}{@{\extracolsep{\fill}} lcccc }
			\hline\hline
			\small    
			Model & MAE & NMAE($\%$) & RMSE & NRMSE($\%$) \\ \hline
			U-Net & 0.3914 ± 0.2068    & 26.4622 ± 2.3516   & 0.4407 ± 0.2325  & 29.8238 ± 2.0372\\
			U-ResNet & \textbf{0.0171 ± 0.0161}    & \textbf{1.0971 ± 0.6110 }   & \textbf{0.0234 ± 0.0204} & \textbf{1.5439 ± 0.8845}\\
			FNO & 0.1090 ± 0.1326   & 6.7926 ± 3.6857   & 0.1434 ± 0.1545    & 9.0119 ± 4.0679\\            
			UFNO & 0.1966 ± 0.1831     & 12.1066 ± 6.3003   & 0.2516 ± 0.2326 &  15.5849 ± 8.7578  \\ \hline\hline
		\end{tabular*}%
	\end{center}
\end{table}

\subsubsection{Prediction of pressure}
Twenty test cases, generated using random stenosis geometries at $Re=200$, were used to evaluate the performance of four neural network architectures: U-Net, U-ResNet, FNO, and UFNO. As summarized in Table~\ref{tab:5}, the U-ResNet achieves superior accuracy in predicting pressure, outperforming other architectures by orders of magnitude. For U-ResNet, $MAE = 0.0171 \pm 0.0161$, representing a 95\% reduction in error compared to FNO and a 98\% improvement over UFNO. While for U-Net, $MAE = 0.3914 \pm 0.2068$, highlighting limitations in resolving steep pressure gradients near stenosis. Spectral methods (FNO/UFNO) exhibit higher errors (FNO: $0.1090 \pm 0.1326$; UFNO: $0.1966 \pm 0.1831$), likely due to spectral bias in capturing localized boundary layer dynamics. U-ResNet achieves the lowest normalized mean absolute error ($NMAE = 1.10 \pm 0.61\%$) and normalized root mean squared error ($NRMSE = 1.54 \pm 0.88\%$), demonstrating its ability to resolve pressure variations with minimal deviation from CFD ground truth. In contrast, FNO and UFNO exhibit significantly higher errors ($NMAE = 6.79 \pm 3.69\%$ and $12.11 \pm 6.30\%$, respectively), indicating challenges in capturing localized pressure gradients near stenosis regions due to spectral smoothing. U-Net performs poorly, with $NMAE$ exceeding 26\%, highlighting its limitations in handling complex boundary layer dynamics. Similar prediction abilities are observed for $RMSE$. U-ResNet’s robustness across metrics and low variance in predictions underscores its suitability for real-time hemodynamic modeling, while FNO and UFNO remain computationally intensive and less accurate for clinical applications. 

\begin{figure}[htbp]
    \centering
    % 第一行
    \begin{subfigure}[b]{0.49\textwidth}
        \begin{overpic}[width=\linewidth]{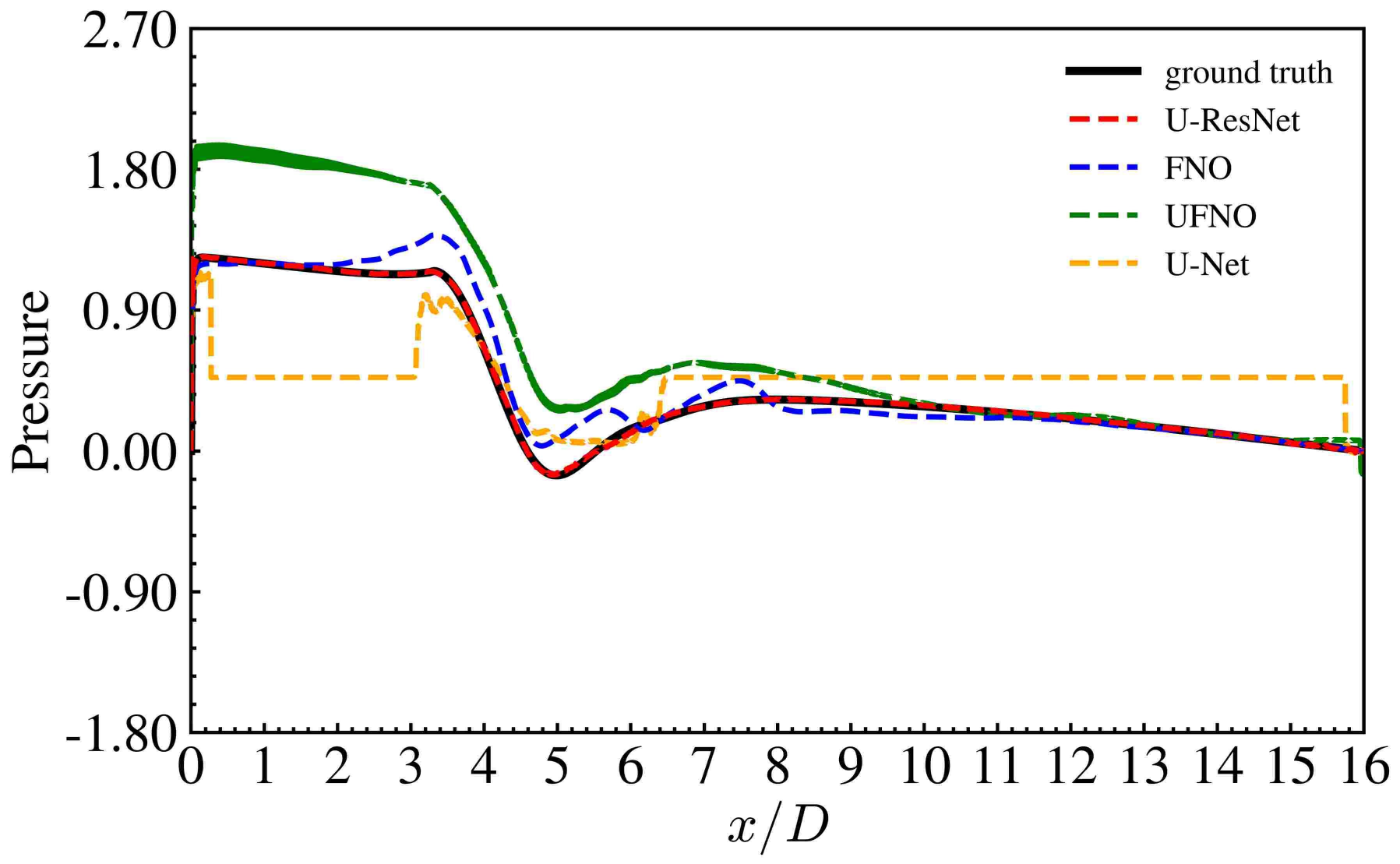}
            \put(-1,62){\small (a)}  
        \end{overpic}
    \end{subfigure}
    \hfill
    \begin{subfigure}[b]{0.49\textwidth}
        \begin{overpic}[width=\linewidth]{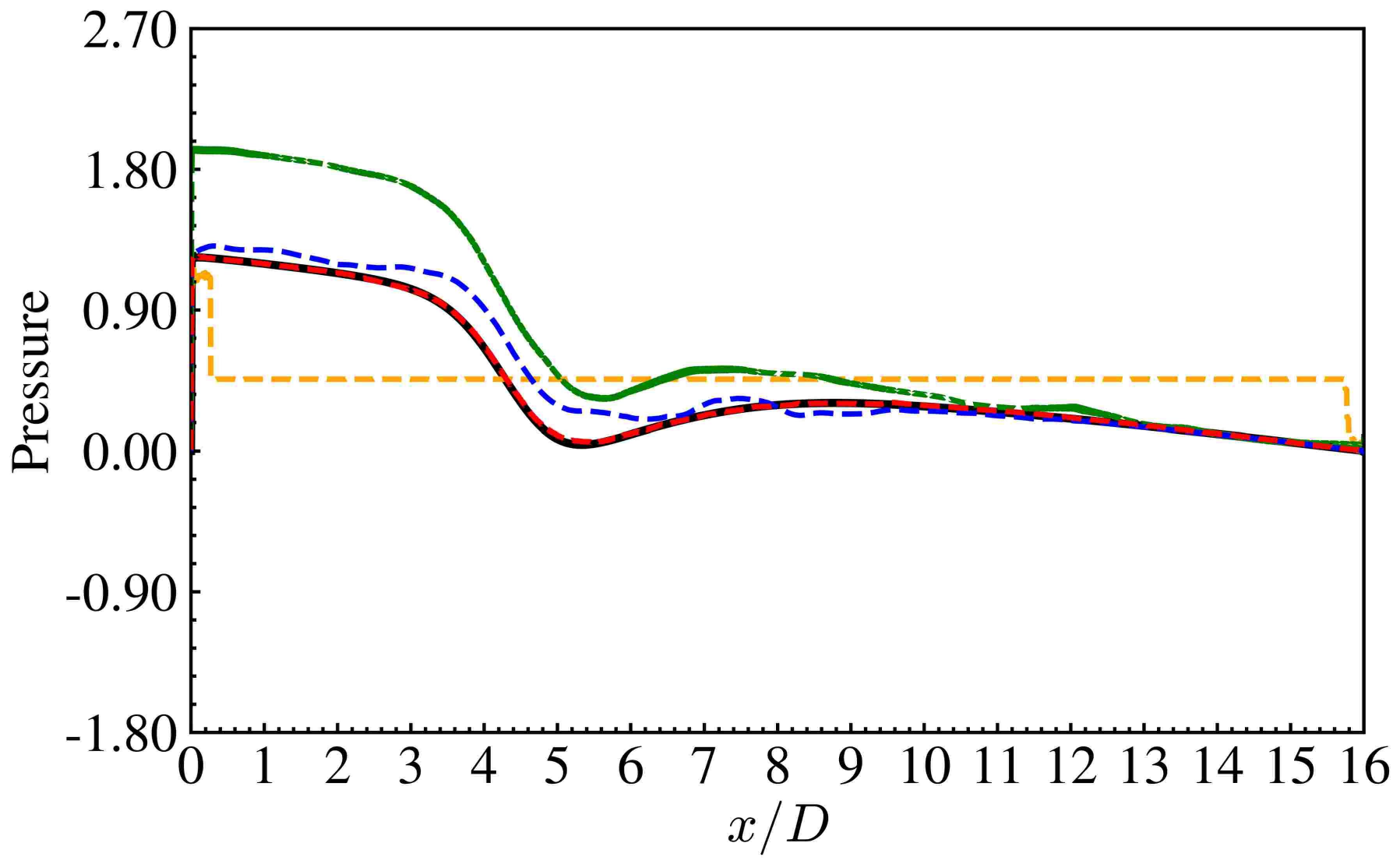}
            \put(-1,62){\small (b)} 
        \end{overpic} 
    \end{subfigure}
    \vspace{0.1cm}

    % 第二行
    \begin{subfigure}[b]{0.49\textwidth}
        \begin{overpic}[width=\linewidth]{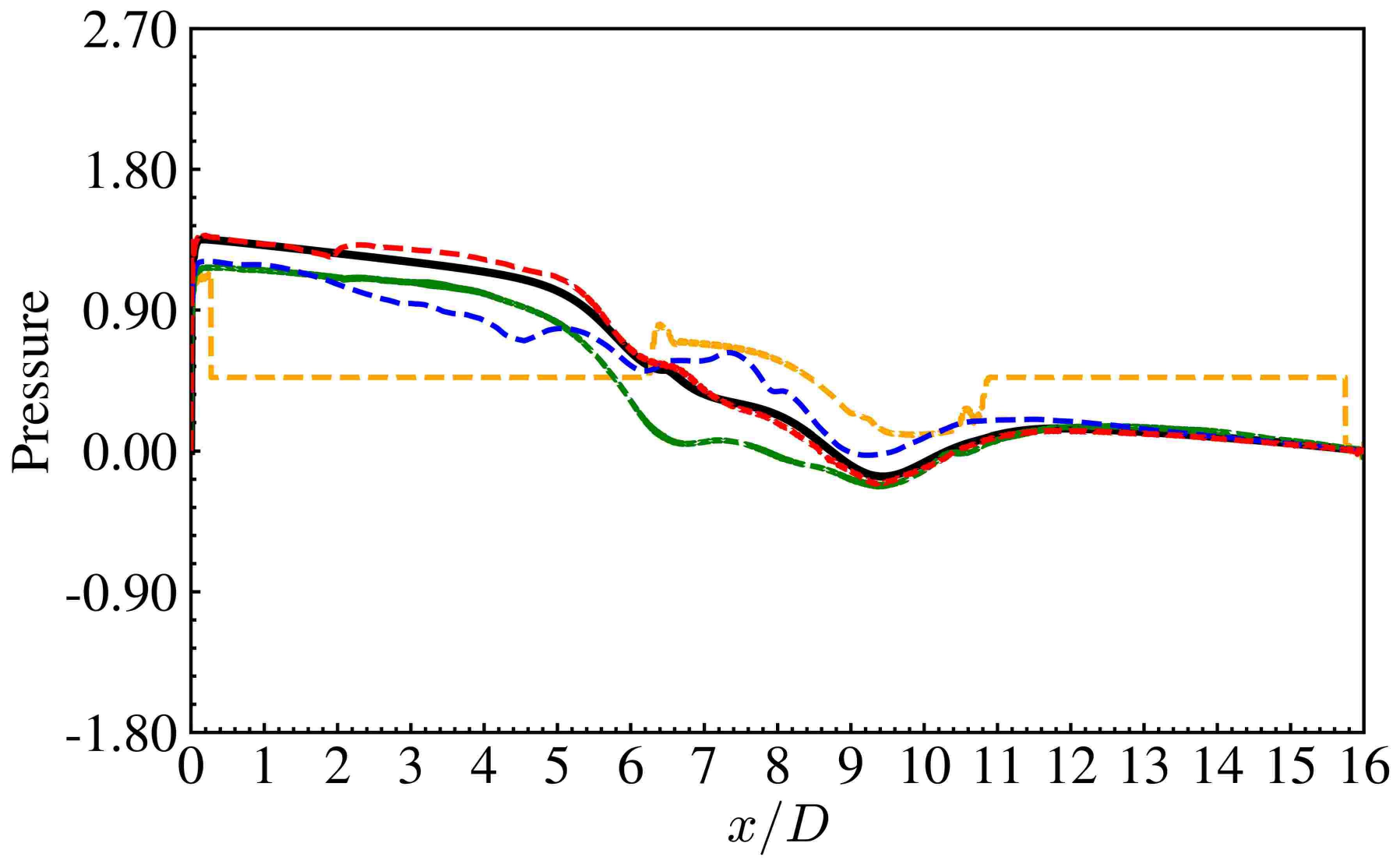}
            \put(-1,62){\small (c)} 
        \end{overpic}
    \end{subfigure}
    \hfill
    \begin{subfigure}[b]{0.49\textwidth}
        \begin{overpic}[width=\linewidth]{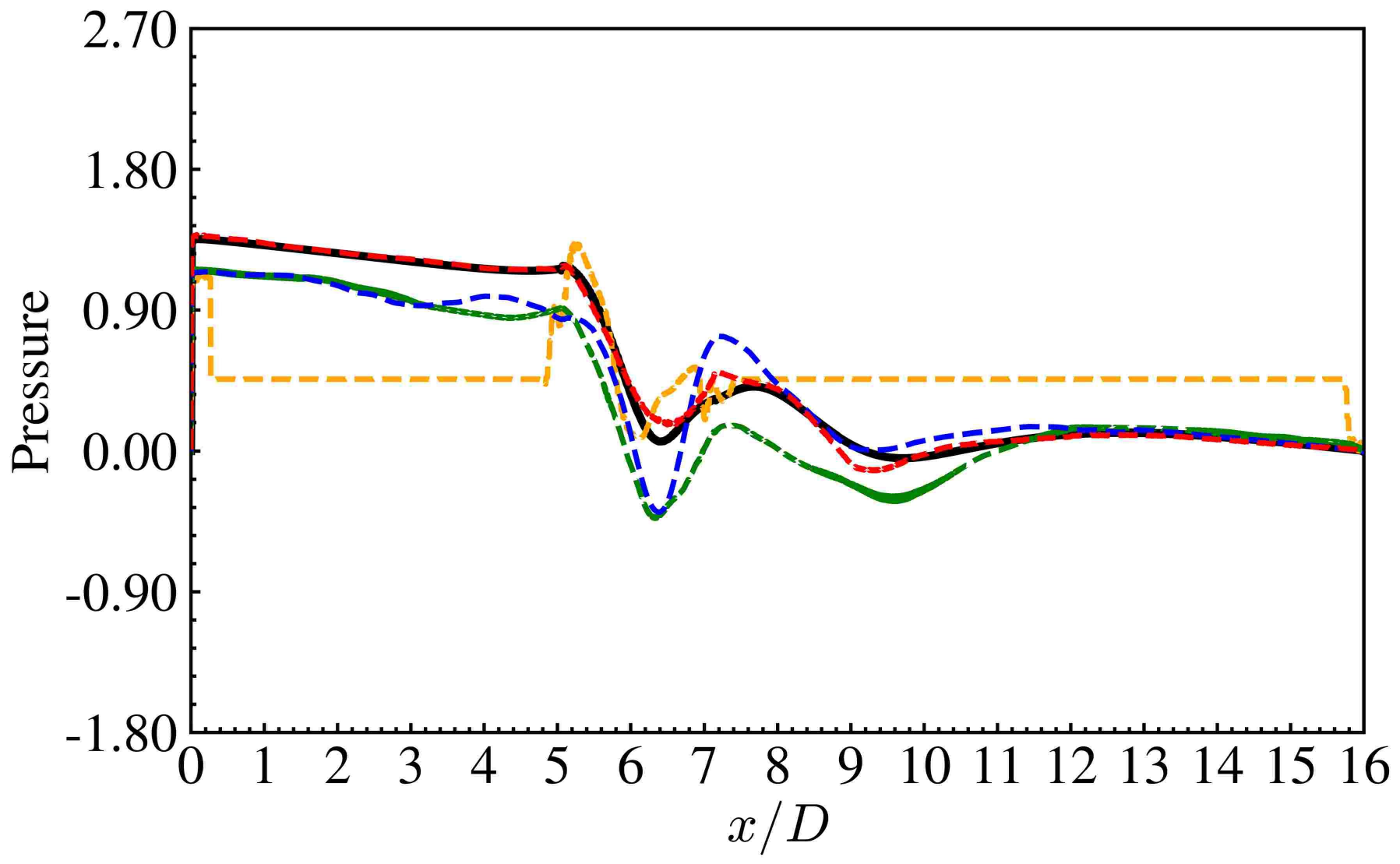}
            \put(-1,62){\small (d)} 
        \end{overpic}
    \end{subfigure}
    \vspace{0.1cm}

    % 第三行
    \begin{subfigure}[b]{0.49\textwidth}
        \begin{overpic}[width=\linewidth]{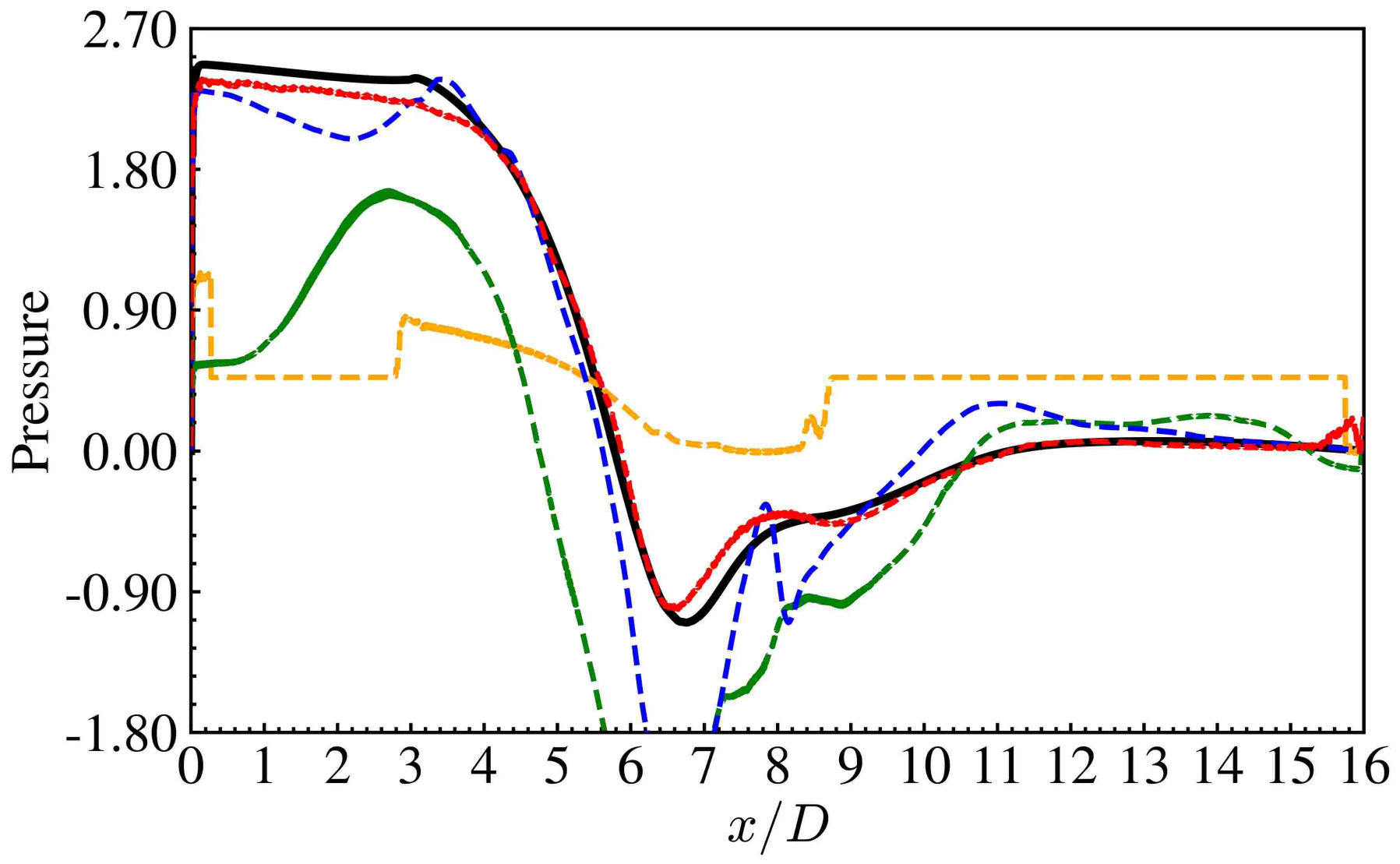}
            \put(-1,62){\small (e)} 
        \end{overpic}
    \end{subfigure}
    \hfill
    \begin{subfigure}[b]{0.49\textwidth}
        \begin{overpic}[width=\linewidth]{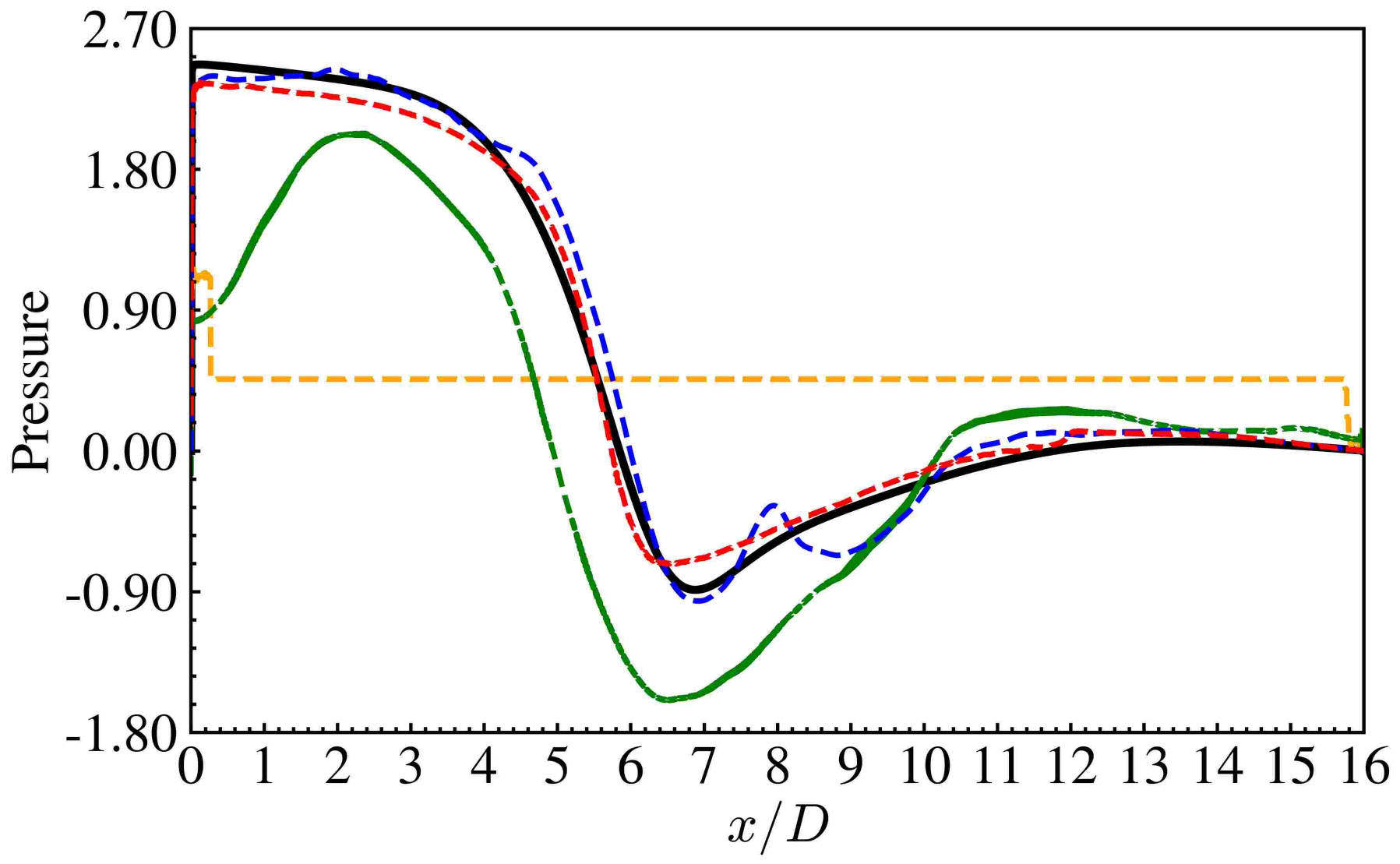}
            \put(-1,62){\small (f)}  
        \end{overpic}
    \end{subfigure}
	\caption{Pressure prediction on the wall for randomly generated stenosis cases at $Re=200$ using four neural network models: (a) Best performance case, upper wall; (b) Best performance case, lower wall; (c) Moderate performance case, upper wall; (d) Moderate performance case, lower wall; (e) Worst performance case, upper wall; (f) Worst performance case, lower wall.}\label{fig:6}
\end{figure}

Fig.~\ref{fig:6} compares the predicted pressure distributions along the upper and lower channel walls using four neural network models (U-Net, U-ResNet, FNO, UFNO) across three representative stenotic cases. Figs.~\ref{fig:6}(a)–(b) show the best-performing case among the 20 test cases—a unilateral stenosis with a single pressure drop. Figs.~\ref{fig:6}(c)–(d) present a moderate case with bilateral stenosis and dual pressure drops, while Figs.~\ref{fig:6}(e)–(f) display the worst-performing case, again with unilateral stenosis and a single pressure drop.
U-ResNet performs well across all three scenarios, demonstrating high consistency and accuracy. In contrast, FNO consistently underperforms in all cases. U-ResNet achieves excellent agreement with CFD ground truth, accurately resolving pressure gradients induced by both unilateral and bilateral stenoses. U-Net, however, fails to capture the complete pressure profile, particularly in regions with high curvature or severe stenosis. FNO and UFNO capture general pressure trends in regions with mild gradients but struggle near stenosis throats, with peak pressure drop errors exceeding 15\%.
Notably, U-ResNet accurately captures both single and double pressure drops (Figs.~\ref{fig:6}(a)–(f)), ensuring reliable predictions even in complex stenotic configurations. Such precision is crucial for clinical applications like fractional flow reserve estimation, where pressure gradients directly inform stenosis severity~\citep{de2008fractional,tu2016diagnostic}.

\begin{figure}[htbp]\centering
	\includegraphics[width=0.7\textwidth]{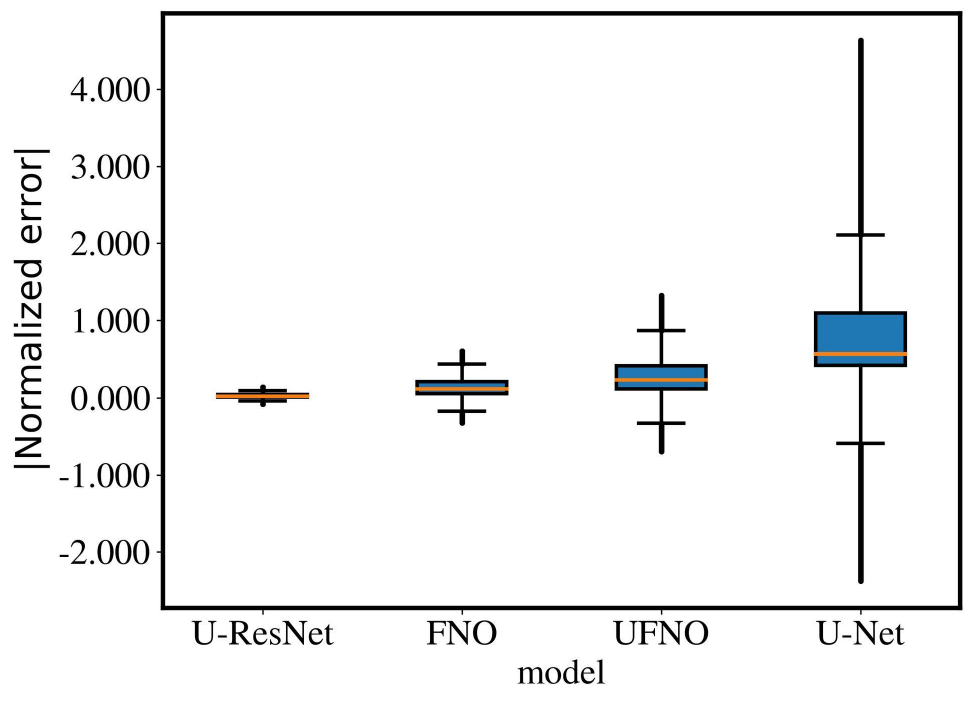}
	\caption{The box plots of pressure prediction error for four neural network models. Here, the errors correspond to interpolation results.}\label{fig:7}
\end{figure}

Statistical analysis of prediction errors across all test cases, presented in Fig.~\ref{fig:7}, quantifies these observations through box plot distributions. U-ResNet exhibits the most concentrated error range with minimal outliers, indicating robust performance across diverse geometric configurations. FNO demonstrates moderately narrow error distributions but still produces significant outliers in complex stenosis cases. UFNO shows broader error ranges despite its hybrid architecture, while U-Net displays the widest error distribution with numerous outliers, confirming its instability for pressure prediction in stenotic flows. U-ResNet's consistent performance across diverse stenotic geometries suggests robust generalization capabilities essential for hemodynamic modeling applications.

\begin{figure}[htbp]\centering
	\includegraphics[width=0.7\textwidth]{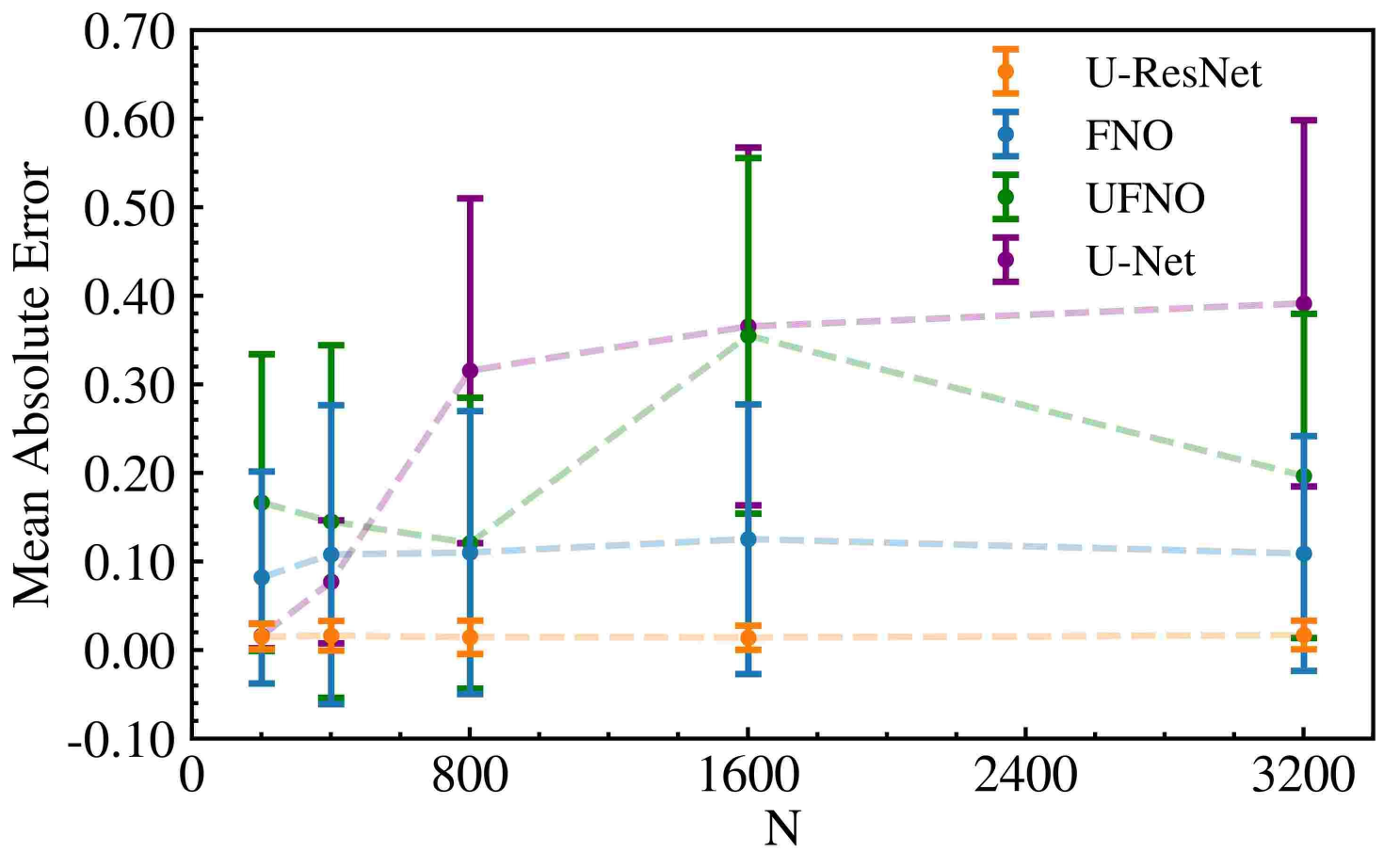}
	\caption{Pressure prediction mean absolute error when choosing different wall grid resolution ($N_g=201$, 401, 801, 1601, 3201). Here, the errors correspond to interpolation results.}\label{fig:8}
\end{figure}

\begin{figure}[htbp]
    \centering
    % 第1行
    \begin{subfigure}[b]{0.49\textwidth}
        \begin{overpic}[width=\linewidth]{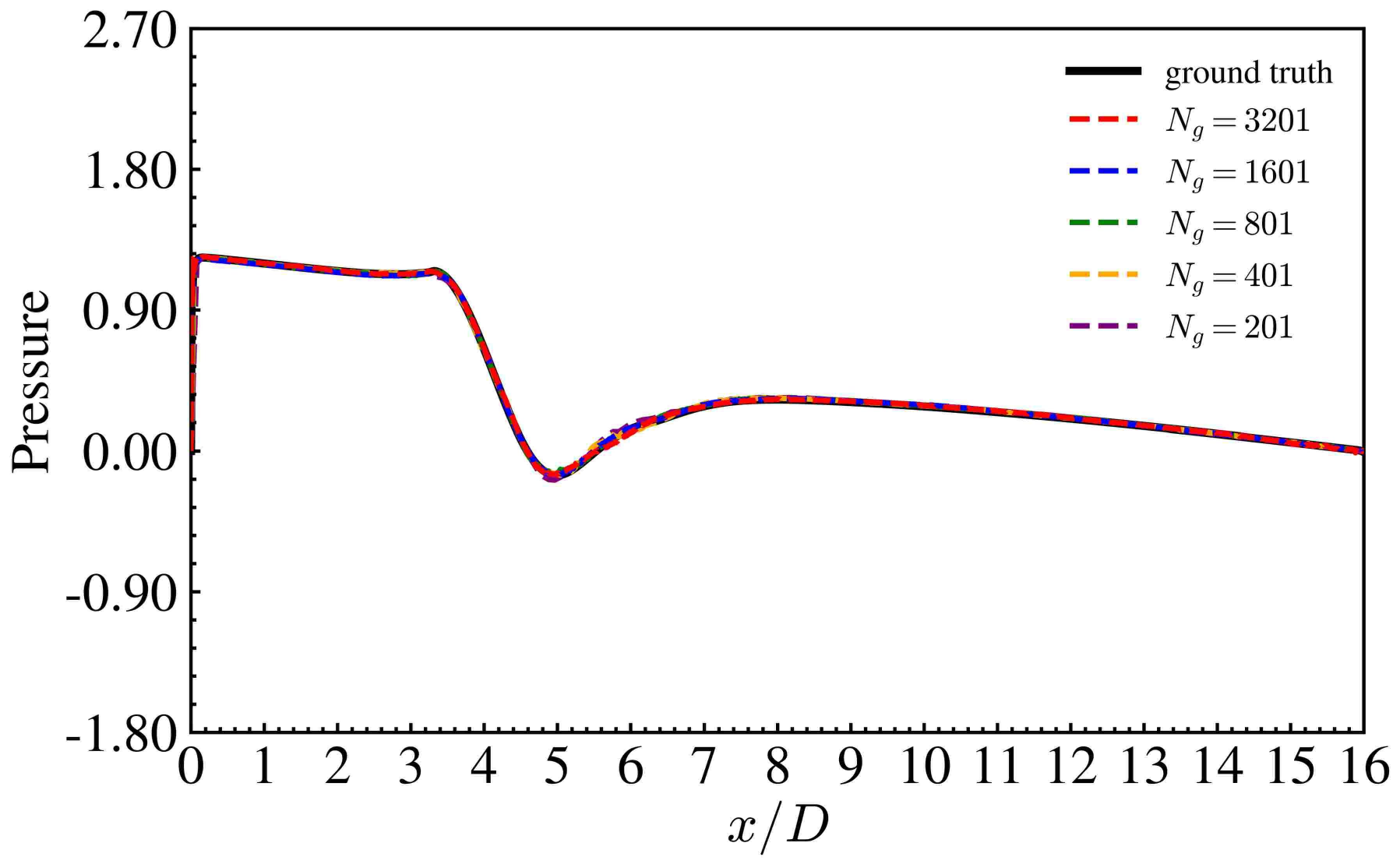}
            \put(-1,62){\small (a)}  
        \end{overpic}
    \end{subfigure}
    \hfill
    \begin{subfigure}[b]{0.49\textwidth}
        \begin{overpic}[width=\linewidth]{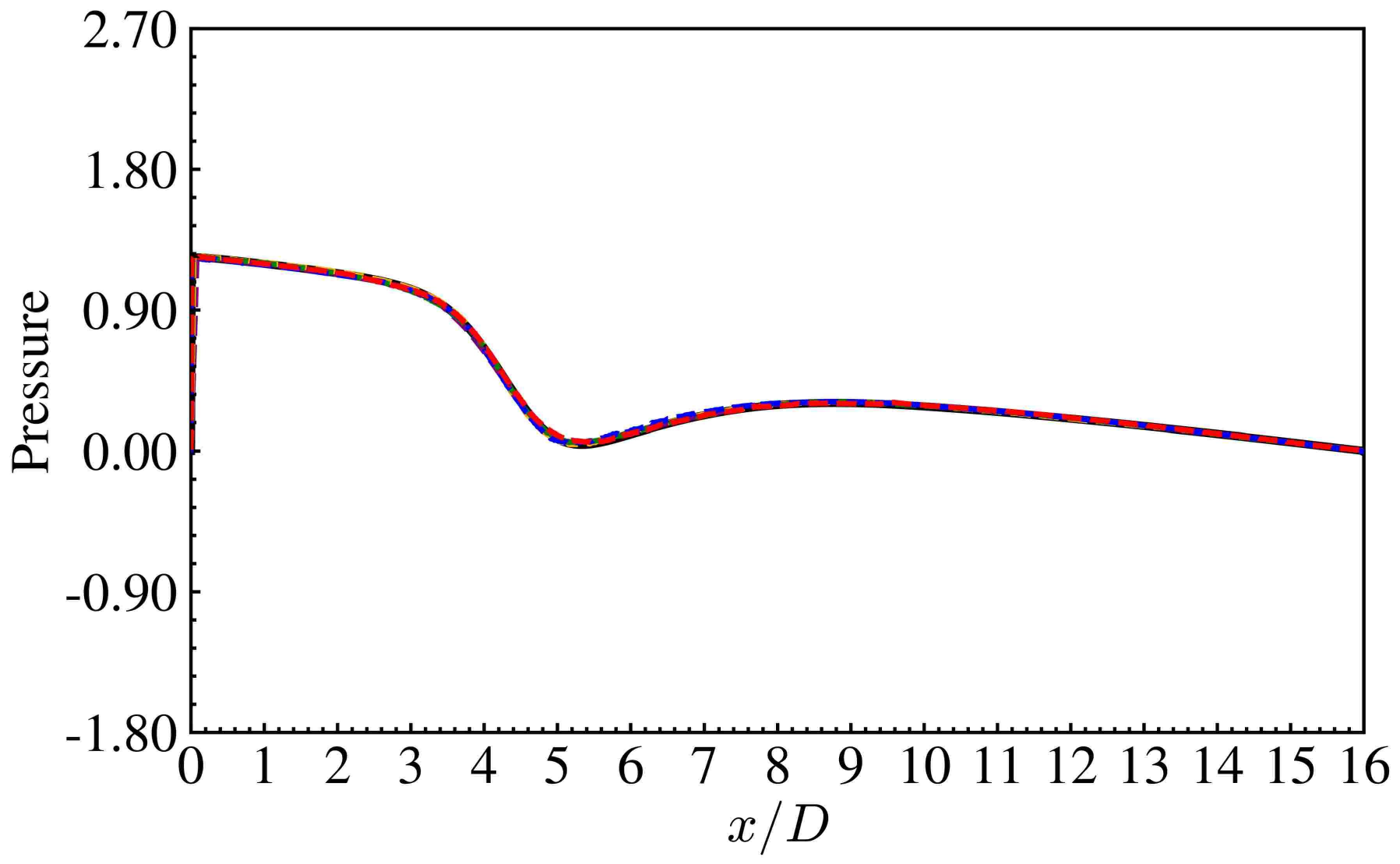}
            \put(-1,62){\small (b)}  
        \end{overpic}
    \end{subfigure}

    \vspace{0.1cm} % 添加垂直间距

    % 第2行
    \begin{subfigure}[b]{0.49\textwidth}
        \begin{overpic}[width=\linewidth]{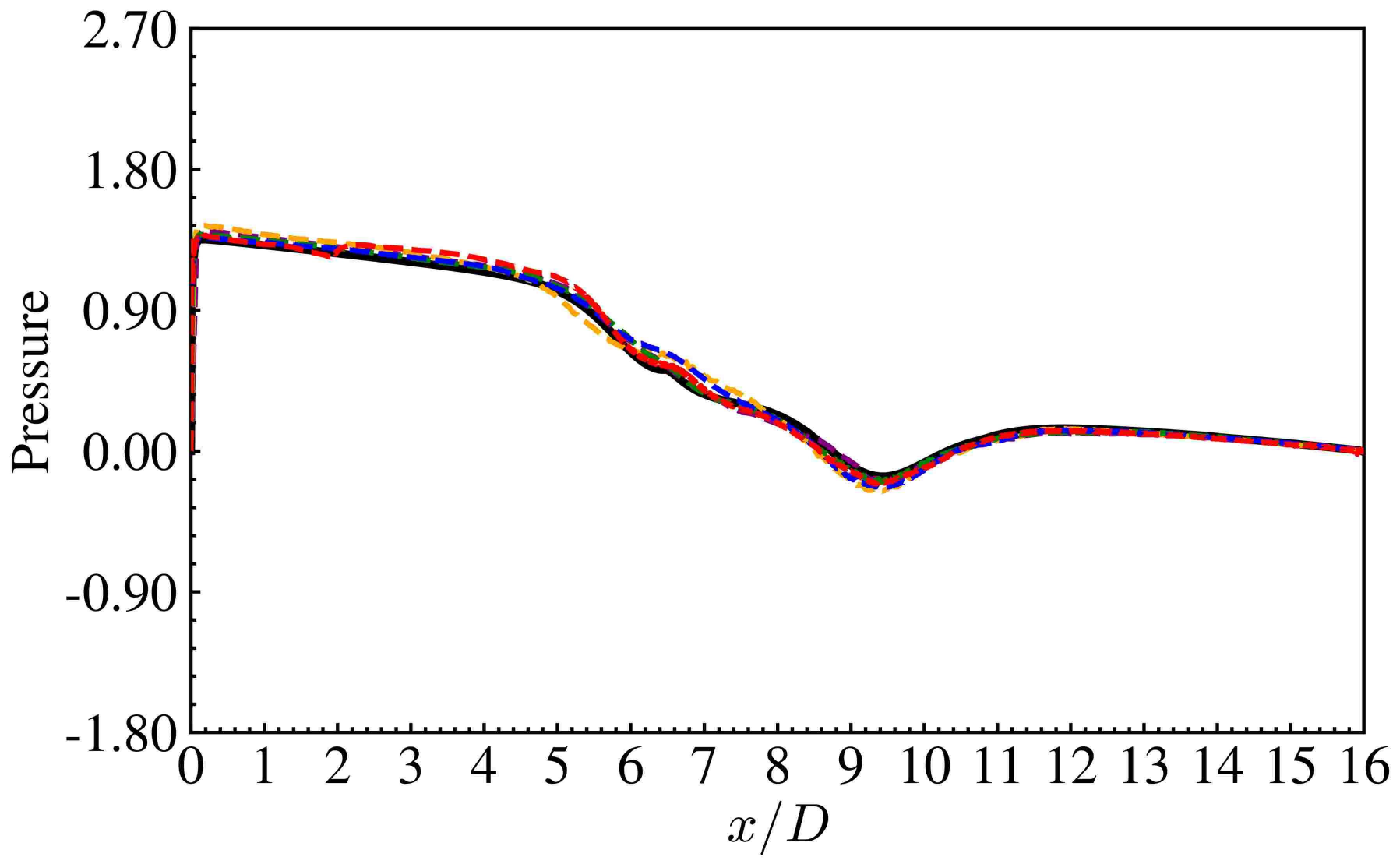}
            \put(-1,62){\small (c)}  
        \end{overpic}
    \end{subfigure}
    \hfill
    \begin{subfigure}[b]{0.49\textwidth}
        \begin{overpic}[width=\linewidth]{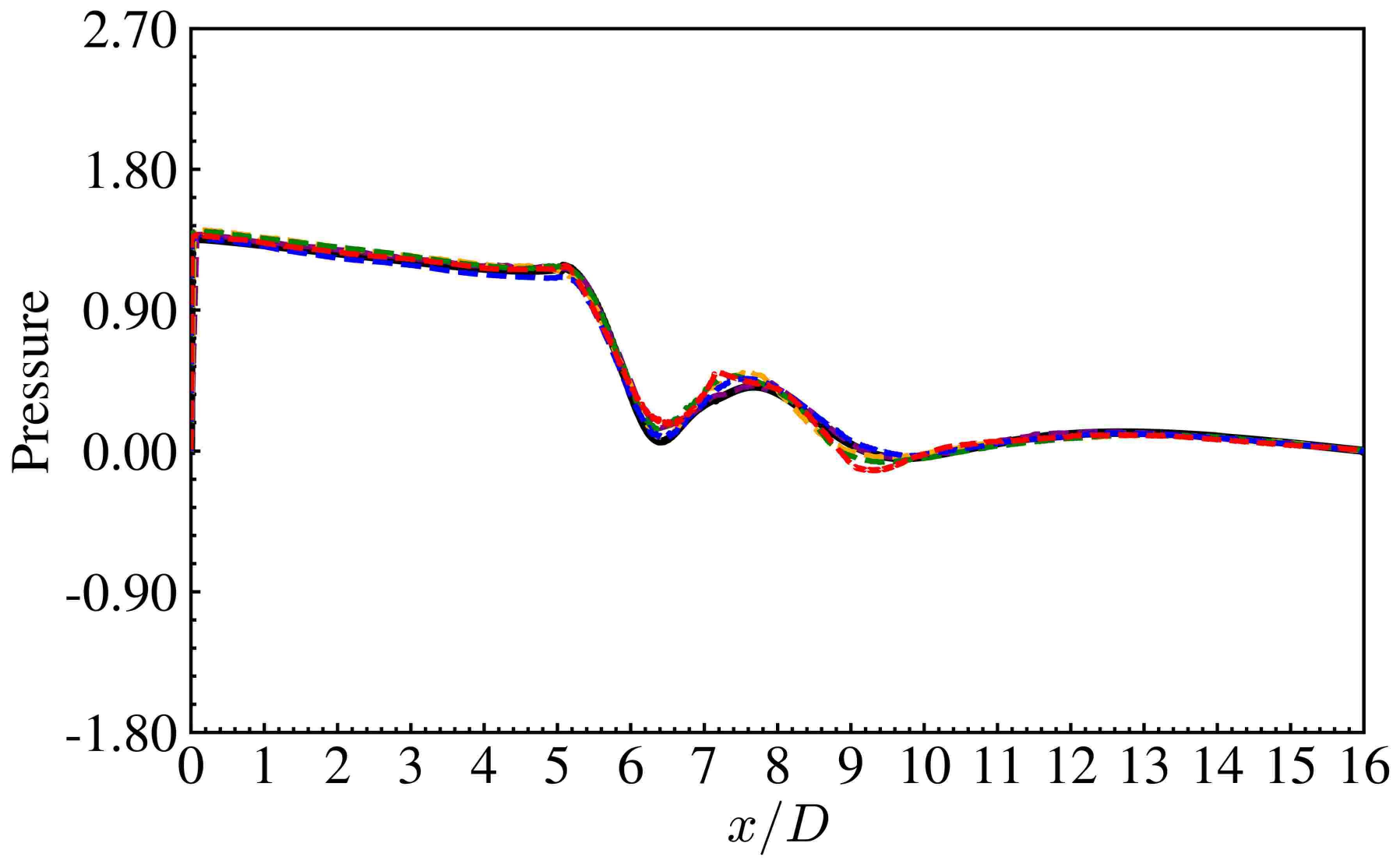}
            \put(-1,62){\small (d)}  
        \end{overpic}
    \end{subfigure}
    \vspace{0.1cm} % 添加垂直间距

    % 第3行
    \begin{subfigure}[b]{0.49\textwidth}
        \begin{overpic}[width=\linewidth]{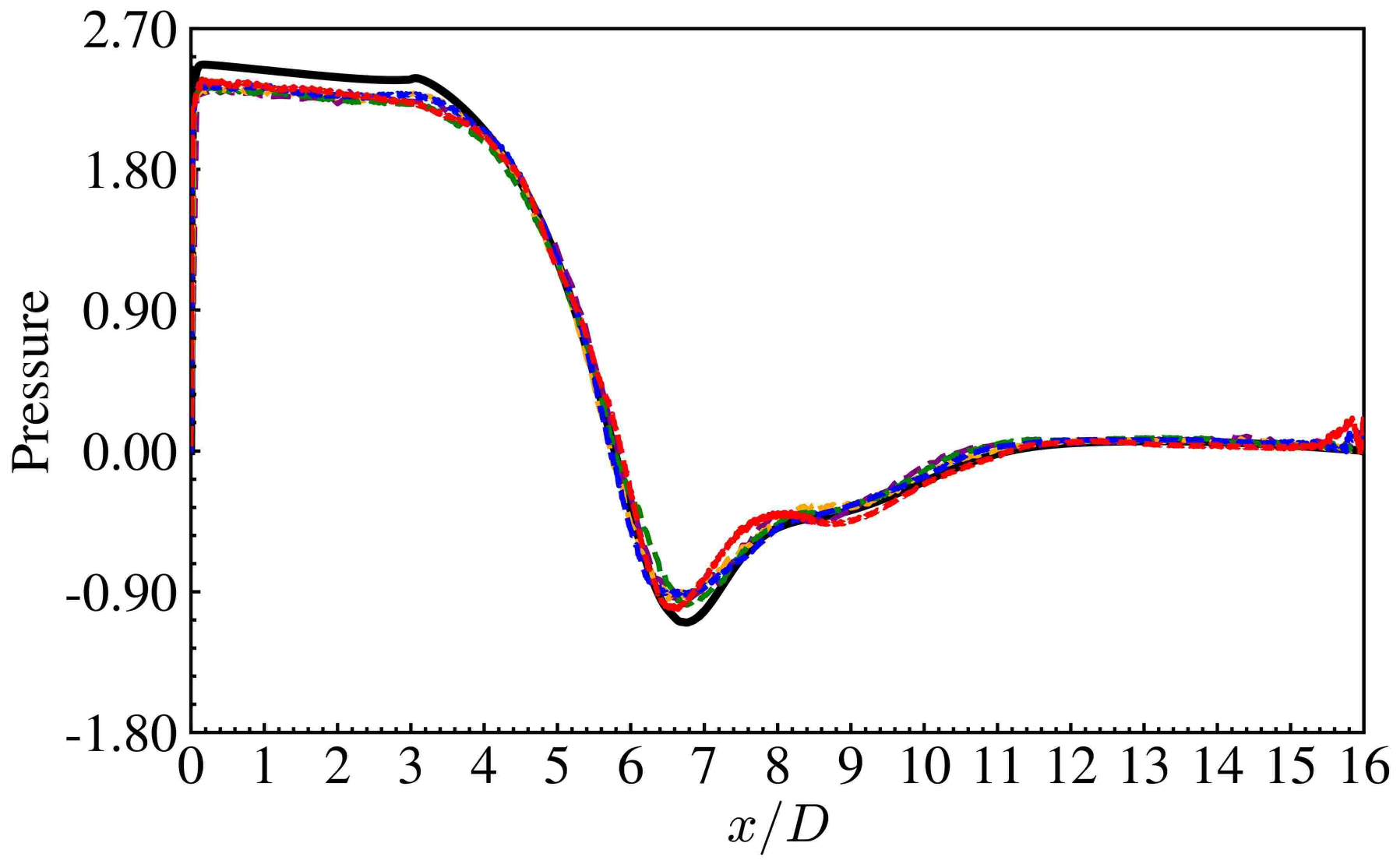}
            \put(-1,62){\small (e)}  
        \end{overpic}
    \end{subfigure}
    \hfill
    \begin{subfigure}[b]{0.49\textwidth}
        \begin{overpic}[width=\linewidth]{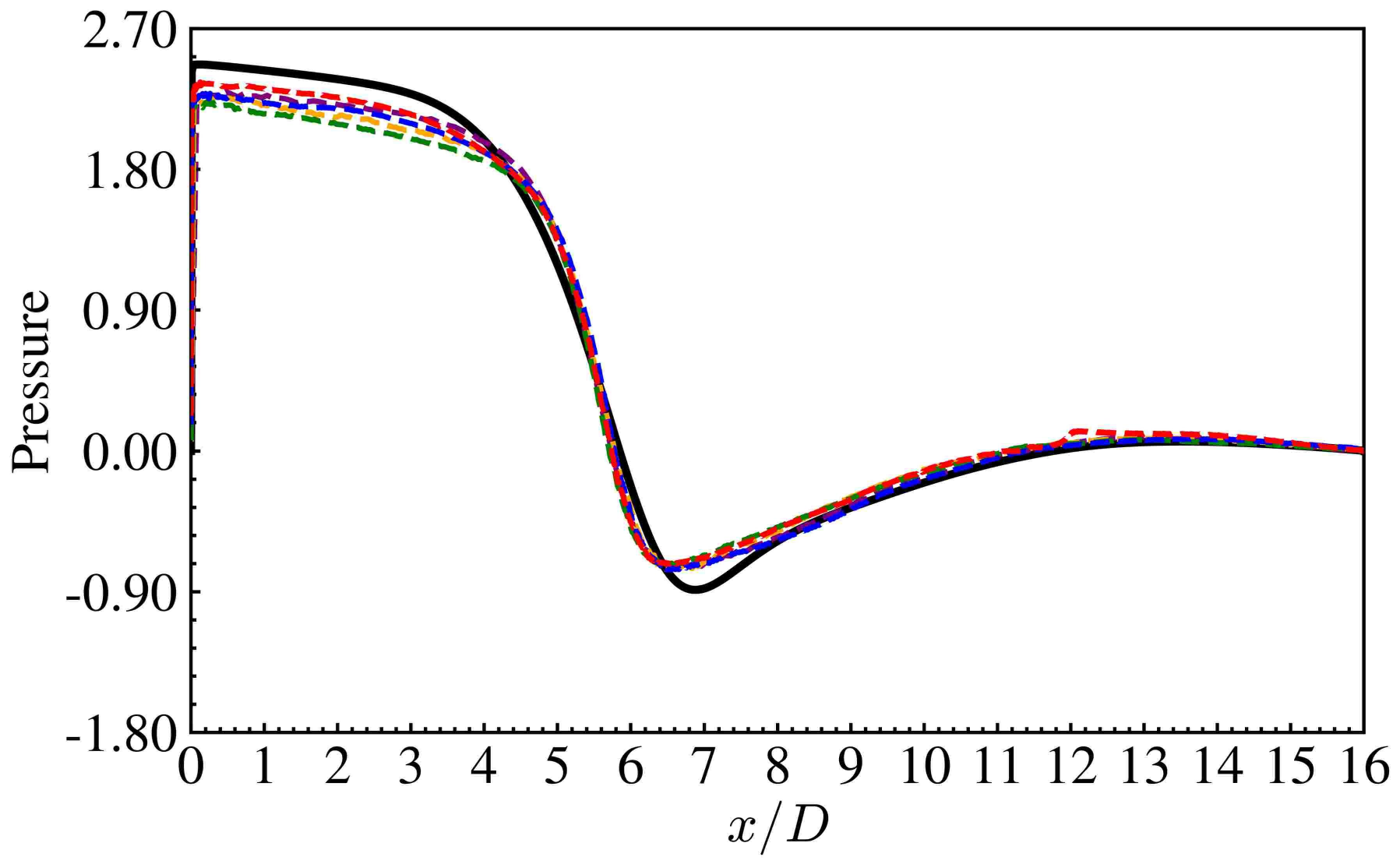}
            \put(-1,62){\small (f)}  
        \end{overpic}
    \end{subfigure}
    
	\caption{Pressure prediction on the walls for U-ResNet when choosing different wall grid point resolutions ($N_g=201$, 401, 801, 1601, 3201): (a) Best performance case: upper wall, (b) Best performance case: lower wall, (c) Moderate performance case: upper wall, (d) Moderate performance case: lower wall, (e) Worst performance case: upper wall, (f) Worst performance case: lower wall.}\label{fig:9}
\end{figure}

To evaluate the impact of wall spatial discretization, four neural network models were tested across wall grid resolutions ranging from $N_g=201$ to $3201$ points per wall. Fig.~\ref{fig:8} illustrates the mean absolute error ($MAE$) of pressure predictions as a function of wall grid resolution. U-ResNet demonstrates exceptional robustness, maintaining $MAE <0.02$ across all resolutions, while competing models (U-Net, FNO, UFNO) exhibit errors increasing by $2-5\times$ as it decreases below $801$. This highlights U-ResNet’s ability to preserve accuracy even on coarser grids, a critical advantage for applications requiring computational efficiency. 
Figs.~\ref{fig:9}(a)–(f) further support this observation by showing pressure distributions for representative unilateral and bilateral stenosis cases. In the best-performing case with unilateral stenosis [Figs.~\ref{fig:9}(a)–(b)], U-ResNet yields peak pressure gradient errors below $0.5\%$. For the moderate and more complex bilateral stenosis case [Figs.~\ref{fig:9}(c)–(d)], the error remains under $1.5\%$, demonstrating U-ResNet’s ability to resolve dual pressure drops with high fidelity. In the worst-performing unilateral case [Figs.~\ref{fig:9}(e)–(f)], the error still stays below $2.5\%$. These results confirm U-ResNet’s grid-invariant behavior and its robustness in capturing pressure gradients across varying stenosis complexities.

\begin{figure}[htbp]\centering
	\includegraphics[width=0.8\textwidth]{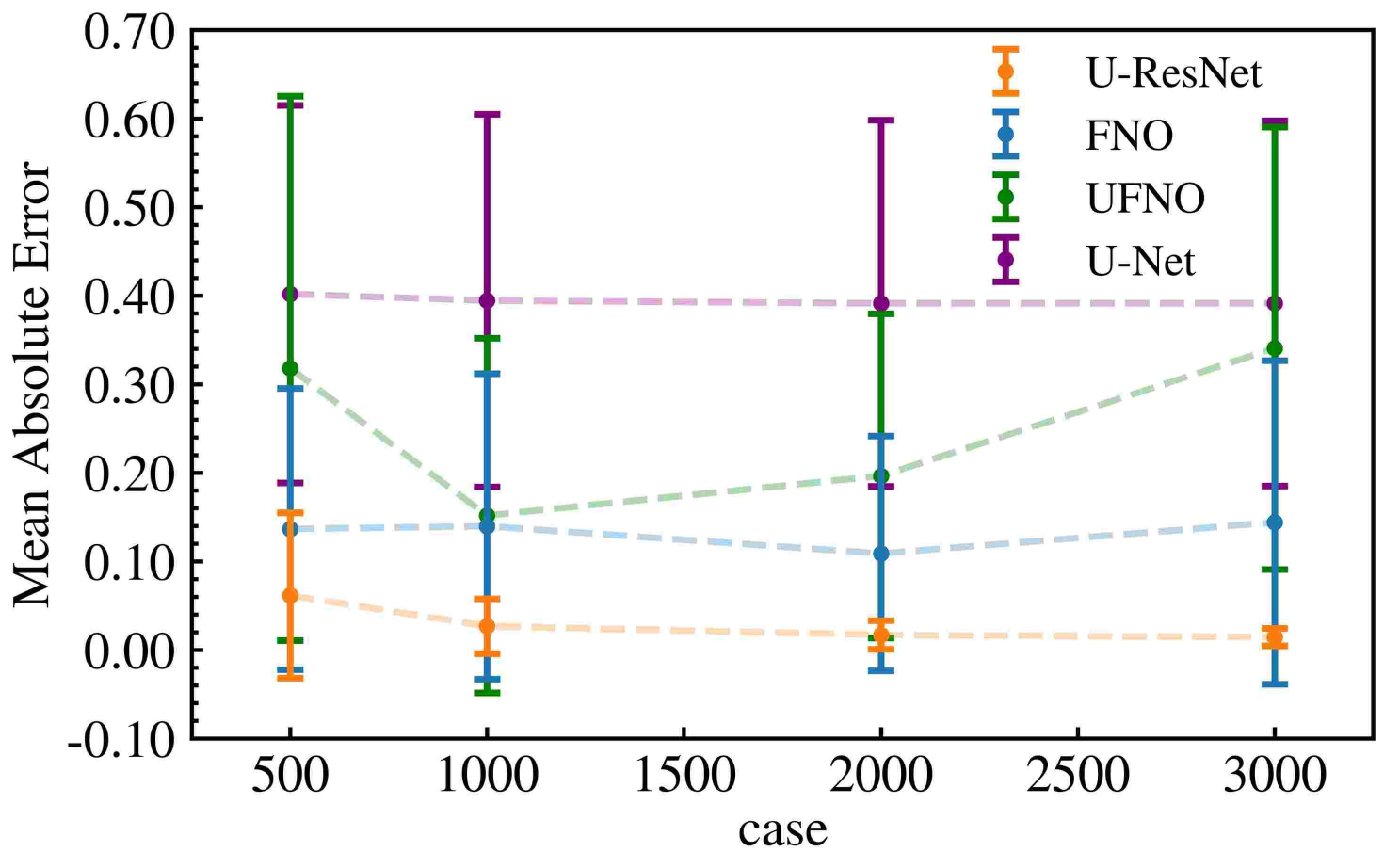}
	\caption{Pressure prediction mean absolute error when choosing different training sample sizes ($N_s=500$, 1000, 2000, 3000). Here, the errors correspond to interpolation results.}\label{fig:10}
\end{figure}

Fig.~\ref{fig:10} compares $MAE$ across training sample sizes ($N_s=500-3000$). U-ResNet achieves $MAE = 0.06 \pm 0.16$ even at $N_s=500$, outperforming FNO ($MAE = 0.14 \pm 0.29$) and UFNO ($MAE = 0.3 \pm 0.62$) by an order of magnitude. With $N_s=500-3000$, U-ResNet attains convergence ($MAE$ reduction $<0.5\%$ per epoch), while other models require $N_s>2000$ to stabilize.

\begin{figure}[htbp]
    \centering
    % 第1行
    \begin{subfigure}[b]{0.49\textwidth}
        \begin{overpic}[width=\linewidth]{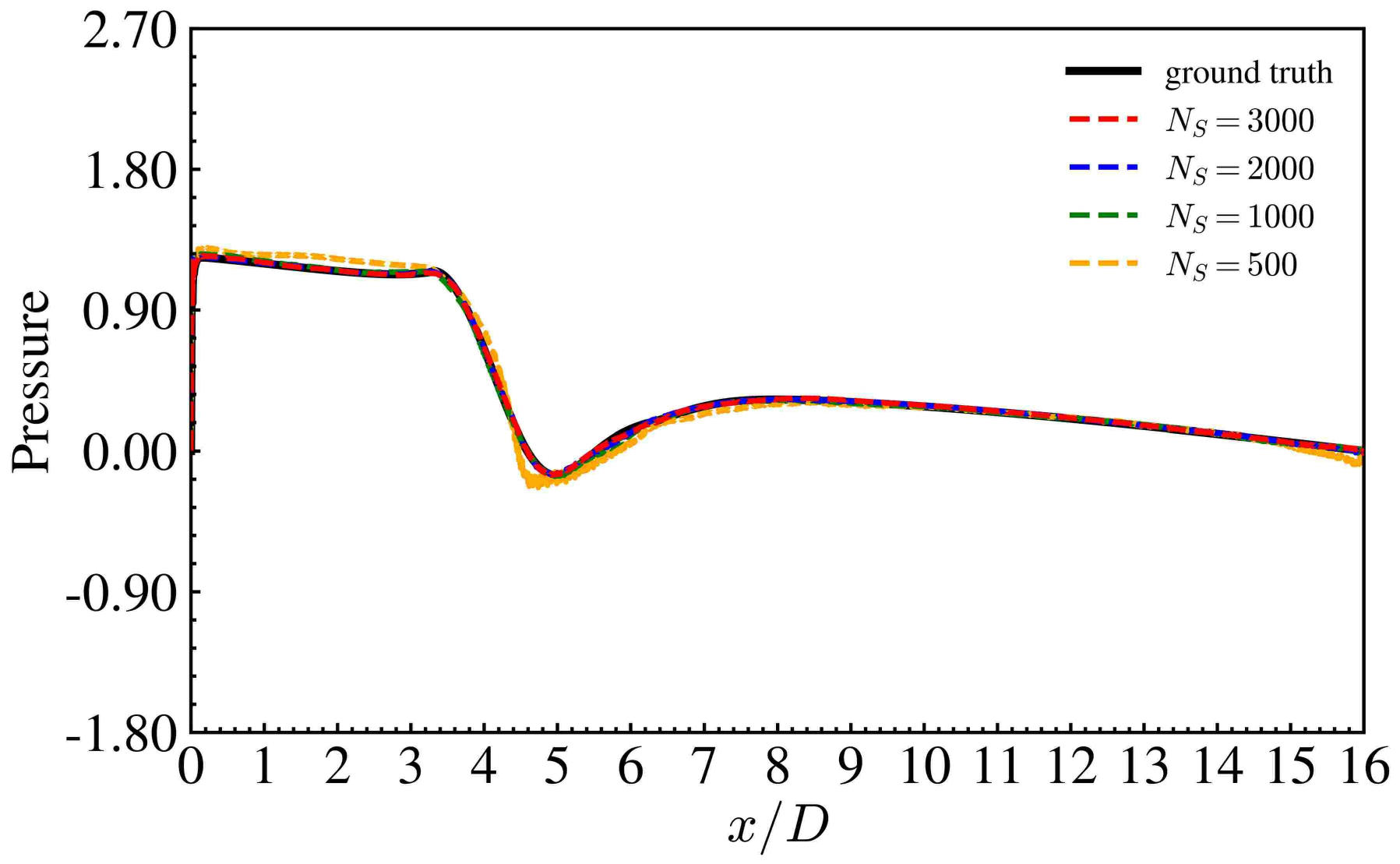}
            \put(-1,62){\small (a)}  
        \end{overpic}
    \end{subfigure}
    \hfill
    \begin{subfigure}[b]{0.49\textwidth}
        \begin{overpic}[width=\linewidth]{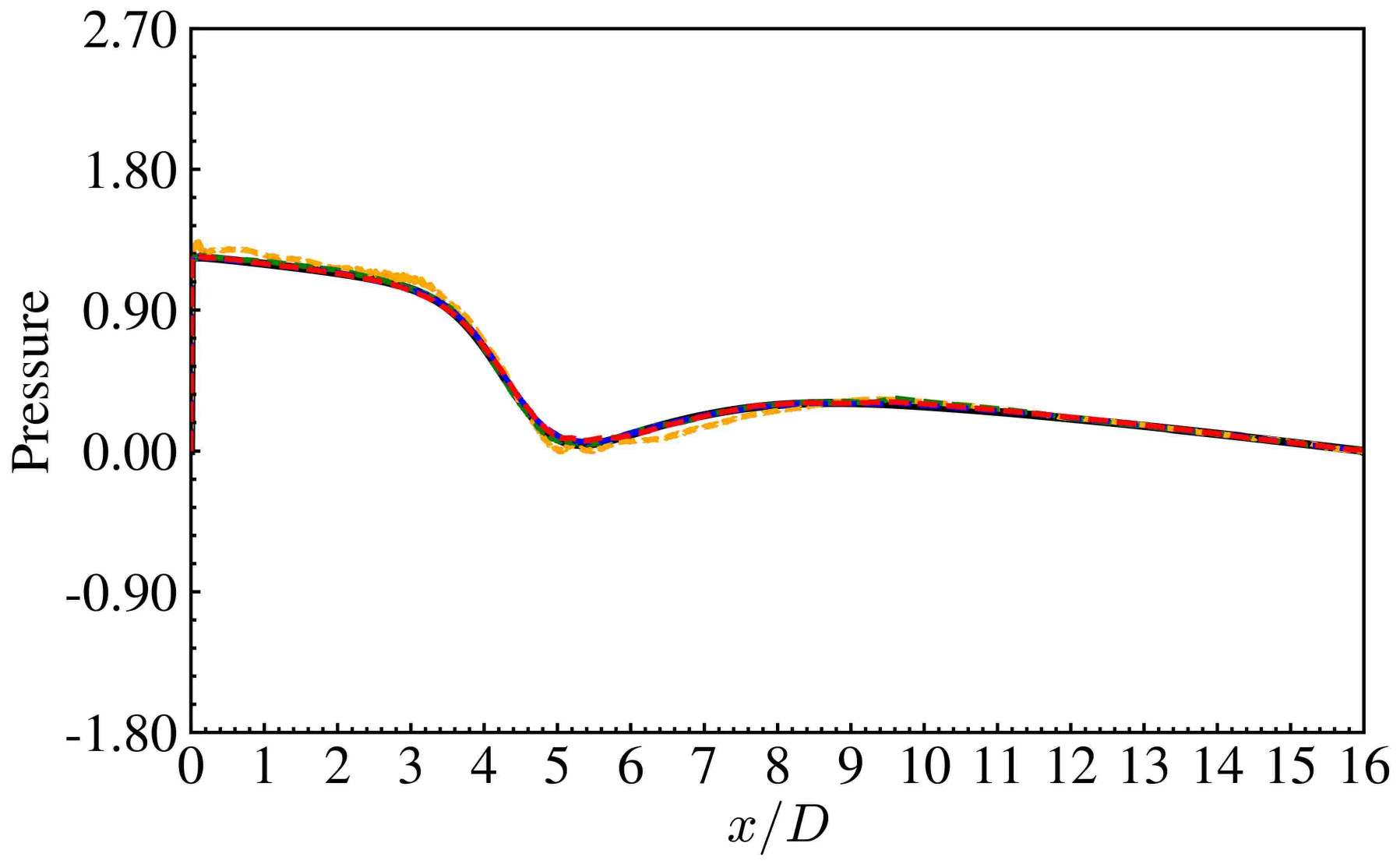}
            \put(-1,62){\small (b)}  
        \end{overpic}
    \end{subfigure}

    \vspace{0.1cm} % 添加垂直间距

    % 第2行
    \begin{subfigure}[b]{0.49\textwidth}
        \begin{overpic}[width=\linewidth]{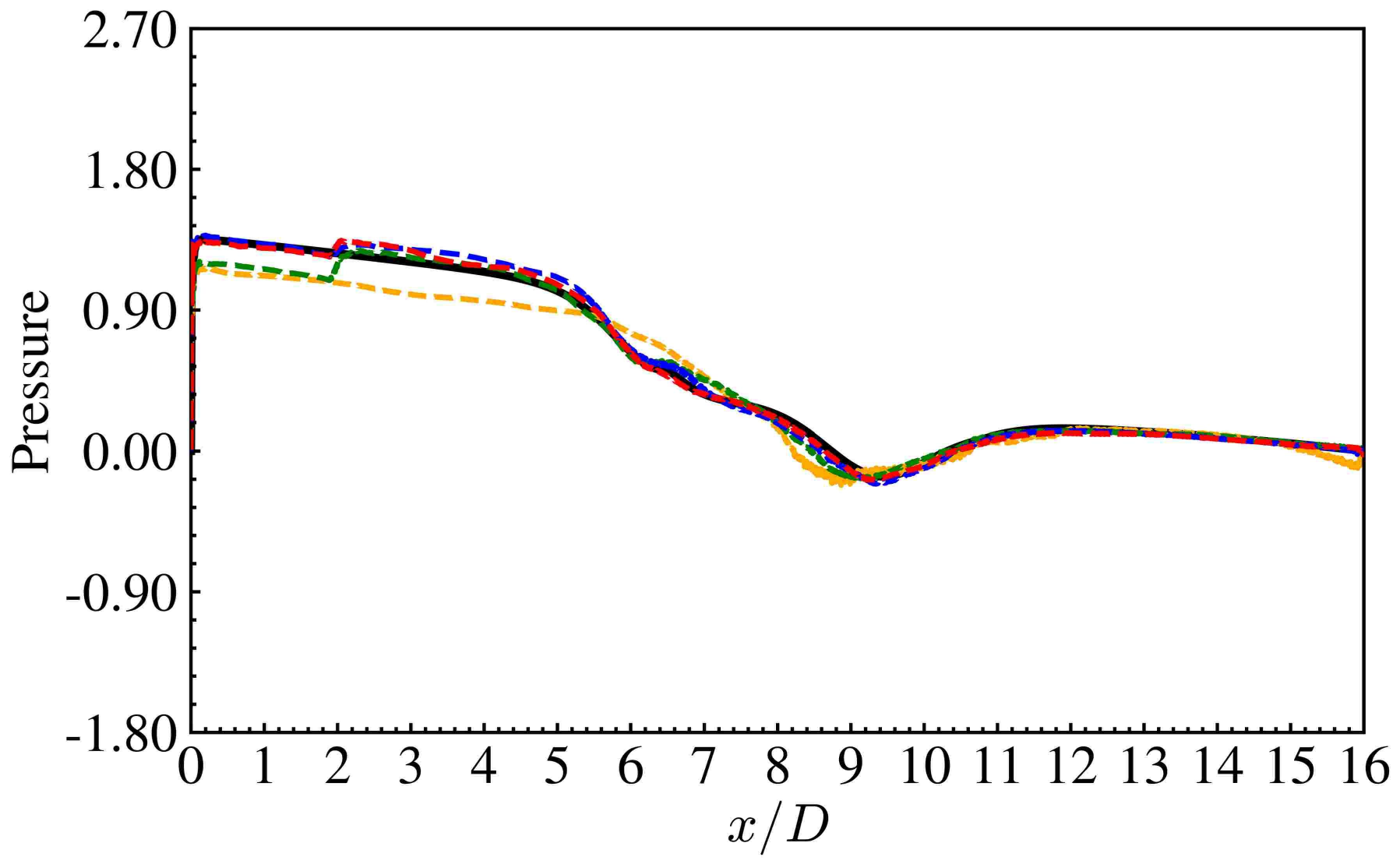}
            \put(-1,62){\small (c)}  
        \end{overpic}
    \end{subfigure}
    \hfill
    \begin{subfigure}[b]{0.49\textwidth}
        \begin{overpic}[width=\linewidth]{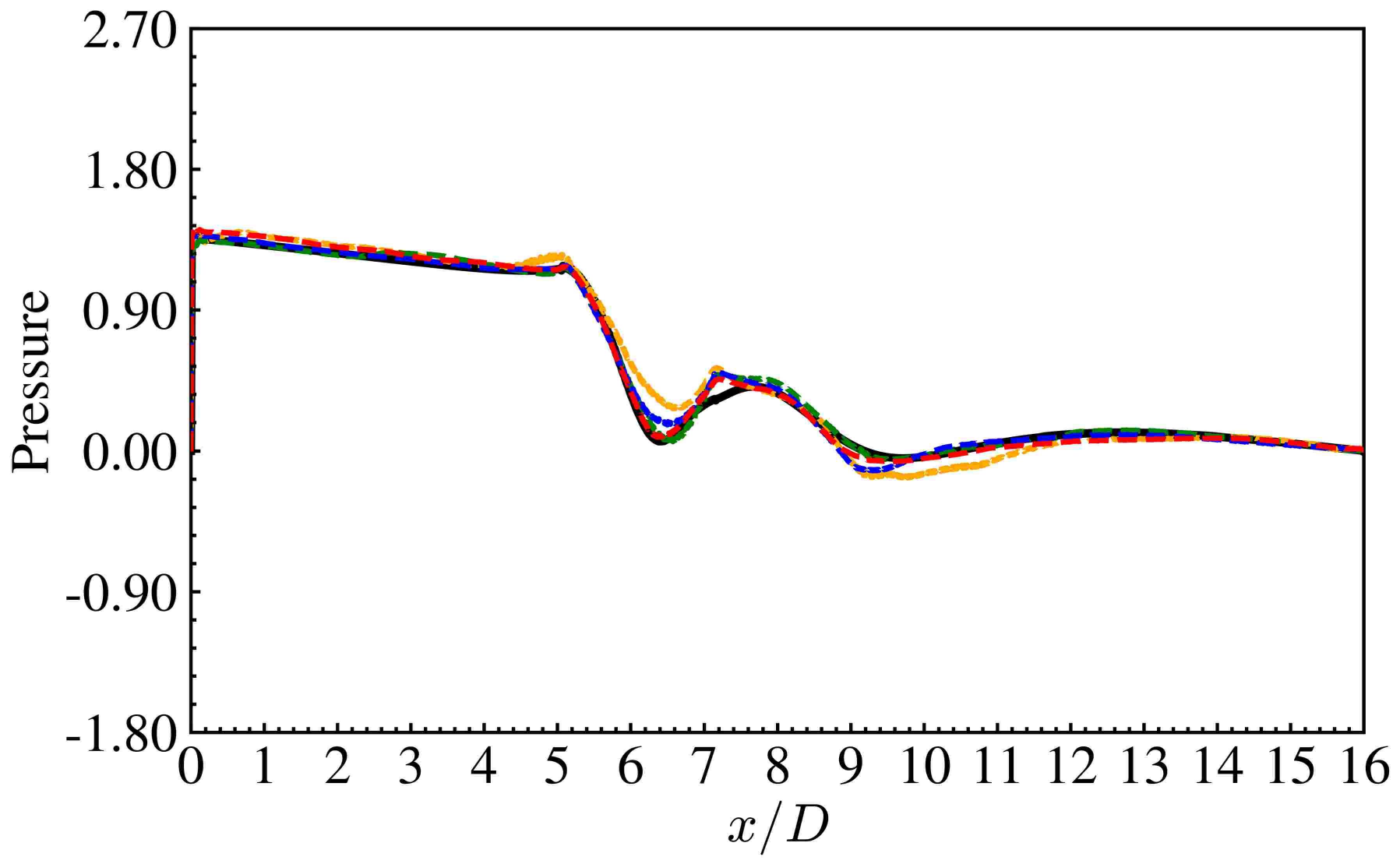}
            \put(-1,62){\small (d)}  
        \end{overpic}
    \end{subfigure}	

    \vspace{0.1cm} % 添加垂直间距

    % 第3行
    \begin{subfigure}[b]{0.49\textwidth}
        \begin{overpic}[width=\linewidth]{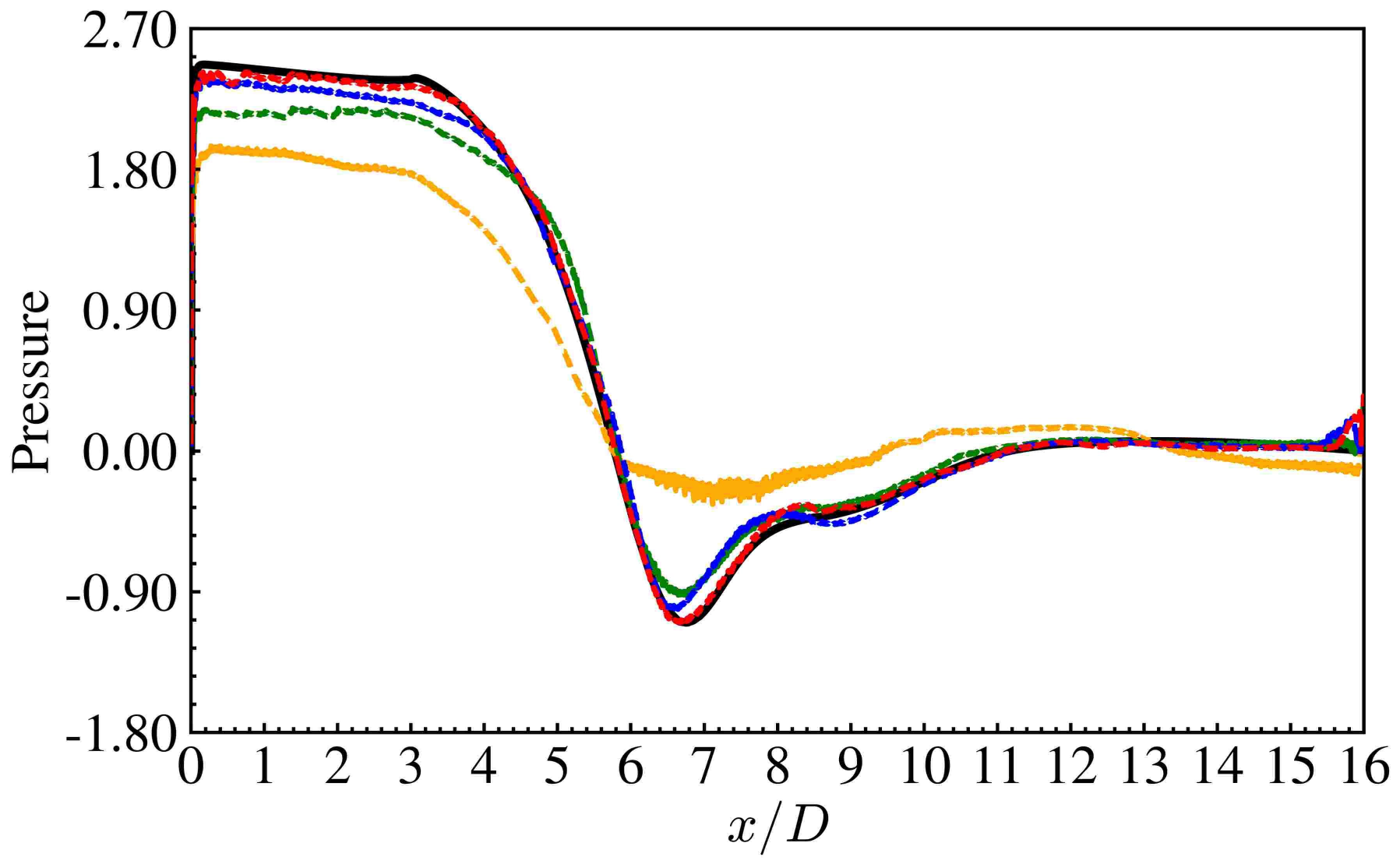}
            \put(-1,62){\small (e)}  
        \end{overpic}
    \end{subfigure}
    \hfill
    \begin{subfigure}[b]{0.49\textwidth}
        \begin{overpic}[width=\linewidth]{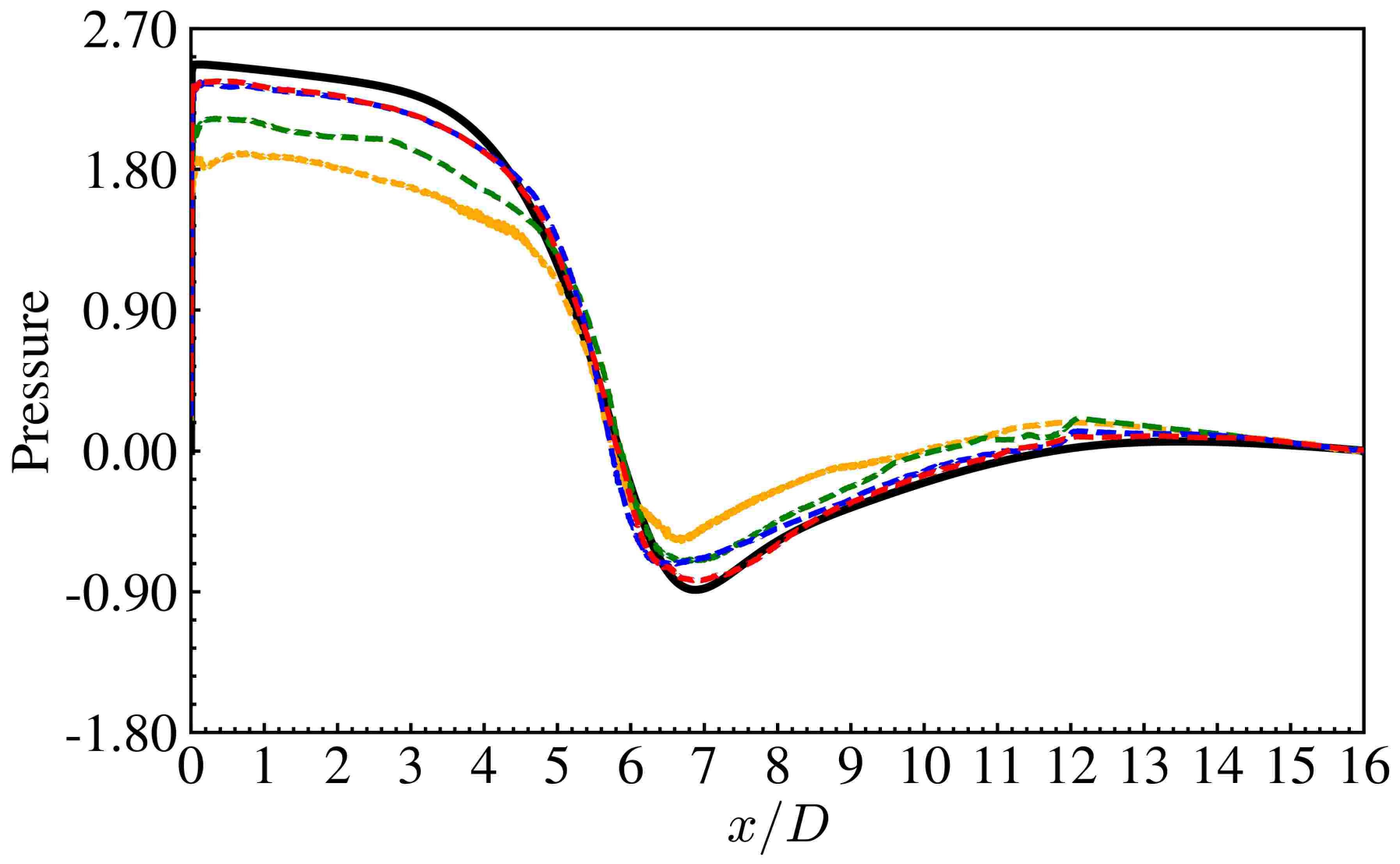}
            \put(-1,62){\small (f)}  
        \end{overpic}
    \end{subfigure}	
    
	\caption{Pressure prediction on the walls for U-ResNet when choosing different training sample sizes (500, 1000, 2000, 3000): (a) Best performance case: upper wall, (b) Best performance case: lower wall, (c) Moderate performance case: upper wall, (d) Moderate performance case: lower wall, (e) Worst performance case: upper wall, (f) Worst performance case: lower wall.}\label{fig:11}
\end{figure}

Figs.~\ref{fig:11}(a) and (b) show the best-performing case with unilateral stenosis under varying wall grid resolutions $N_g$. Figs.~\ref{fig:11}(c) and (d) depict a moderate case with bilateral stenosis, while Figs.~\ref{fig:11}(e) and (f) correspond to the worst-performing unilateral case. In the best case, U-ResNet achieves high accuracy with only minor deviations from the CFD ground truth.
In the moderate and worst cases [Figs.~\ref{fig:11}(c)–(f)], U-ResNet trained on only $500$ samples shows noticeable discrepancies, especially in regions with steep pressure gradients. However, when trained with $2000$ samples, it achieves excellent agreement with CFD results, maintaining prediction errors below $2\%$ across the entire domain. These findings suggest that a minimum of approximately $2000$ training samples is required to ensure reliable pressure prediction in stenotic geometries, with further increases in dataset size yielding marginal performance gains.


\subsubsection{Prediction of WSS}
Table~\ref{tab:6} presents the wall shear stress (WSS) prediction accuracy for four neural network architectures, evaluated across $20$ randomly generated stenotic geometries at $Re = 200$. U-ResNet significantly outperforms all other models, achieving exceptional accuracy with a mean absolute error ($MAE$) of $0.0008 \pm 0.0009$, representing an $86\%$ reduction in error compared to FNO ($0.0057 \pm 0.0075$) and an $80\%$ improvement over UFNO ($0.0040 \pm 0.0023$). The normalized mean absolute error ($NMAE$) further highlights U-ResNet's superior performance, with an $NMAE$ of $0.56 \pm 0.44\%$ compared to FNO ($3.99 \pm 3.77\%$), UFNO ($3.01 \pm 1.52\%$), and U-Net ($5.70 \pm 2.47\%$). U-ResNet's $RMSE$ ($0.0013 \pm 0.0014$) is $4$ times lower than UFNO ($0.0052 \pm 0.0037$). U-ResNet achieves an $NRMSE$ of $0.8852 \pm 0.6673\%$, compared to $3.7910 \pm 1.6792\%$ for UFNO.

\begin{table}[tbp]
	\begin{center}
		\caption{Performance evaluation of predicted wall shear stress on the wall for 20 randomly generated stenosis at $Re=200$ using four neural network architectures, comparing mean absolute error ($MAE$), normalized mean absolute error ($NMAE$), root mean squared error ($RMSE$), and normalized root mean squared error ($NRMSE$). Here, these errors correspond to interpolation results.}\label{tab:6}
		\begin{tabular*}{1\textwidth}{@{\extracolsep{\fill}} lcccc }
			\hline\hline
			\small    
			Model & MAE & NMAE($\%$) & RMSE & NRMSE($\%$) \\ \hline
			U-Net & 0.0079 ± 0.0052    & 5.7000 ± 2.4734   & 0.0144 ± 0.009  & 10.1976 ± 4.2038\\
			U-ResNet & \textbf{0.0008 ± 0.0009}    & \textbf{0.5582 ± 0.4384 }   & \textbf{0.0013 ± 0.0014} & \textbf{0.8852 ± 0.6673}\\
			FNO & 0.0057 ± 0.0075   & 3.9851 ± 3.7693   & 0.0084 ± 0.0096    & 5.9032 ± 4.7887\\            
			UFNO & 0.0040 ± 0.0023     & 3.0097 ± 1.5177  & 0.0052 ± 0.0037 &  3.7910 ± 1.6792  \\ \hline\hline
		\end{tabular*}%
	\end{center}
\end{table}

\begin{figure}[htbp]
    \centering

    % 第一行
    \begin{subfigure}[b]{0.49\textwidth}
        \begin{overpic}[width=\linewidth]{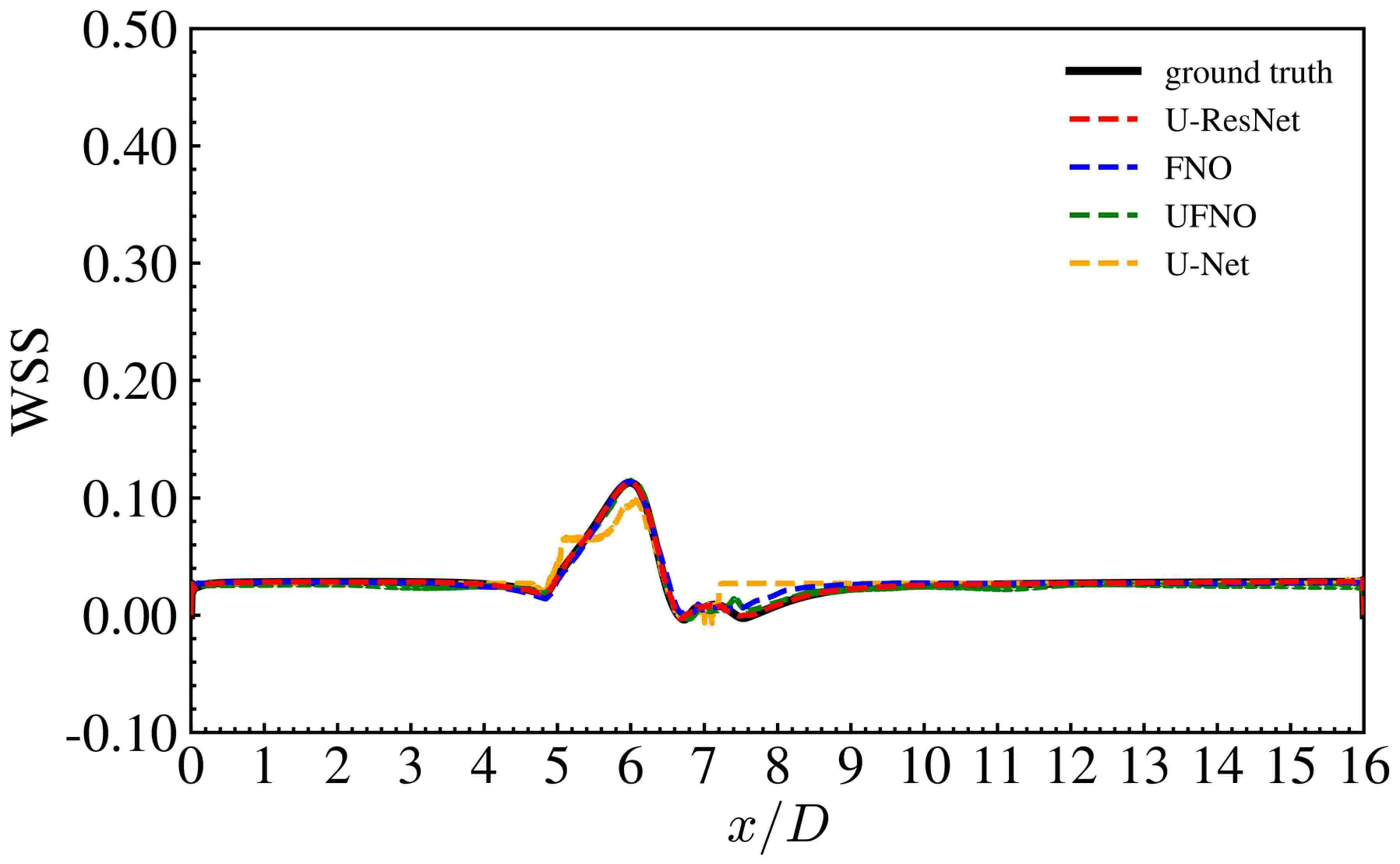}
            \put(-1,62){\small (a)}  
        \end{overpic}
    \end{subfigure}
    \hfill
    \begin{subfigure}[b]{0.49\textwidth}
        \begin{overpic}[width=\linewidth]{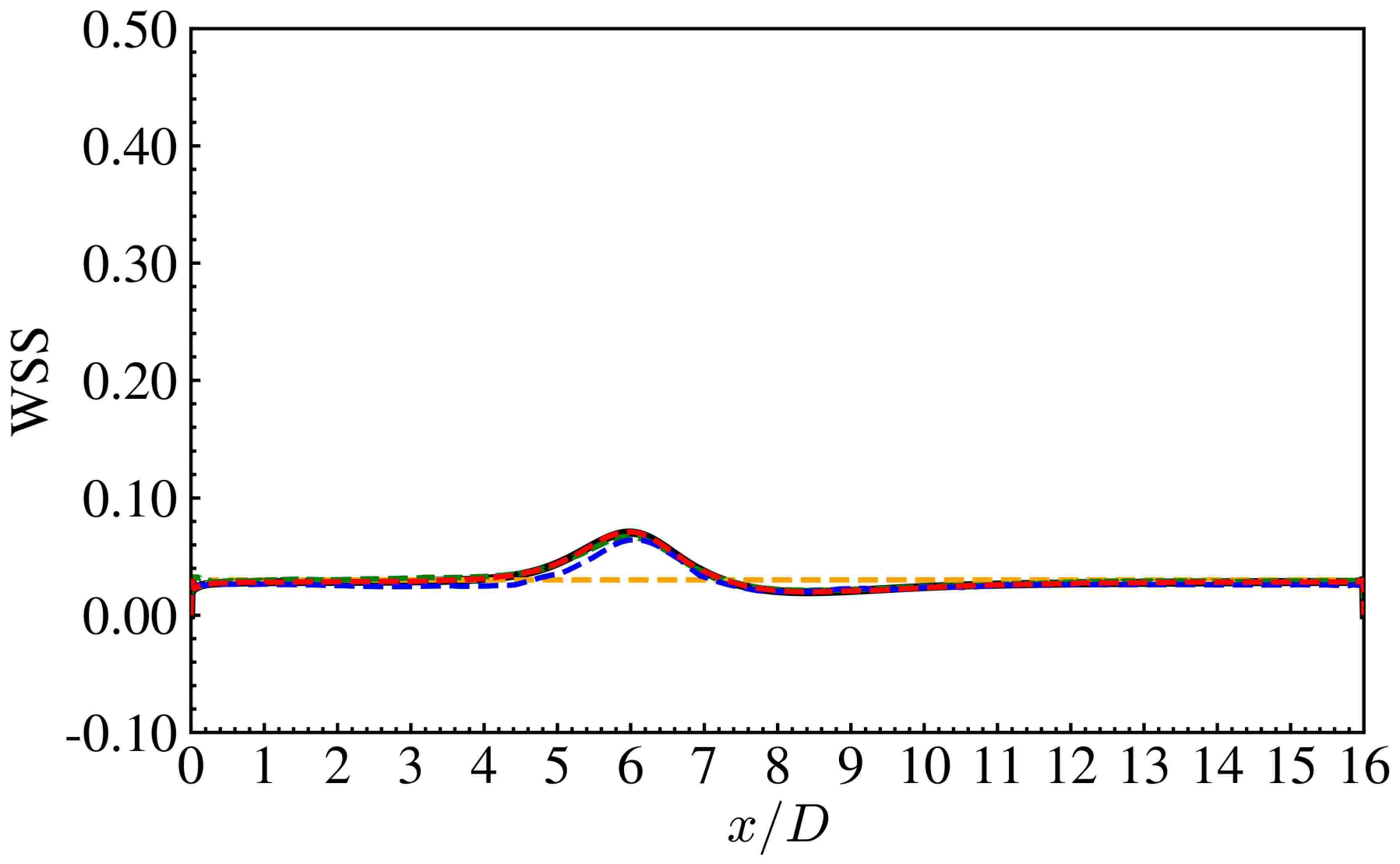}
            \put(-1,62){\small (b)} 
        \end{overpic} 
    \end{subfigure}
    \vspace{0.1cm}

    % 第二行
    \begin{subfigure}[b]{0.49\textwidth}
        \begin{overpic}[width=\linewidth]{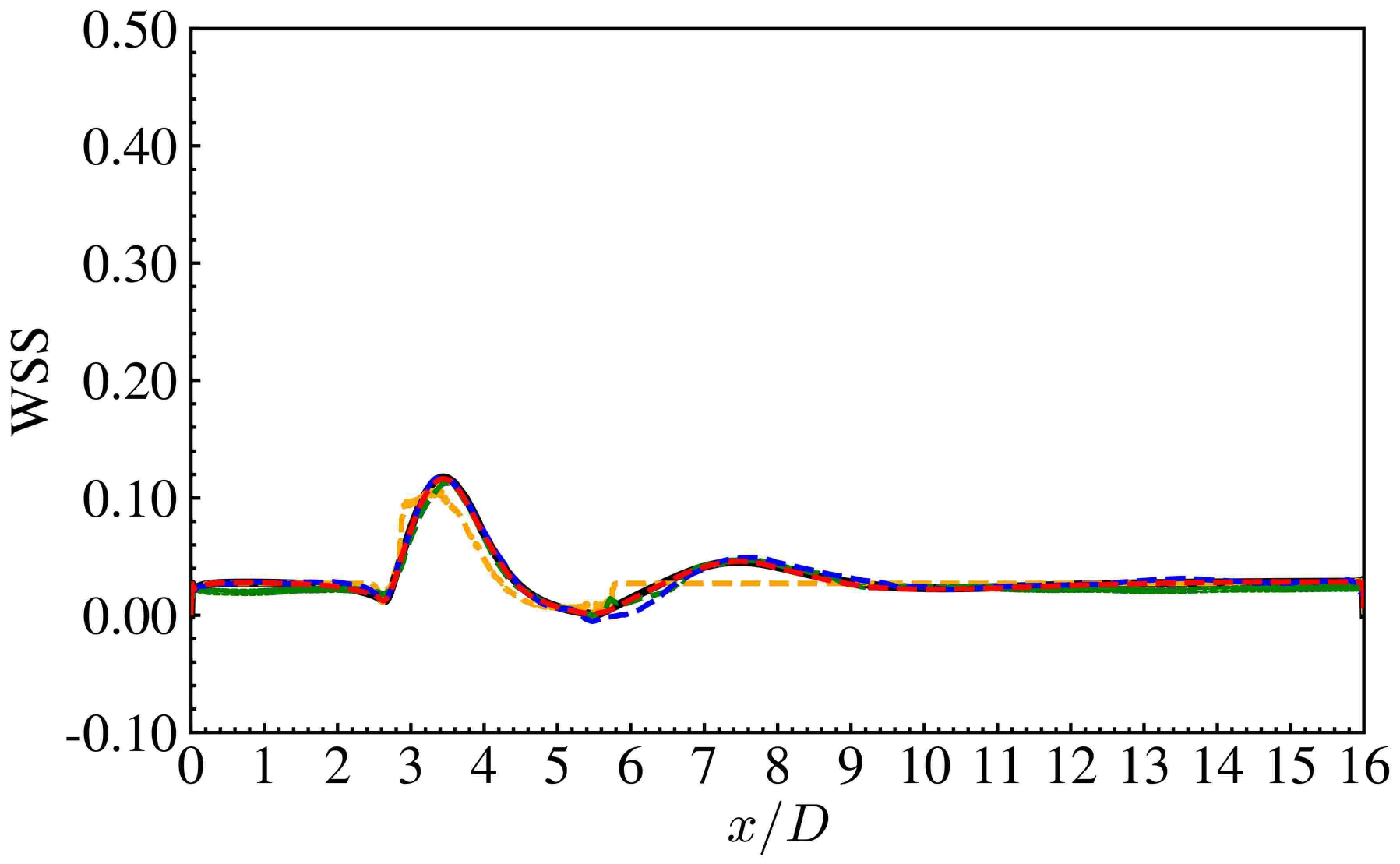}
            \put(-1,62){\small (c)} 
        \end{overpic}
    \end{subfigure}
    \hfill
    \begin{subfigure}[b]{0.49\textwidth}
        \begin{overpic}[width=\linewidth]{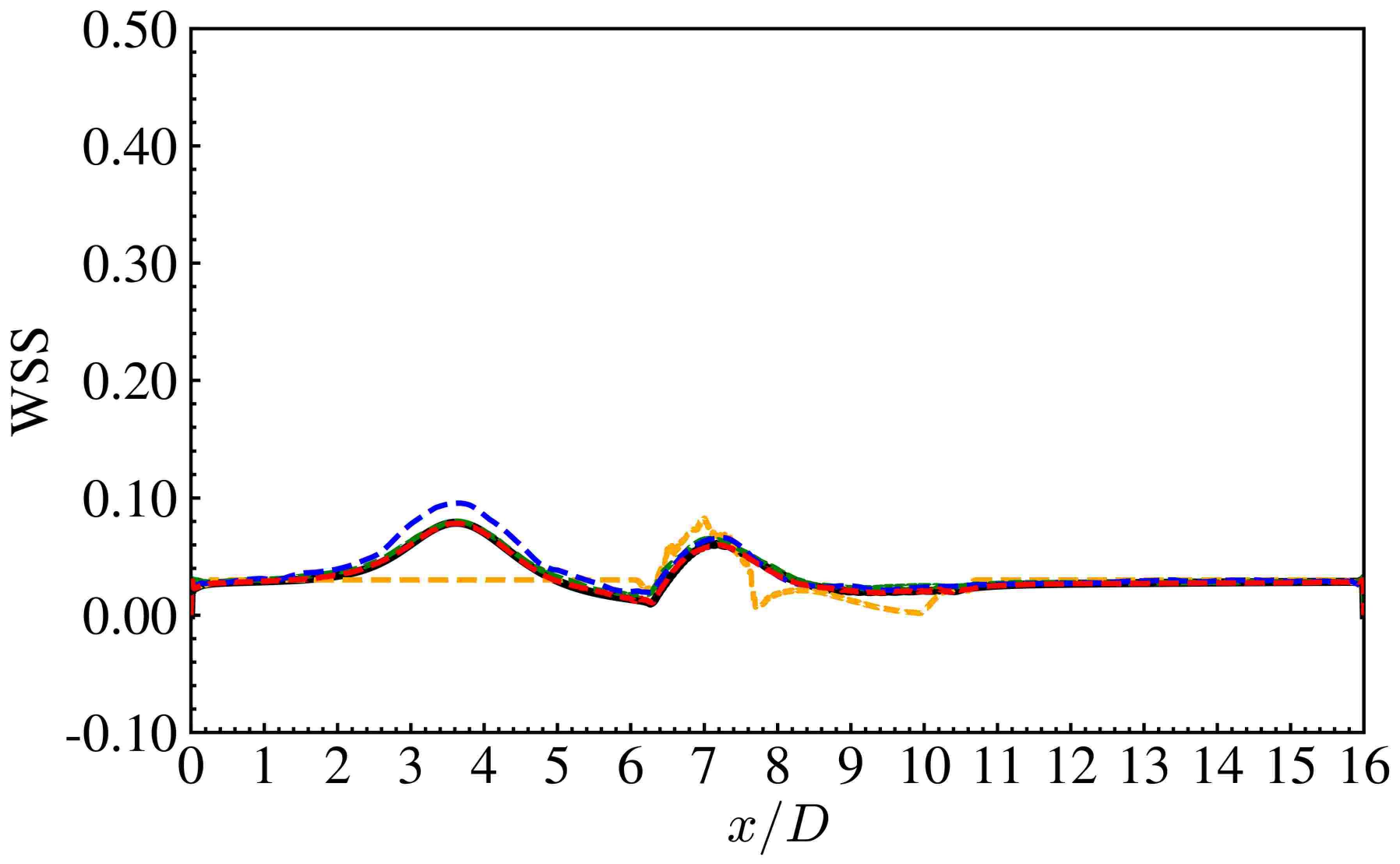}
            \put(-1,62){\small (d)} 
        \end{overpic}
    \end{subfigure}
    \vspace{0.1cm}

    % 第三行
    \begin{subfigure}[b]{0.49\textwidth}
        \begin{overpic}[width=\linewidth]{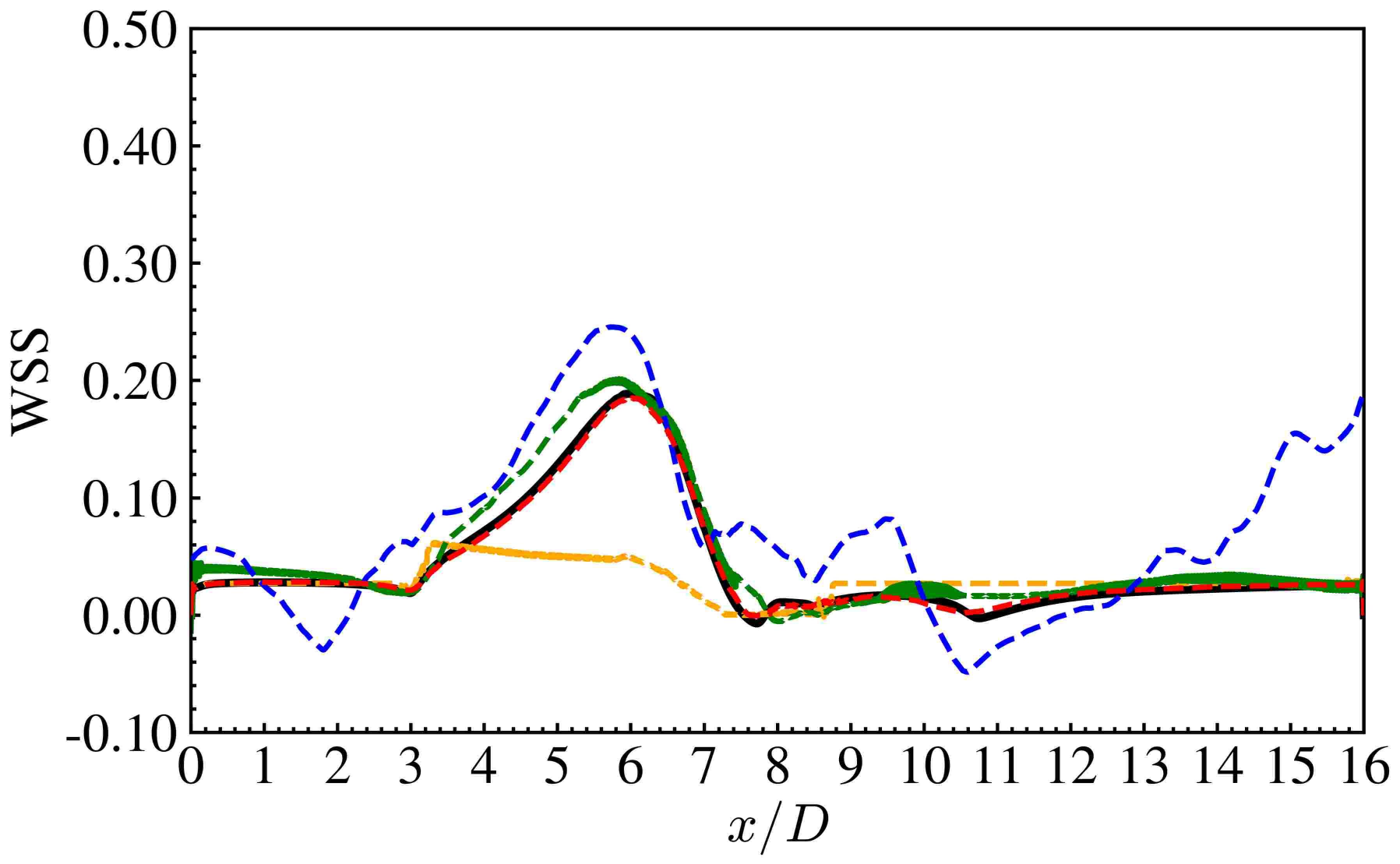}
            \put(-1,62){\small (e)} 
        \end{overpic}
    \end{subfigure}
    \hfill
    \begin{subfigure}[b]{0.49\textwidth}
        \begin{overpic}[width=\linewidth]{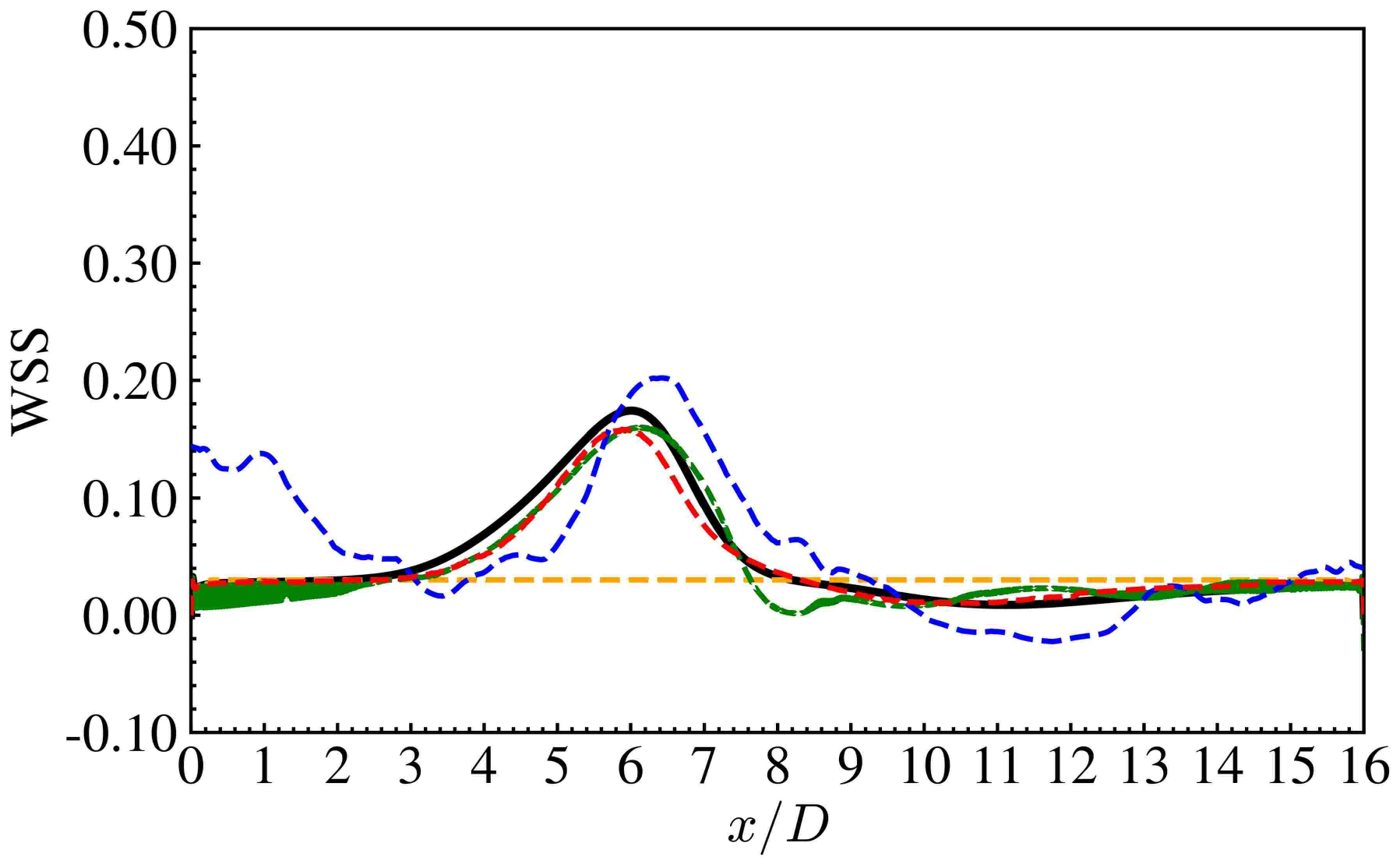}
            \put(-1,62){\small (f)}  
        \end{overpic}
    \end{subfigure}
	\caption{Wall shear stress (WSS) prediction on the walls for randomly generated stenosis cases at $Re=200$ using four neural network models: (a) Best performance case: upper wall, (b) Best performance case: lower wall, (c) Moderate performance case: upper wall, (d) Moderate performance case: lower wall, (e) Worst performance case: upper wall, (f) Worst performance case: lower wall.}\label{fig:12}
\end{figure}

Fig.~\ref{fig:12} presents WSS predictions for three representative cases from 20 test cases. U-ResNet accurately captures both the magnitude and spatial distribution of WSS peaks, which is crucial for assessing endothelial dysfunction and atherosclerosis progression. In contrast, other models exhibit notable deviations in peak intensity and location.
Figs.~\ref{fig:12}(a)–(b) show the best-performing case with unilateral stenosis and a single WSS peak; Figs.~\ref{fig:12}(c)–(d) represent a moderate case with bilateral stenosis and two WSS peaks; and Figs.~\ref{fig:12}(e)–(f) correspond to the worst case with unilateral stenosis.
In the worst case, only U-ResNet successfully resolves peak WSS with high fidelity, while U-Net, FNO, and UFNO fail to capture accurate magnitude or position. In the moderate and best cases, U-ResNet, FNO, and UFNO perform comparably well, whereas U-Net consistently underperforms. Across all scenarios, U-ResNet maintains deviation from CFD results below $0.5\%$, highlighting its superior accuracy and robustness.

\begin{figure}[htbp]\centering
	\includegraphics[width=0.7\textwidth]{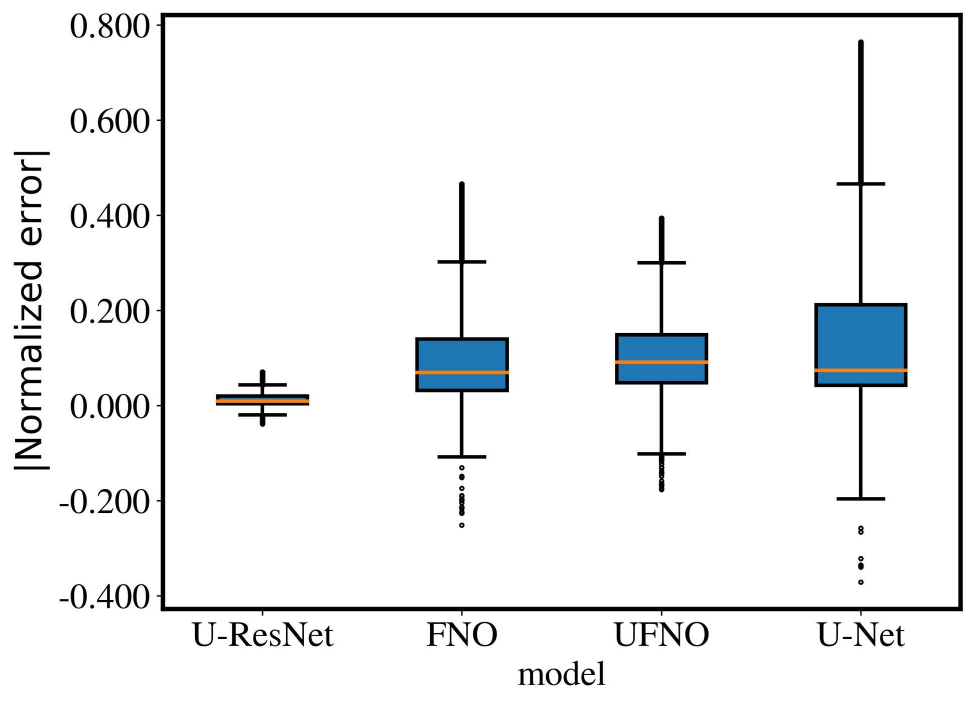}
	\caption{The box plots of wall shear stress prediction error for four neural network models. Here, the errors correspond to interpolation results.}\label{fig:13}
\end{figure}

Statistical analysis of prediction errors, presented in Fig.~\ref{fig:13}, confirms U-ResNet's superior performance through box plot distributions. U-ResNet exhibits the most concentrated error range with minimal outliers, indicating consistent performance across diverse stenotic geometries. This robust prediction capability is essential for clinical applications where accurate WSS quantification serves as a critical hemodynamic descriptor for assessing stenosis severity and predicting atherosclerotic disease progression.

\begin{figure}[htbp]\centering
	\includegraphics[width=0.7\textwidth]{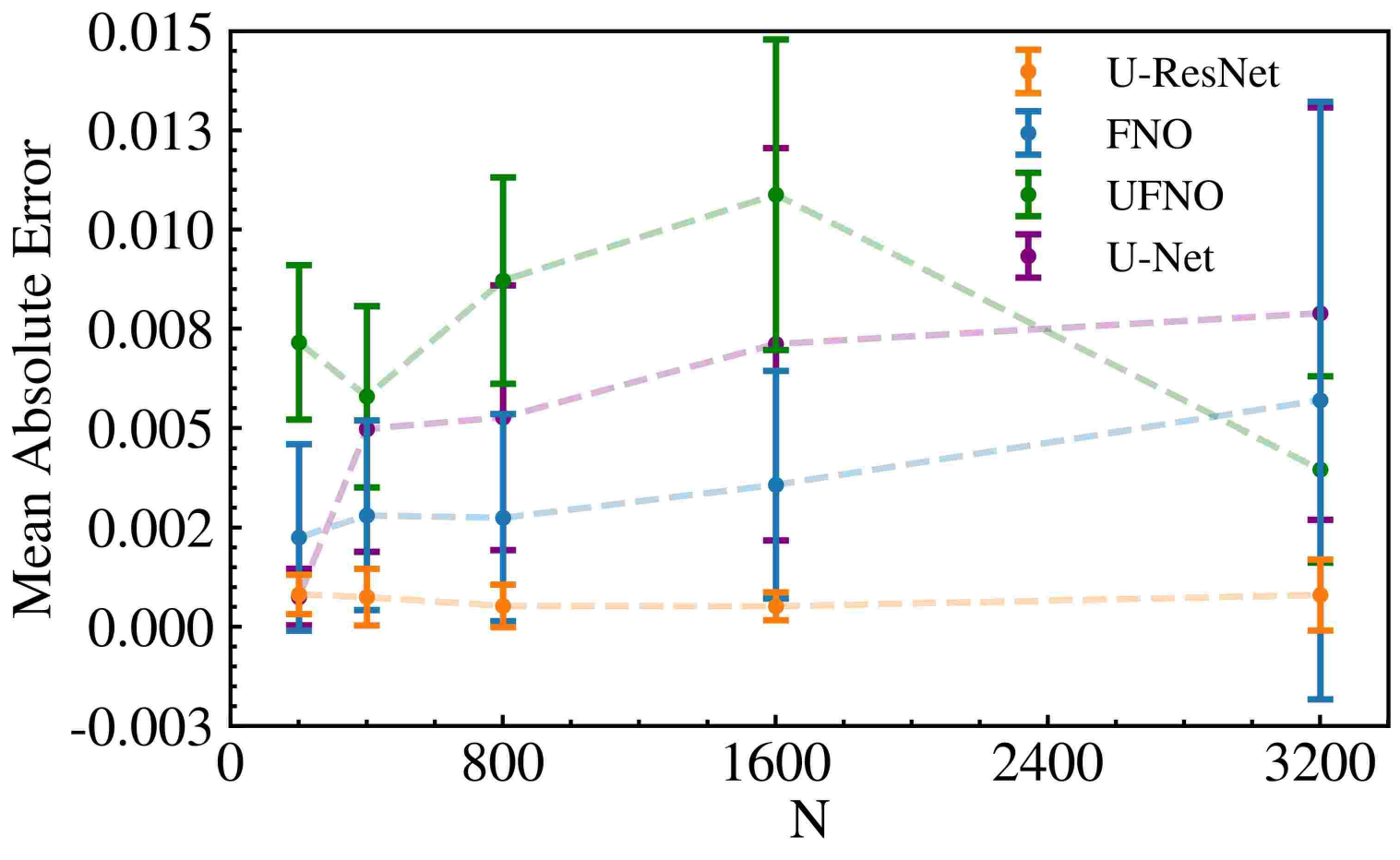}
	\caption{WSS prediction mean absolute error when choosing different wall grid resolutions ($N_g=201$, 401, 801, 1601, 3201). Here, the errors correspond to interpolation results.}\label{fig:14}
\end{figure}

Fig.~\ref{fig:14} evaluates the wall shear stress (WSS) prediction accuracy of four neural network architectures (U-Net, U-ResNet, FNO, UFNO) across varying grid resolutions. U-ResNet demonstrates exceptional grid invariance, maintaining a mean absolute error ($MAE$) below $0.0015$ for wall grid resolutions ranging from $N_g=201$ to $N_g=3201$.  In contrast, UFNO exhibit significant error amplification at coarser resolutions, while FNO and U-Net fail to converge reliably at any resolution. It is noted that the MAE for U-ResNet increases slightly as the resolution increases from 1600 to 3200. The resolution-dependent MAE trend reflects U-ResNet’s architectural constraints when extrapolating to finer grids. While the model excels at its trained resolution (1600 grids), higher resolutions (3200 grids) introduce unresolved high-frequency features due to the fixed receptive field and training data limitations.

\begin{figure}[htbp]
    \centering
    % 第1行	
    \begin{subfigure}[b]{0.49\textwidth}
        \begin{overpic}[width=\linewidth]{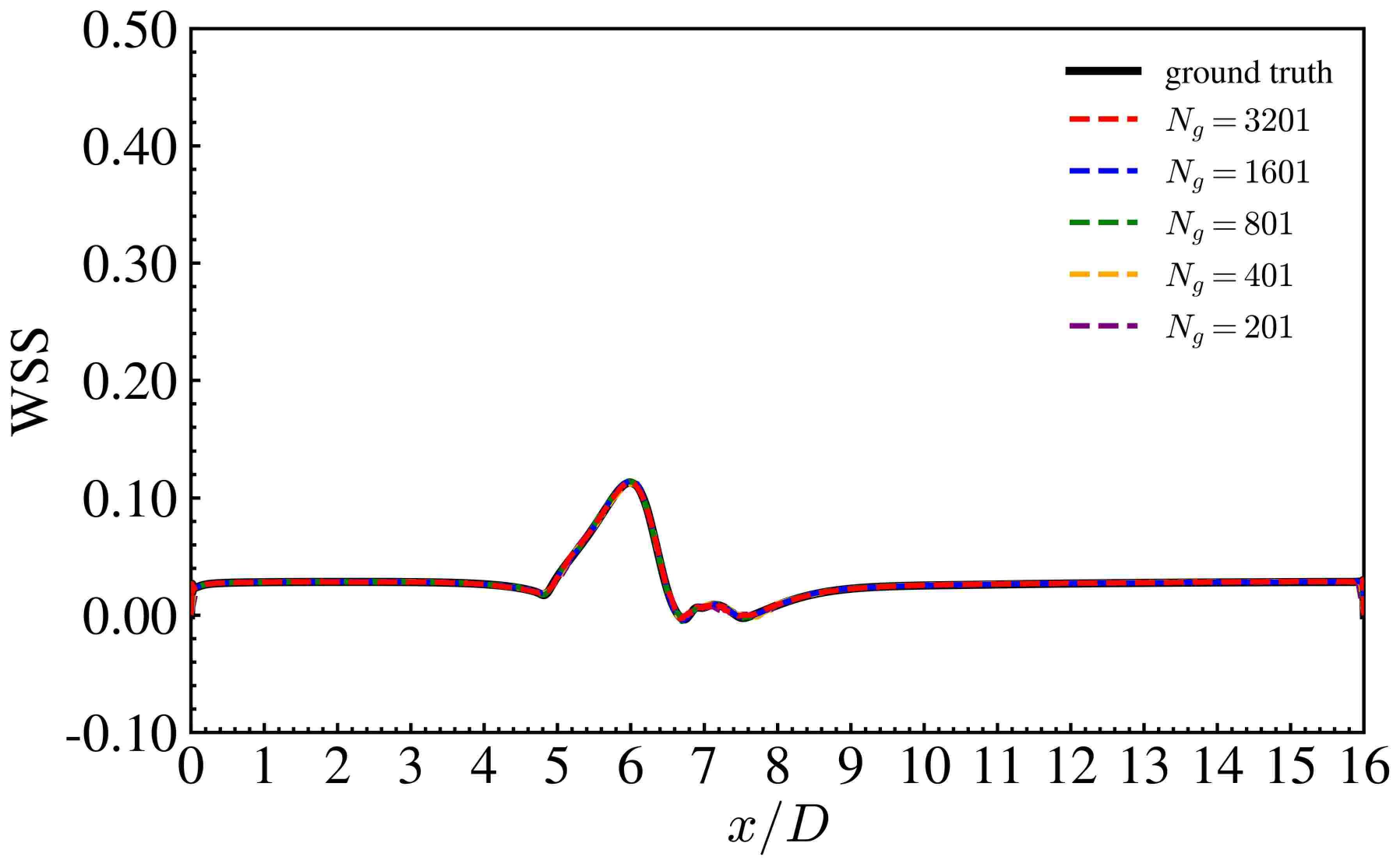}
            \put(-1,62){\small (a)}  
        \end{overpic}
    \end{subfigure}
    \hfill
    \begin{subfigure}[b]{0.49\textwidth}
        \begin{overpic}[width=\linewidth]{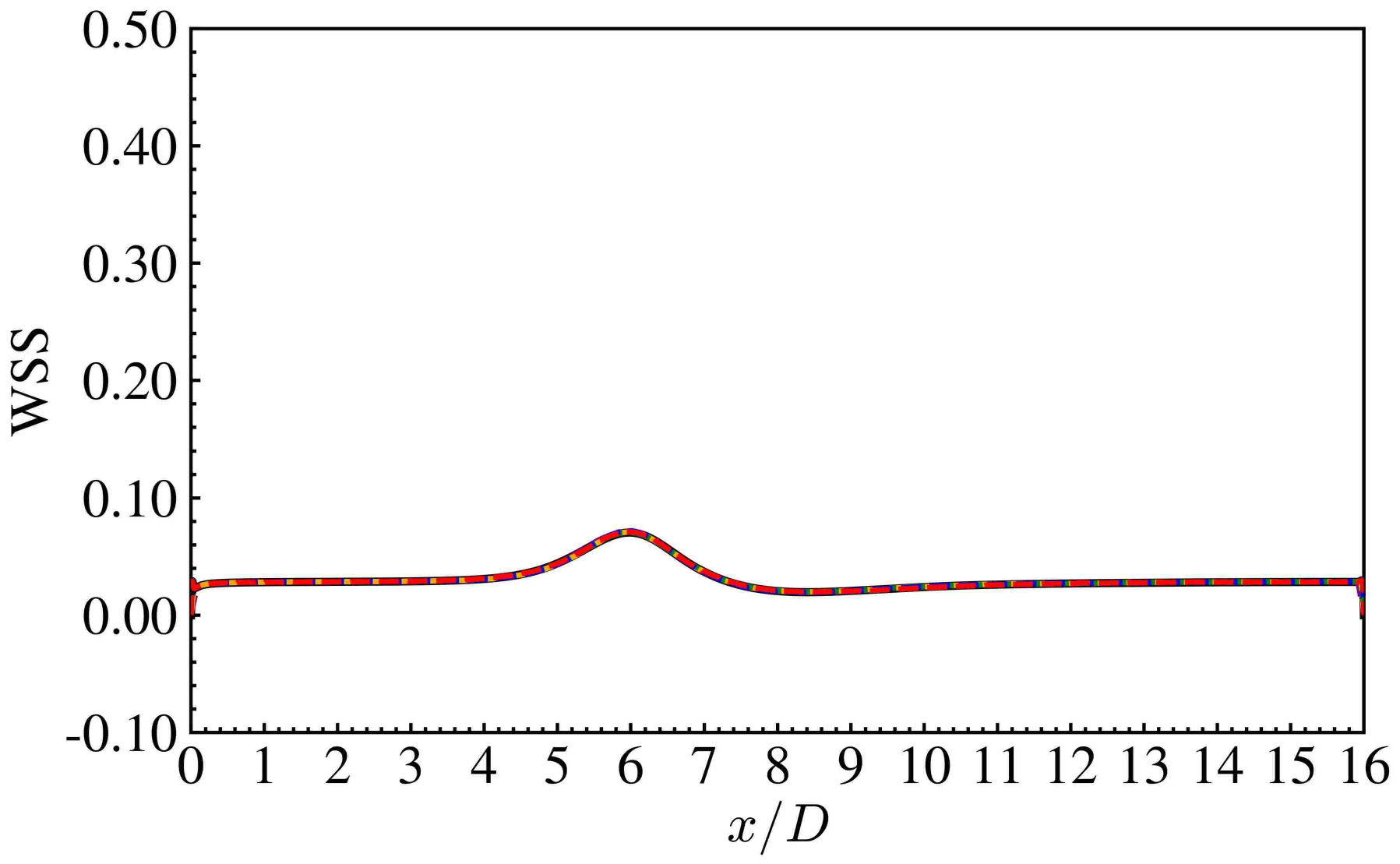}
            \put(-1,62){\small (b)}  
        \end{overpic}
    \end{subfigure}

    \vspace{0.1cm} % 添加垂直间距

    % 第2行
    \begin{subfigure}[b]{0.49\textwidth}
        \begin{overpic}[width=\linewidth]{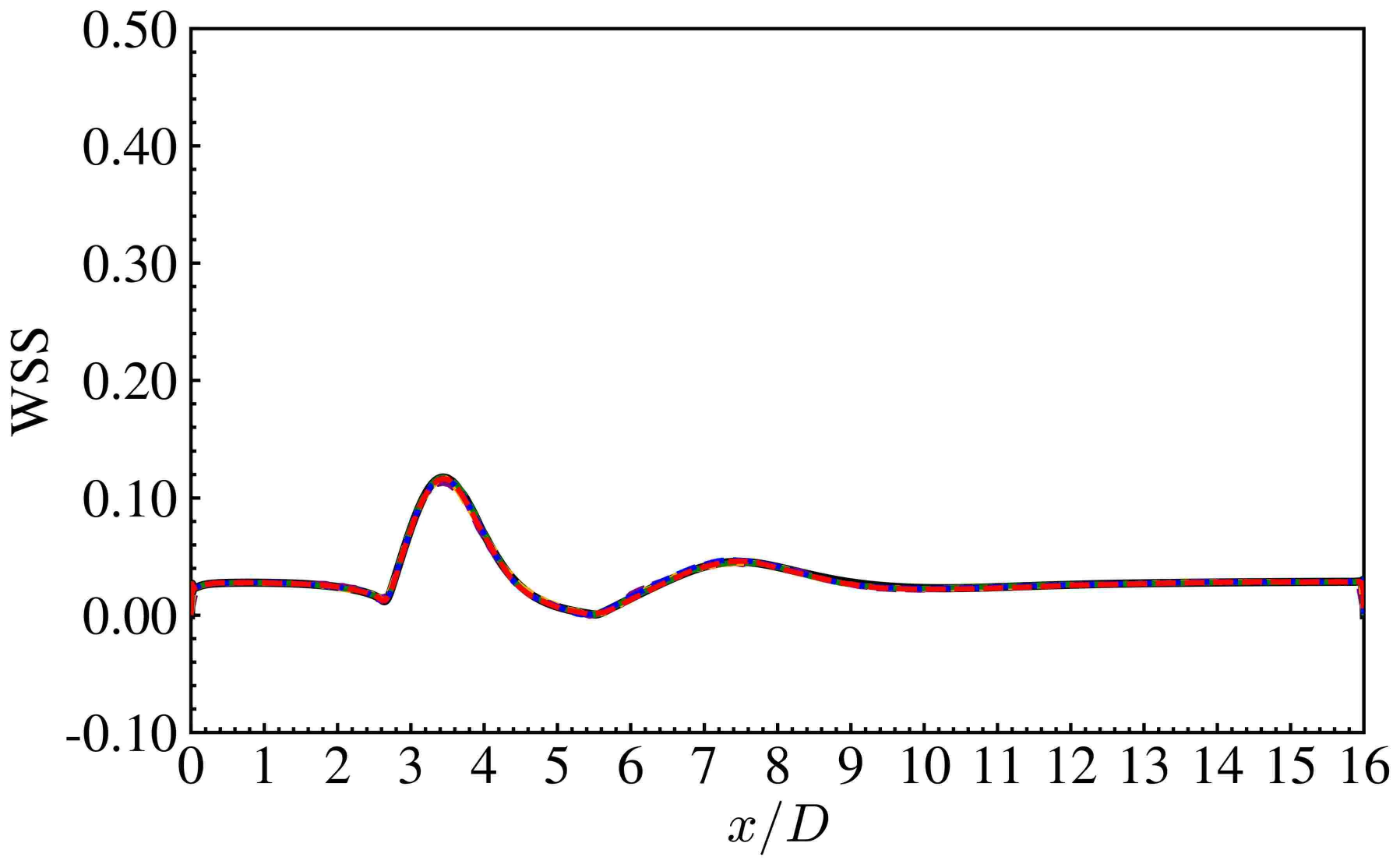}
            \put(-1,62){\small (c)}  
        \end{overpic}
    \end{subfigure}
    \hfill
    \begin{subfigure}[b]{0.49\textwidth}
        \begin{overpic}[width=\linewidth]{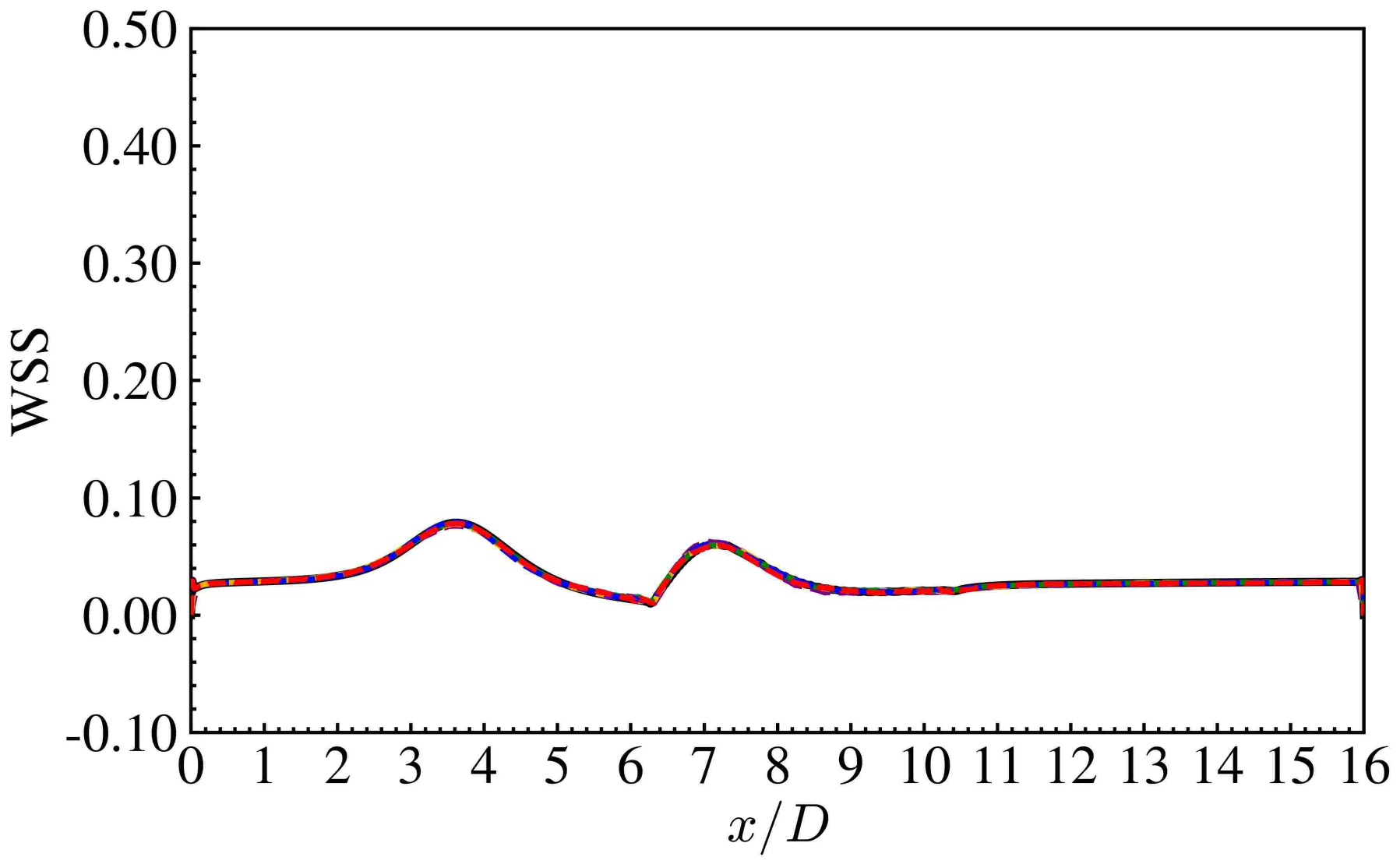}
            \put(-1,62){\small (d)}  
        \end{overpic}
    \end{subfigure}

    \vspace{0.1cm} % 添加垂直间距

    % 第3行
    \begin{subfigure}[b]{0.49\textwidth}
        \begin{overpic}[width=\linewidth]{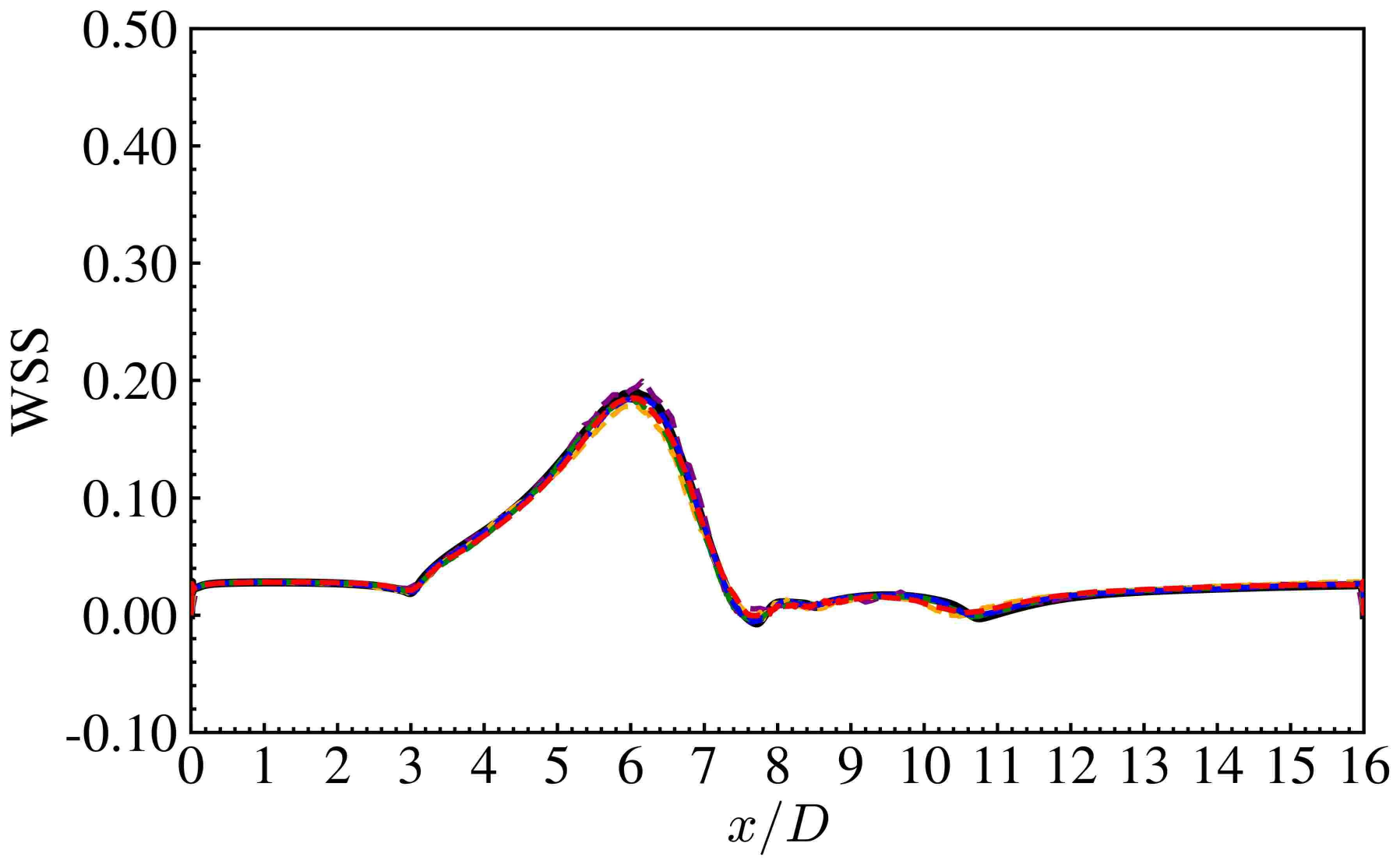}
            \put(-1,62){\small (e)}  
        \end{overpic}
    \end{subfigure}
    \hfill
    \begin{subfigure}[b]{0.49\textwidth}
        \begin{overpic}[width=\linewidth]{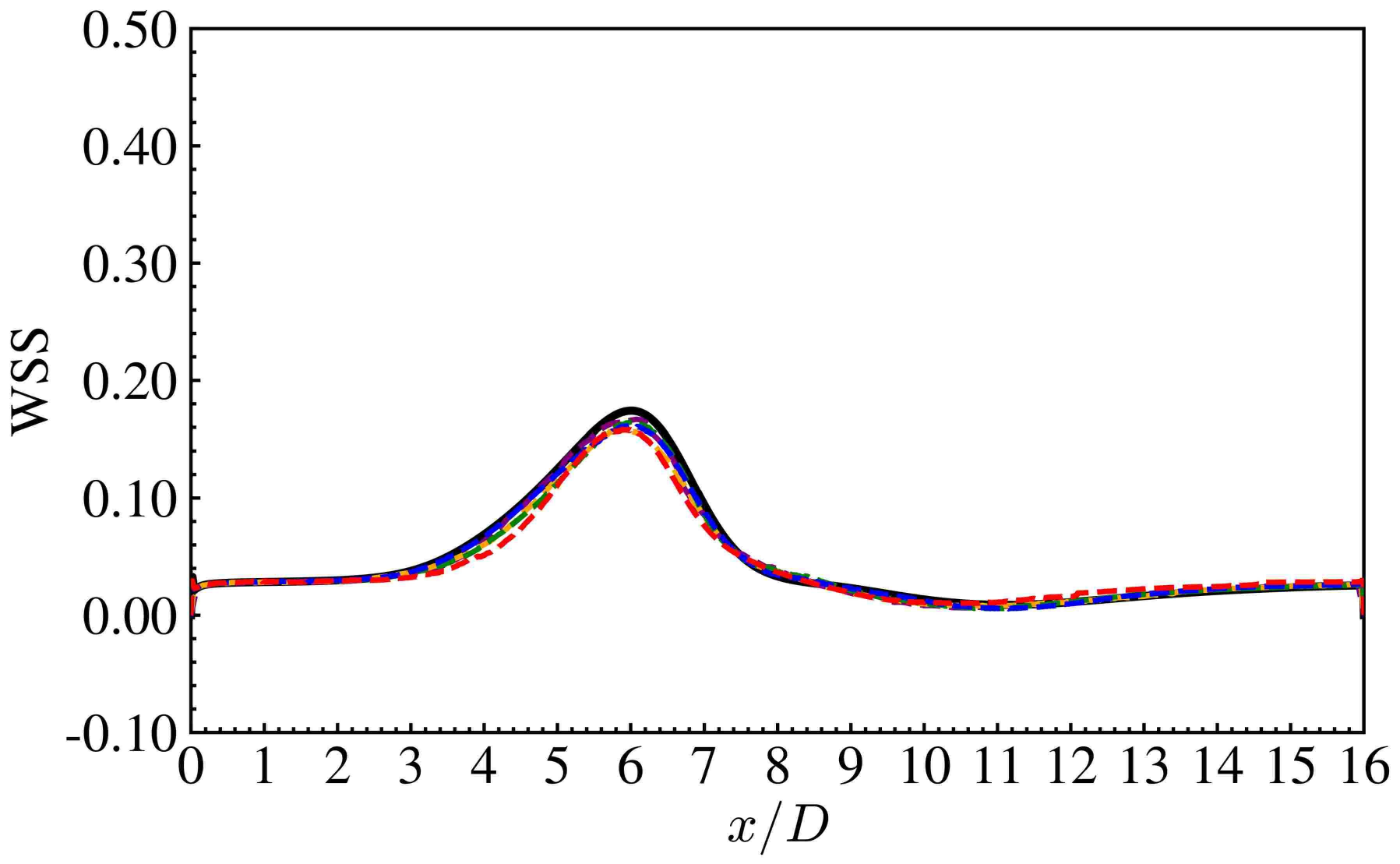}
            \put(-1,62){\small (f)}  
        \end{overpic}
    \end{subfigure}
    
	\caption{WSS prediction on the walls for U-ResNet when choosing different grid point counts ($N_g=201$, 401, 801, 1601, 3201): (a) Best performance case: upper wall, (b) Best performance case: lower wall, (c) Moderate performance case: upper wall, (d) Moderate performance case: lower wall, (e) Worst performance case: upper wall, (f) Worst performance case: lower wall.}\label{fig:15}
\end{figure}

Fig.~\ref{fig:15} demonstrates U-ResNet’s grid robustness by comparing WSS distributions across three representative cases: the best-performing case with unilateral stenosis [Figs.~\ref{fig:15}(a)–(b)], a moderate case with bilateral stenosis [Figs.~\ref{fig:15}(c)–(d)], and the worst-performing case with unilateral stenosis [Figs.~\ref{fig:15}(e)–(f)]. 
Across all scenarios, U-ResNet maintains high spatial accuracy despite a $16\times$ reduction in wall grid resolution (from $N_g = 3201$ to $201$), with less than $3\%$ deviation in WSS peak magnitude and under $2\%$ spatial error in recirculation zone prediction. This highlights its robustness and reliability in capturing key hemodynamic features essential for clinical evaluation of endothelial shear stress.

\begin{figure}[htbp]\centering
	\includegraphics[width=0.7\textwidth]{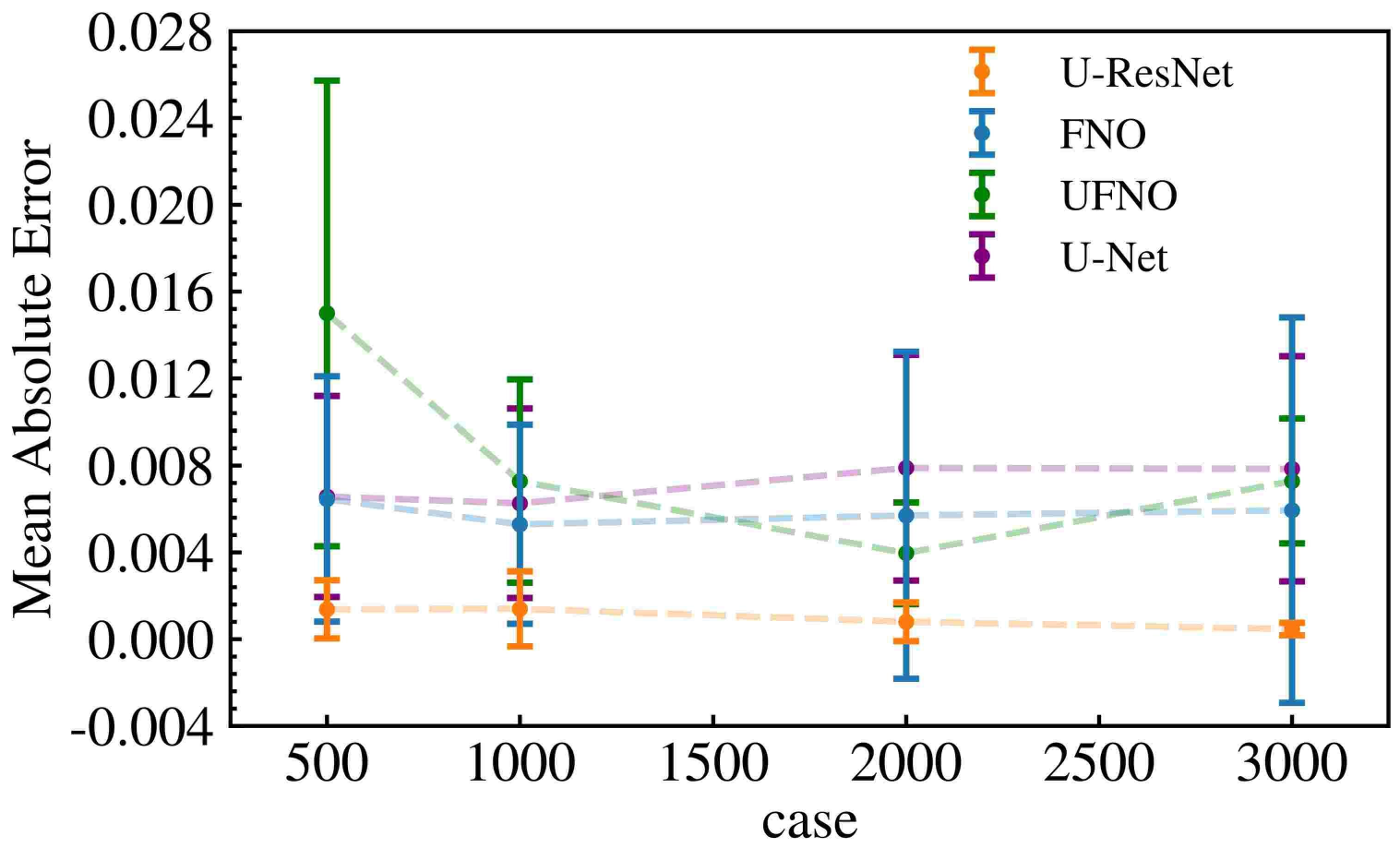}
	\caption{WSS prediction mean absolute error when choosing different training sample sizes ($N_s=500$, 1000, 2000, 3000). Here, the errors correspond to interpolation results.}\label{fig:16}
\end{figure}

Fig.~\ref{fig:16} quantifies the relationship between training sample size ($N_s=500$,1000,2000,3000) and WSS prediction $MAE$. U-ResNet achieves asymptotic error convergence at $N_s=1000$ ($MAE$ = 0.0011±0.0008), with marginal improvements (<5\%) at larger sample sizes. U-Net, FNO, and UFNO fail to converge even at $N_s=2000$.

\begin{figure}[htbp]
    \centering
    % 第一行	
    \begin{subfigure}[b]{0.49\textwidth}
        \begin{overpic}[width=\linewidth]{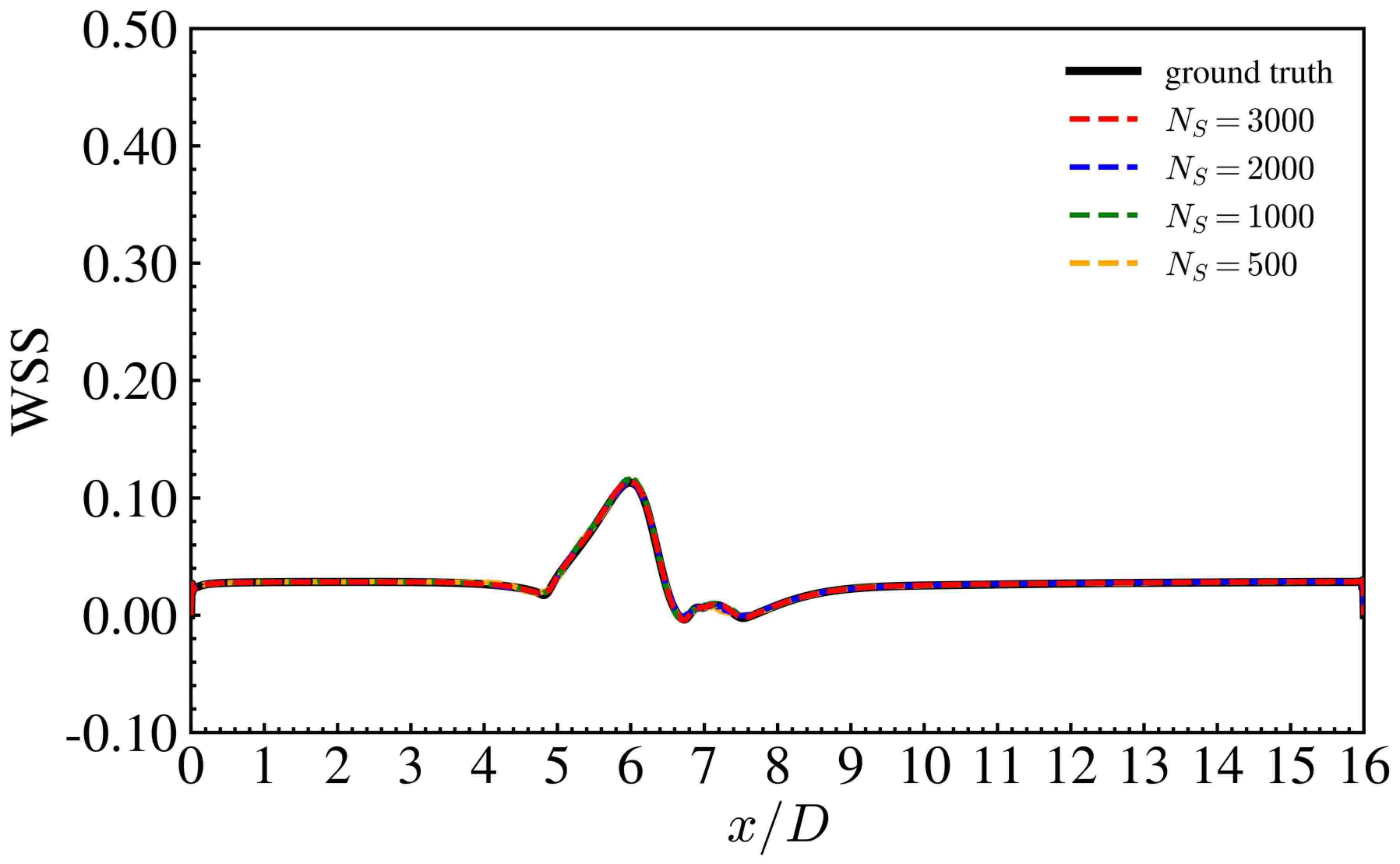}
            \put(-1,62){\small (a)}  
        \end{overpic}
    \end{subfigure}
    \hfill
    \begin{subfigure}[b]{0.49\textwidth}
        \begin{overpic}[width=\linewidth]{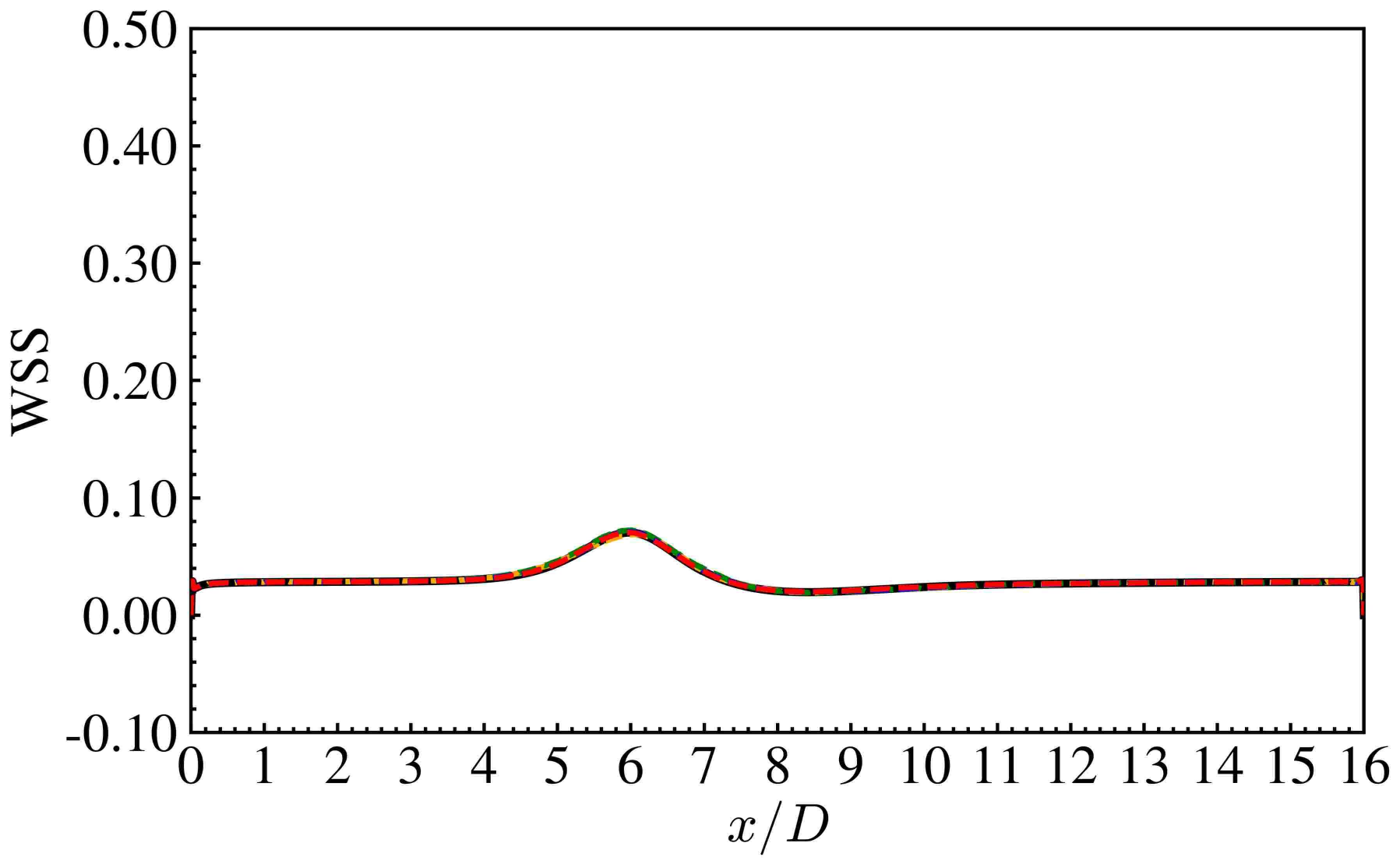}
            \put(-1,62){\small (b)}  
        \end{overpic}
    \end{subfigure}

    \vspace{0.1cm} % 添加垂直间距

    % 第2行
    \begin{subfigure}[b]{0.49\textwidth}
        \begin{overpic}[width=\linewidth]{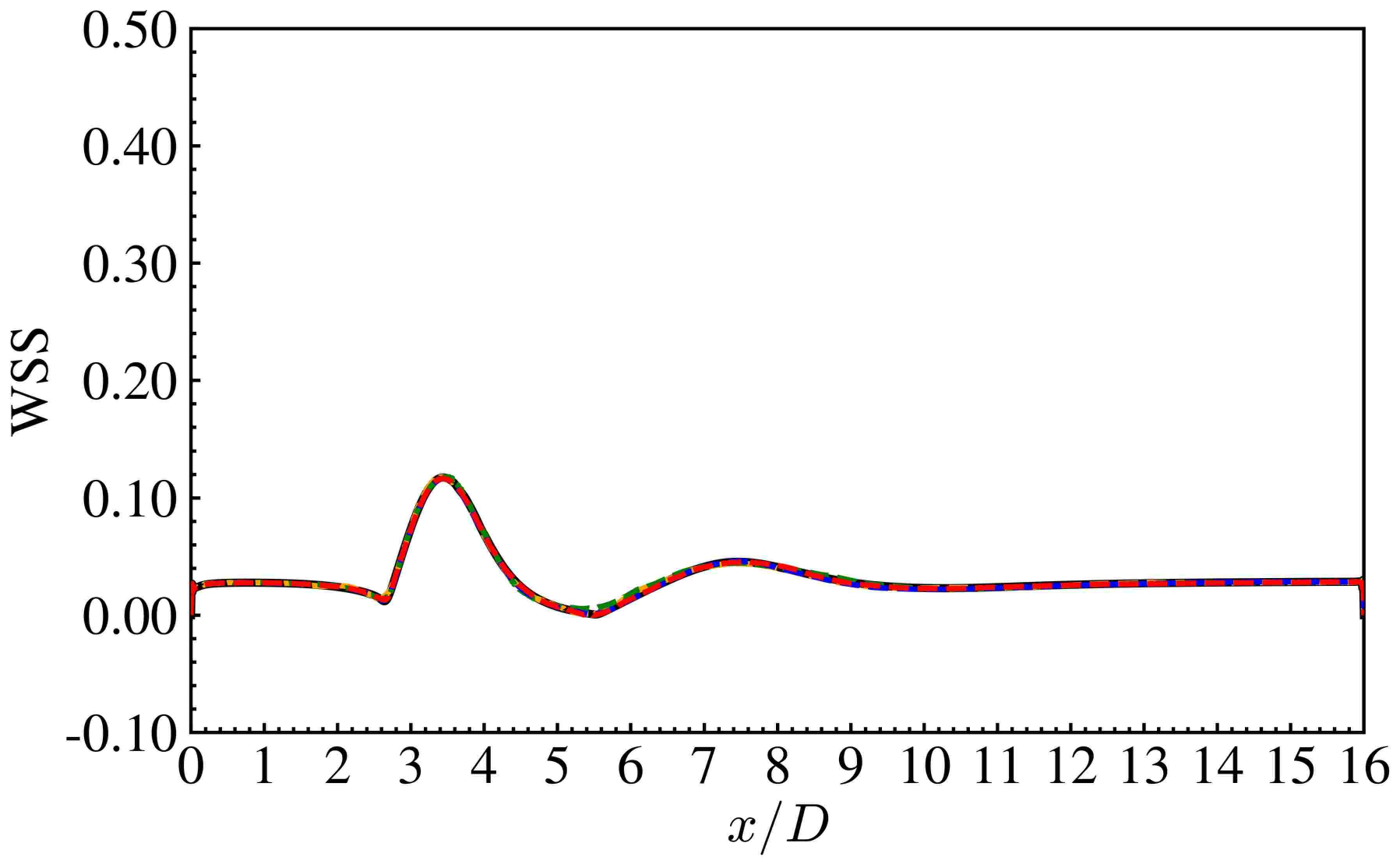}
            \put(-1,62){\small (c)}  
        \end{overpic}
    \end{subfigure}
    \hfill
    \begin{subfigure}[b]{0.49\textwidth}
        \begin{overpic}[width=\linewidth]{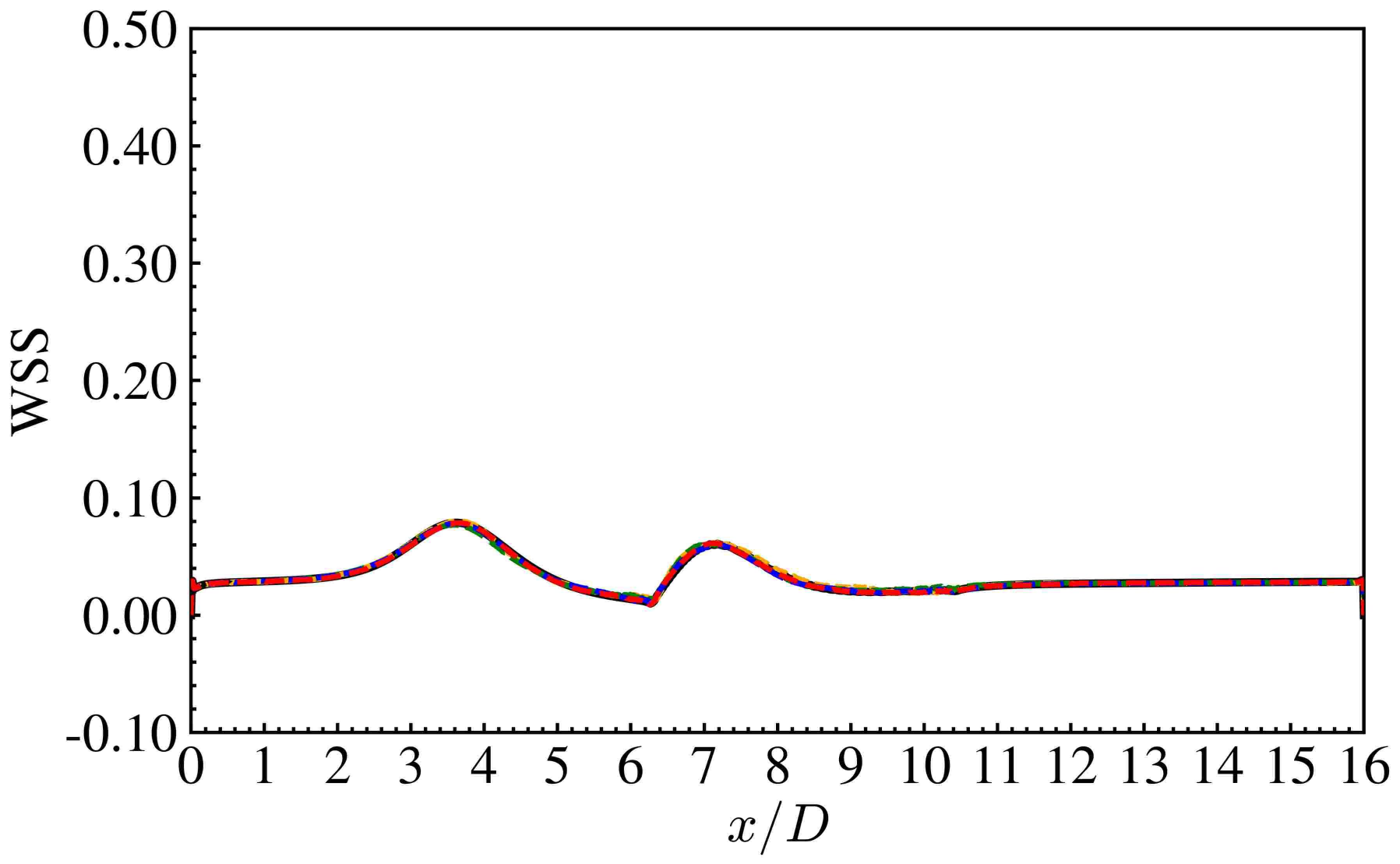}
            \put(-1,62){\small (d)}  
        \end{overpic}
    \end{subfigure}	

    \vspace{0.1cm} % 添加垂直间距

    % 第3行
    \begin{subfigure}[b]{0.49\textwidth}
        \begin{overpic}[width=\linewidth]{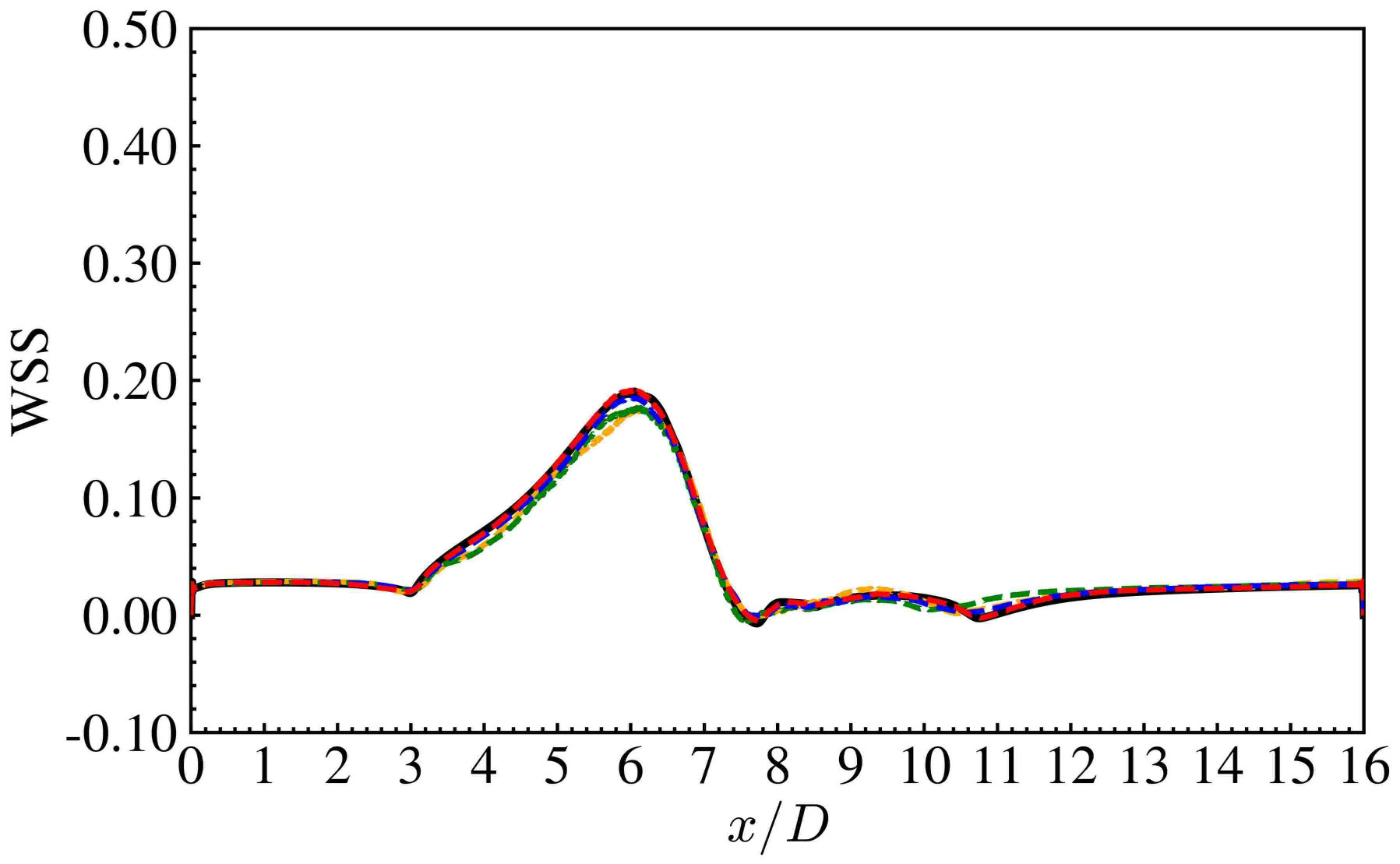}
            \put(-1,62){\small (e)}  
        \end{overpic}
    \end{subfigure}
    \hfill
    \begin{subfigure}[b]{0.49\textwidth}
        \begin{overpic}[width=\linewidth]{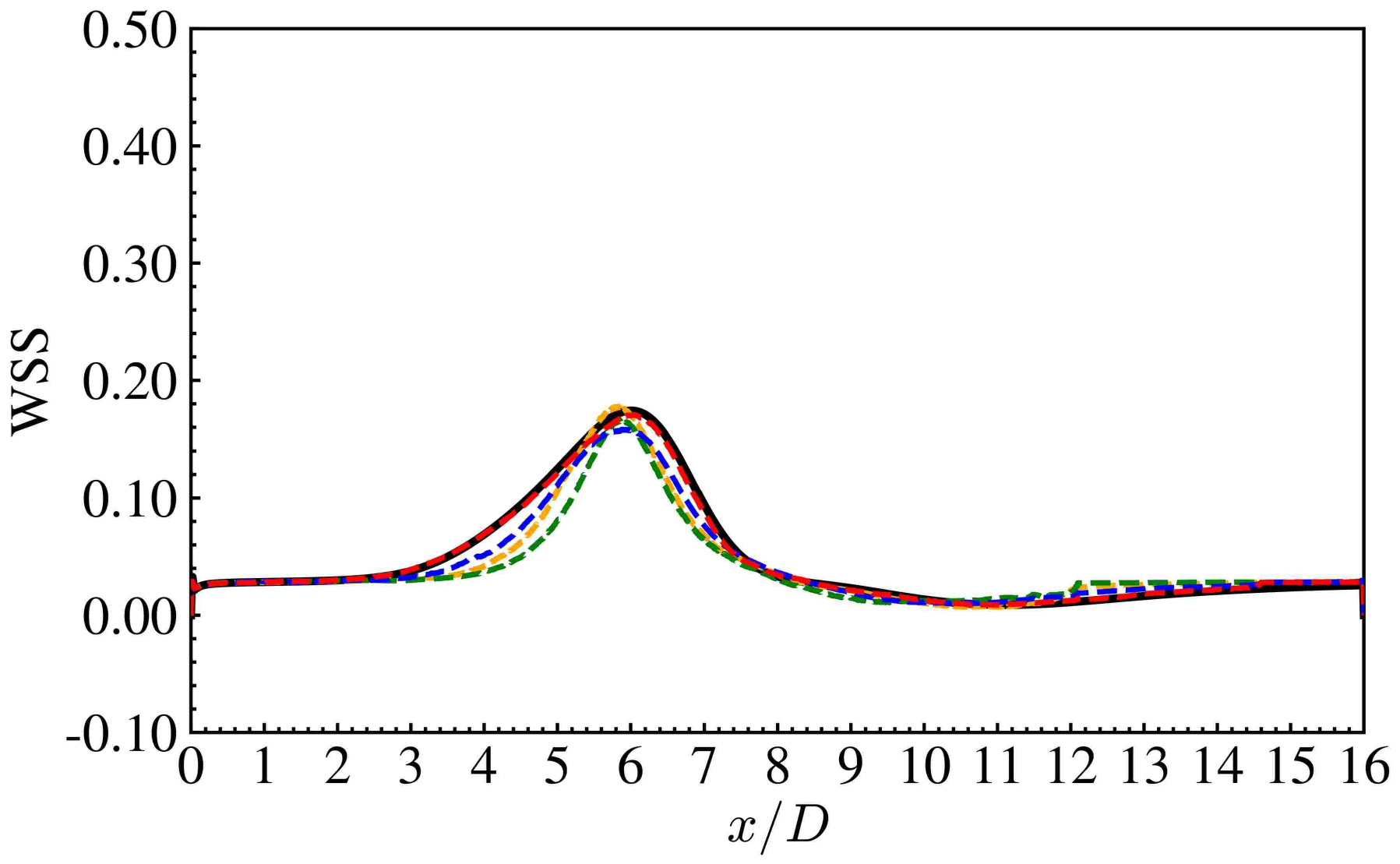}
            \put(-1,62){\small (f)}  
        \end{overpic}
    \end{subfigure}	    
    
	\caption{WSS prediction on the walls for U-ResNet when choosing different training sample sizes ($N_s=500$, 1000, 2000, 3000): (a) Best performance case: upper wall, (b) Best performance case: lower wall, (c) Moderate performance case: upper wall, (d) Moderate performance case: lower wall, (e) Worst performance case: upper wall, (f) Worst performance case: lower wall.}\label{fig:17}
\end{figure}

Fig.~\ref{fig:17} investigates the impact of training sample size $N_s$ on WSS prediction across three representative cases: the best-performing case with unilateral stenosis [Figs.~\ref{fig:17}(a)–(b)], a moderate case with bilateral stenosis [Figs.~\ref{fig:17}(c)–(d)], and the worst case with unilateral stenosis [Figs.~\ref{fig:17}(e)–(f)].
In the best and moderate cases, U-ResNet trained with as few as $1000$ samples yields WSS profiles that closely match those obtained with larger training sets, maintaining error deviations below $0.5\%$. In contrast, the worst case requires approximately $3000$ samples to achieve comparable accuracy. These results suggest that $1000$ training samples are sufficient for most geometries to capture the complex mapping between stenosis morphology and WSS distribution, while more challenging cases may necessitate larger datasets for convergence.

\subsubsection{Prediction of velocity field }

Table~\ref{tab:7} compares the prediction accuracy of U-ResNet and FNO for the streamwise velocity component across 20 stenotic flow cases at $Re=200$. U-ResNet achieves superior performance, with a mean absolute error ($MAE$) of $0.0227\pm0.0142$, representing an 83\% reduction compared to FNO ($0.1328\pm0.0206$). Similarly, U-ResNet’s normalized $MAE$ ($NMAE$) of $1.0572\pm0.4755\%$ is 84\% lower than FNO’s $6.4840\pm0.6415\%$. The root mean squared error ($RMSE$) and normalized $RMSE$ ($NRMSE$) further highlight U-ResNet’s precision, with errors $6\times$ smaller than FNO in both metrics. These results demonstrate U-ResNet’s ability to resolve bulk flow dynamics and streamwise velocity gradients with high fidelity, critical for assessing flow rate and momentum transfer in stenotic regions.

\begin{table}[tbp]
	\begin{center}
		\caption{Performance evaluation of predicted velocity field in the streamwise direction $u_x$ for 20 randomly generated stenosis at $Re=200$ using two neural network models. Here, these errors correspond to interpolation results.}\label{tab:7}
		\begin{tabular*}{1\textwidth}{@{\extracolsep{\fill}} lcccc }
			\hline\hline
			\small    
			Model & MAE & NMAE($\%$) & RMSE & NRMSE($\%$) \\ \hline
			U-ResNet & \textbf{0.0227 ± 0.0142}    & \textbf{1.0572 ± 0.4755 }   & \textbf{0.0418 ± 0.0249} & \textbf{1.9513 ± 0.8171}\\          
			FNO & 0.1328 ± 0.0206     & 6.4840 ± 0.6415  & 0.2514 ± 0.0492 & 12.2173 ± 1.2425  \\ \hline\hline
		\end{tabular*}%
	\end{center}
\end{table}

For the vertical velocity component (Table~\ref{tab:8}), U-ResNet maintains its performance advantage, though the margin narrows due to the inherently smaller magnitude of normal velocities in stenotic flows. U-ResNet achieves an $MAE$ of $0.0100\pm0.0044$, 31\% lower than FNO ($0.0144\pm0.0053$). The $NMAE$ ($2.8780\pm1.0718\%$ vs $4.0364\pm1.1569\%$) and $NRMSE$ ($5.8309\pm1.8024\%$ vs $8.9901\pm2.0905\%$) further confirm U-ResNet’s superiority in capturing secondary flow patterns and recirculation zones, which are essential for evaluating wall shear stress distributions and vortex formation.

\begin{table}[tbp]
	\begin{center}
		\caption{Performance evaluation of velocity field in the vertical direction $u_y$ for 20 randomly generated stenosis at $Re=200$ using two neural network models. Here, these errors correspond to interpolation results.}\label{tab:8}
		\begin{tabular*}{1\textwidth}{@{\extracolsep{\fill}} lcccc }
			\hline\hline
			\small    
			Model & MAE & NMAE($\%$) & RMSE & NRMSE($\%$) \\ \hline
			U-ResNet & \textbf{0.0100 ± 0.0044}    & \textbf{2.8780 ± 1.0718 }   & \textbf{0.0207 ± 0.0090} & \textbf{5.8309 ± 1.8024}\\          
			FNO & 0.0144 ± 0.0053     & 4.0364 ± 1.1569  & 0.0327 ± 0.0123 &  8.9901 ± 2.0905 \\ \hline\hline
		\end{tabular*}%
	\end{center}
\end{table}

\begin{figure}[htbp]\centering
	\includegraphics[width=1.0\textwidth]{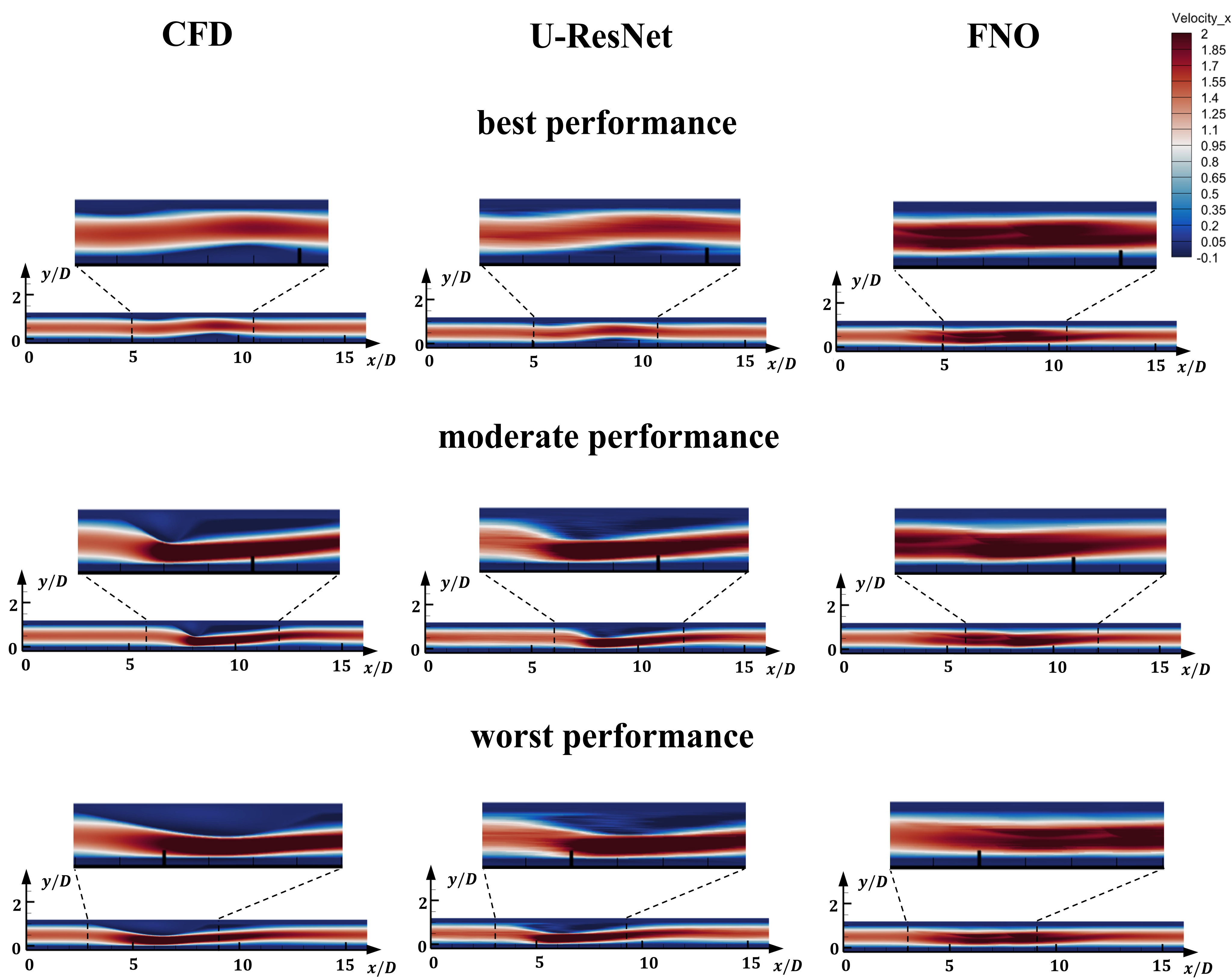}
	\caption{Comparison of streamwise velocity contours between CFD, U-ResNet and FNO for three representative (best, moderate, worst) stenotic flow cases at $Re=200$.}\label{fig:18}
\end{figure}

Fig.~\ref{fig:18} compares streamwise velocity predictions from U-ResNet and FNO against high-fidelity CFD results across three representative cases from 20 test cases at $Re=200$. The best-performing case features bilateral stenosis with dual constrictions, while the moderate and worst cases involve unilateral stenosis with single constrictions.
U-ResNet accurately reconstructs velocity profiles in all cases, closely matching CFD results. Even in the worst-performing case, U-ResNet recovers the majority of the velocity field with only minor discrepancies in regions of steep gradients. In contrast, FNO fails to provide accurate predictions across all three scenarios, significantly deviating from ground truth. The discrepancies between the deep learning models and CFD results in the worst case can be attributed to the inherent limitations of deep learning models. These models, while capable of learning general patterns from the training data, may struggle with outlier or under-represented cases, particularly when the geometry or flow conditions deviate significantly from those encountered during training. As a result, the prediction accuracy varies, performing well on cases similar to the training distribution, while showing degraded performance on more complex or rare configurations.

Moreover, the geometric configurations corresponding to each case in Fig.~\ref{fig:18} are detailed in Table~\ref{tab:geom1}. For unilateral stenosis (with the stenosis located on the upper wall), the table provides the peak height, length, tilt angle, and starting position of stenosis. For bilateral stenosis (with stenoses on both the upper and lower walls), these parameters are provided for each side. The detailed definitions of stenosis shape and size for both unilateral and bilateral cases are provided in Section~\ref{model}, where $D$ denotes the diameter of the healthy (non-stenosed) vessel.

\begin{table}[tbp]
	\begin{center}
		\caption{Geometric configurations for the best, moderate, and worst performance cases in streamwise velocity field prediction.}\label{tab:geom1}
		\begin{tabular*}{1\textwidth}{@{\extracolsep{\fill}} lcccc }
			\hline\hline
			\small    
			Case & Peak height($D$) & Length($D$) & Tilt($D$) & Starting position($D$) \\ \hline
			Best: bilateral(up,down) & (0.124,0.234) & (3.591,3.629) & (-0.688,-0.566) & (3.521,6.470)\\         
			Moderate: unilateral(up) & (0.529)   & (1.723) & (-0.542) & (6.998) \\
			Worst: unilateral(up) & (0.525)   & (5.499)  & (-0.362) &  (3.020) \\ \hline\hline         
		\end{tabular*}%
	\end{center}
\end{table}

\begin{figure}[htbp]\centering
	\includegraphics[width=0.7\textwidth]{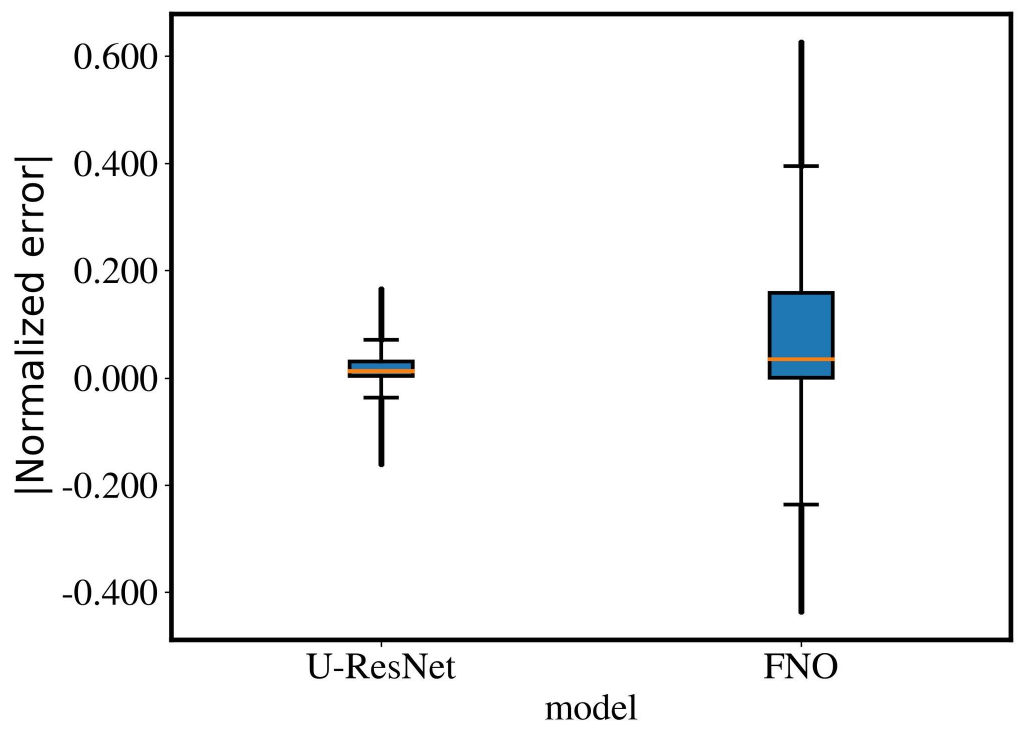}
    \caption{The box plots of streamwise velocity prediction error for U-ResNet and FNO. Here, the errors correspond to interpolation results.}\label{fig:19}
\end{figure}

To quantify model consistency, Fig.~\ref{fig:19} presents box plots of prediction errors for U-ResNet and FNO across all test cases. U-ResNet exhibits a tightly concentrated error distribution (approximately $-0.15$ to $0.15$), reflecting its robustness to geometric variability. FNO, however, shows a broader spread (approximately $-0.45$ to $0.6$). U-ResNet has a significantly smaller interquartile range, indicating more consistent predictions across test cases. FNO exhibits more extreme outliers, particularly in the positive direction, suggesting occasional substantial overprediction of velocity values. This visualization demonstrates U-ResNet's superior performance in streamwise velocity prediction, with approximately $3-4\times$ better error consistency compared to FNO.

\begin{figure}[htbp]
    \centering
    % 第一行
    \begin{subfigure}[b]{0.49\textwidth}
        \begin{overpic}[width=\linewidth]{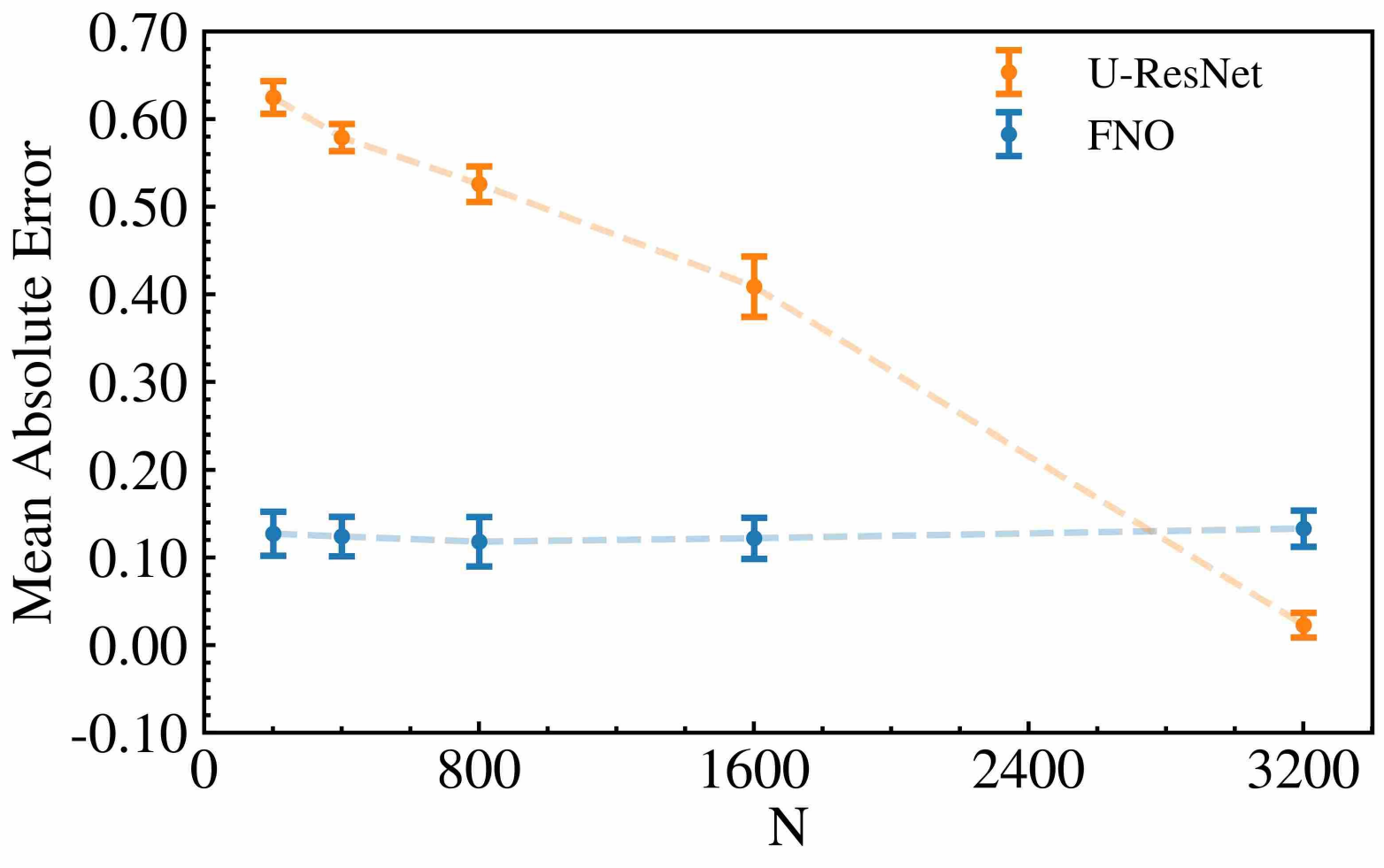}
            \put(-1,62){\small (a)}  
        \end{overpic}
    \end{subfigure}
    \hfill
    \begin{subfigure}[b]{0.49\textwidth}
        \begin{overpic}[width=\linewidth]{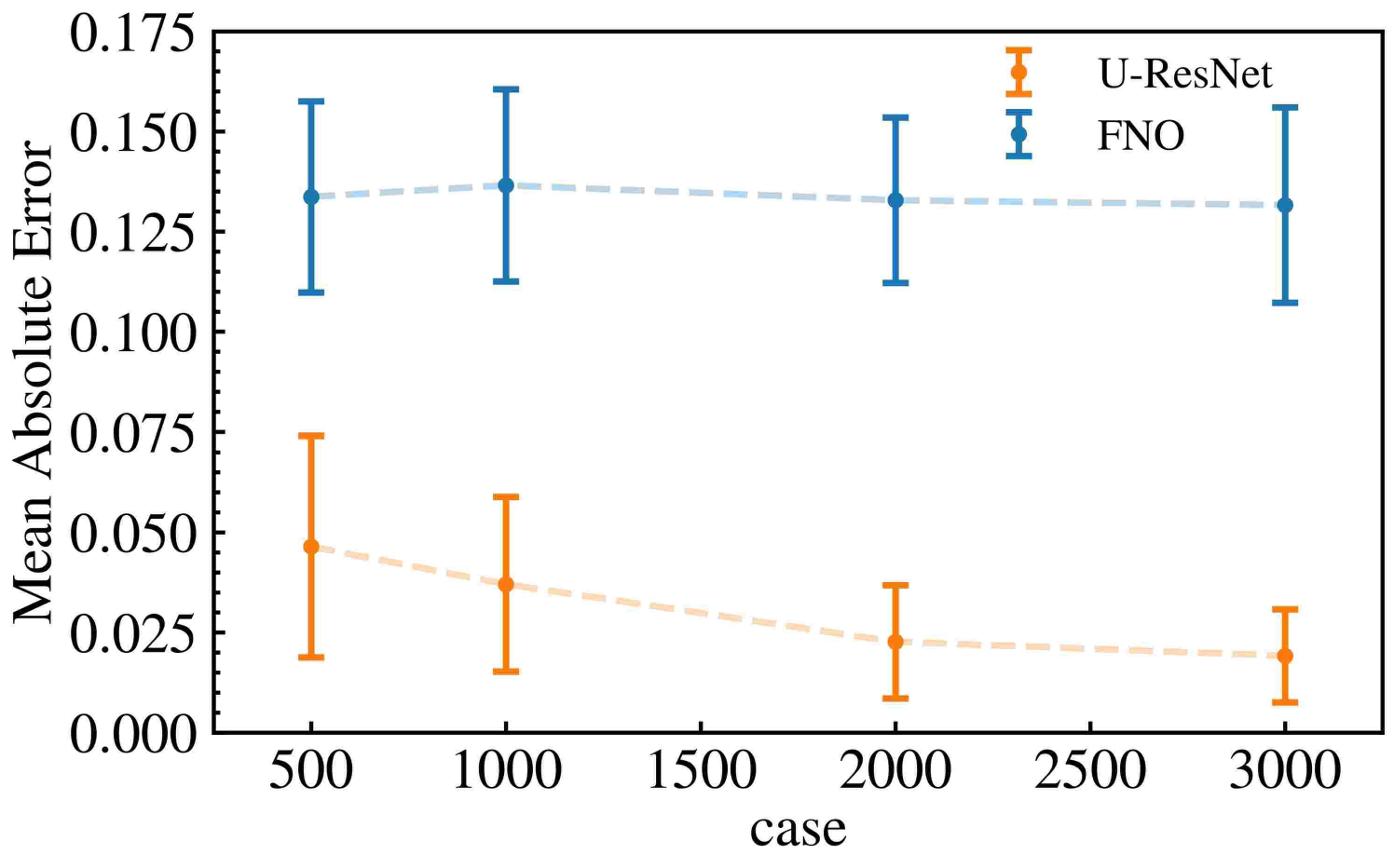}
            \put(-1,62){\small (b)}  
        \end{overpic}
    \end{subfigure}
	\caption{Mean absolute error in streamwise velocity predictions under varying (a) wall grid resolutions ($N_g=201$, 401, 801, 1601, 3201) and (b) training sample sizes ($N_s=500$, 1000, 2000, 3000). Here, the errors correspond to interpolation results.}\label{fig:20}
\end{figure}

Fig.~\ref{fig:20}(a) illustrates the sensitivity of streamwise velocity predictions to wall grid resolution for U-ResNet and FNO. U-ResNet demonstrates strong grid dependence, achieving accurate predictions ($MAE < 0.0025$) only at the finest resolution ($N_g=3201$). Coarser grids ($N_g\leq1601$) result in $MAE$ increases significantly, primarily due to under-resolved boundary layer dynamics. In contrast, FNO exhibits grid-invariant errors ($MAE \approx 0.13$ across resolutions), reflecting its spectral bias in capturing localized velocity gradients. This divergence highlights U-ResNet’s reliance on high-resolution spatial features for boundary layer accuracy, whereas FNO’s global Fourier basis struggles with geometric nonlinearities irrespective of discretization.

Fig.~\ref{fig:20}(b) evaluates model performance as a function of training sample size ($N_s$). U-ResNet achieves convergence at $N_s=2000$, with $MAE$ stabilizing at $0.0227 \pm 0.0142$ - a $83\%$ reduction compared to FNO ($MAE = 0.1328 \pm 0.0206$). The convergence threshold aligns with established deep learning principles where $N_s \propto \mathcal{O}\left(10^3\right)$ ensures robust feature extraction for complex flows. FNO’s slower convergence and higher asymptotic error ($MAE > 0.13$) underscore its limitations in learning multi-scale velocity dynamics without explicit geometric encoding.

It is observed that U-ResNet exhibits higher mean absolute error ($MAE$) when predicting velocity fields, as seen in Fig.~\ref{fig:20}(a) ($MAE \approx 0.5$ for N=800), compared to the lower MAEs observed for pressure and WSS in Figs. ~\ref{fig:8} and ~\ref{fig:14}. This discrepancy arises from fundamental differences in evaluation context and the physical quantities being assessed, rather than inconsistencies in model performance. For pressure and WSS, both of which are evaluated exclusively on the vessel wall, involving a relatively limited number of spatial points and typically smoother distributions. In contrast, for the velocity field, which is predicted throughout the entire computational domain and thus encompasses a much larger set of data points, including regions with complex flow features and sharp gradients. Domain-wide velocity predictions are susceptible to cumulative error accumulation across spatial regions, particularly in areas with high gradients or recirculation zones, whereas wall-based quantities represent localized phenomena with reduced error propagation. Unlike pressure and WSS predictions - where U-ResNet maintains sub-2\% errors across grid resolutions (Figs.~\ref{fig:8} and \ref{fig:14}) - streamwise velocity predictions exhibit pronounced grid sensitivity. This disparity arises from velocity’s stronger dependence on boundary layer resolution and shear layer advection, which require finer grids to resolve inertial subrange turbulence. FNO’s consistent errors across variables further emphasize its inability to adapt to localized flow features. 

\subsubsection{Prediction of vorticity field}
Table~\ref{tab:9} summarizes the vorticity field prediction accuracy of U-ResNet and FNO models across 20 stenotic flow cases at $Re=200$. U-ResNet demonstrates exceptional performance, achieving a mean absolute error ($MAE$) of $0.2562\pm0.1445$, representing a 91.7\% reduction compared to FNO ($3.1018\pm0.2324$). Similarly significant improvements are observed across all error metrics. These order-of-magnitude improvements highlight U-ResNet's superior capacity to resolve complex vortical structures in stenotic flows, particularly in regions of flow separation and recirculation.

\begin{table}[tbp]
	\begin{center}
		\caption{Performance evaluation of vorticity field for 20 randomly generated stenosis at $Re=200$ using two neural network models. Here, these errors correspond to interpolation results.}\label{tab:9}
		\begin{tabular*}{1\textwidth}{@{\extracolsep{\fill}} lcccc }
			\hline\hline
			\small    
			Model & MAE & NMAE($\%$) & RMSE & NRMSE($\%$) \\ \hline
			U-ResNet & \textbf{0.2562 ± 0.1445}    & \textbf{0.6880 ± 0.2907}   & \textbf{0.6380 ± 0.2884} & \textbf{1.6909 ± 0.4249}\\     
			FNO & 3.1018 ± 0.2324    & 8.9259 ± 2.3032  & 6.2270 ± 0.7242 &  17.7971 ± 4.4709 \\ \hline\hline
		\end{tabular*}%
	\end{center}
\end{table}

\begin{figure}[htbp]\centering
	\includegraphics[width=1.0\textwidth]{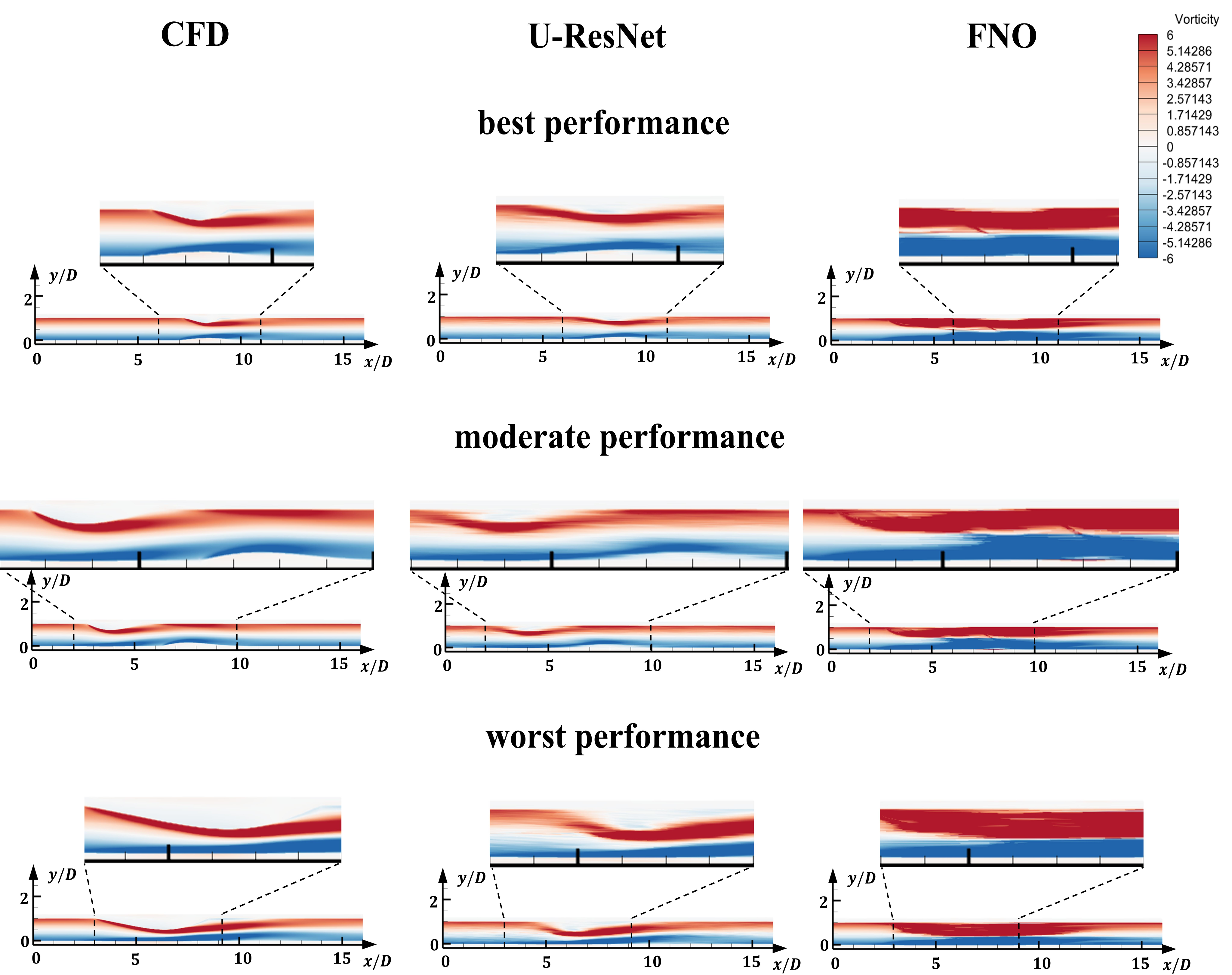}
	\caption{Comparison of vorticity contours between CFD, U-ResNet and FNO for three representative (best, moderate, worst) stenotic flow cases at $Re=200$.}\label{fig:21}
\end{figure}

Fig.~\ref{fig:21} presents vorticity field predictions for three representative stenotic flow cases among 20 test cases at $Re=200$, comparing U-ResNet and FNO against high-fidelity CFD results. The best and moderate cases correspond to bilateral stenosis geometries with more complex flow patterns, while the worst case features a unilateral stenosis.
For the best case, U-ResNet accurately captures the complex vorticity interactions between sequential constrictions, including the amplification and merging of vortical structures within the inter-stenotic region. In the moderate case, it resolves most of the key vorticity features, with only minor discrepancies near the stenotic wall. In the worst case, U-ResNet successfully reconstructs characteristic vorticity structures, such as sharp gradients along the core jet boundary and post-stenotic recirculation zones with counter-rotating vortices, although larger errors are observed near the stenotic wall.
Across all cases, U-ResNet maintains strong predictive accuracy despite increasing flow complexity. In contrast, FNO fails to produce accurate vorticity predictions under all configurations. These results, as visualized in the contour plots, highlight U-ResNet’s superior capability in capturing both the qualitative structures and quantitative magnitudes of the vorticity field across various stenotic geometries.

Furthermore, Table~\ref{tab:geom2} presents the detailed geometric configurations for each case shown in Fig.~\ref{fig:21}. In the case of unilateral stenosis (affecting only the upper wall), the recorded parameters include the peak height, length, tilt angle, and starting position of stenosis. For bilateral stenosis (involving both upper and lower walls), these characteristics are listed separately for each wall.

\begin{table}[tbp]
	\begin{center}
		\caption{Geometric configurations for the best, moderate, and worst performance cases in vorticity field prediction.}\label{tab:geom2}
		\begin{tabular*}{1\textwidth}{@{\extracolsep{\fill}} lcccc }
			\hline\hline
			\small    
			Case & Peak height($D$) & Length($D$) & Tilt($D$) & Starting position($D$) \\ \hline
			Best: bilateral(up,down) & (0.247,0.114)   & (1.789,2.753)  & (-0.418,-0.039) & (7.195,6.988)\\         
			Moderate: bilateral(up,down) & (0.296,0.170) & (2.876,4.140) & (0.336,0.761) & (2.666,6.291) \\
			Worst: unilateral(up) & (0.525) & (5.499)  & (-0.362) &  (3.020) \\ \hline\hline         
		\end{tabular*}%
	\end{center}
\end{table}

\begin{figure}[htbp]\centering
 	\includegraphics[width=0.7\textwidth]{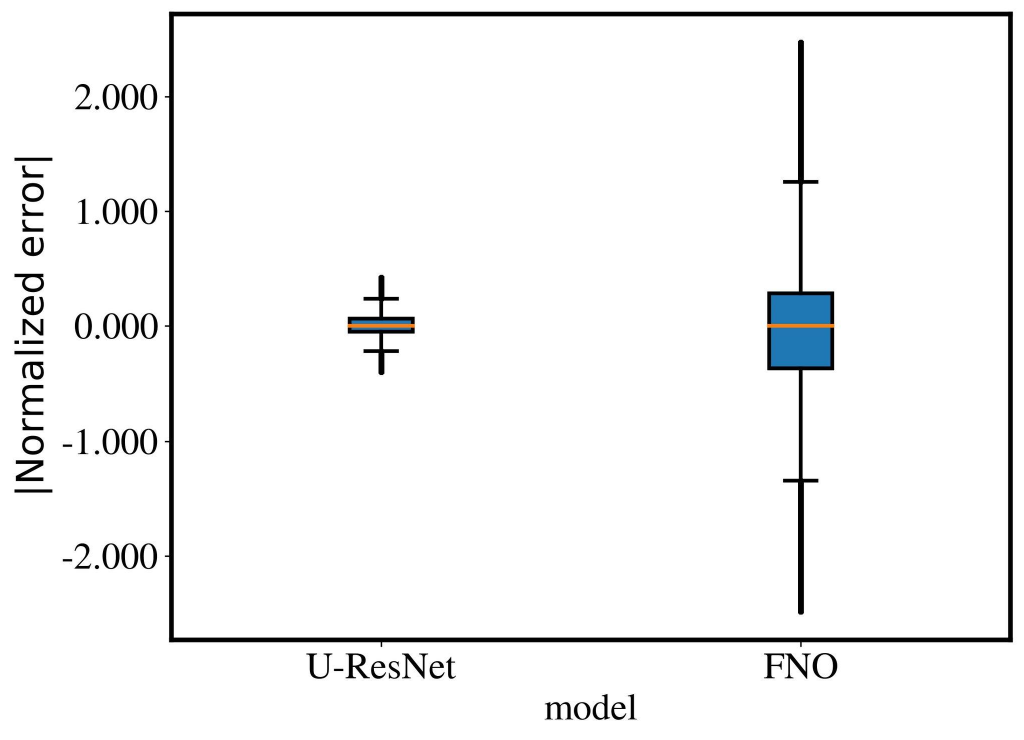}
 	\caption{The box plots of vorticity prediction error 
  for U-ResNet and FNO. Here, the errors correspond to interpolation results.}\label{fig:22}
\end{figure}

Fig.~\ref{fig:22} compares vorticity prediction errors using box plots. U-ResNet shows a tighter error range ($\sim-0.15$ to $0.15$) and smaller interquartile spread, indicating stronger robustness and consistency. In contrast, FNO exhibits a wider distribution ($\sim-0.25$ to $0.25$). Overall, U-ResNet achieves $1.5$–$2\times$ better error consistency.

\begin{figure}[htbp]
    \centering
    % 第一行
    \begin{subfigure}[b]{0.49\textwidth}
        \begin{overpic}[width=\linewidth]{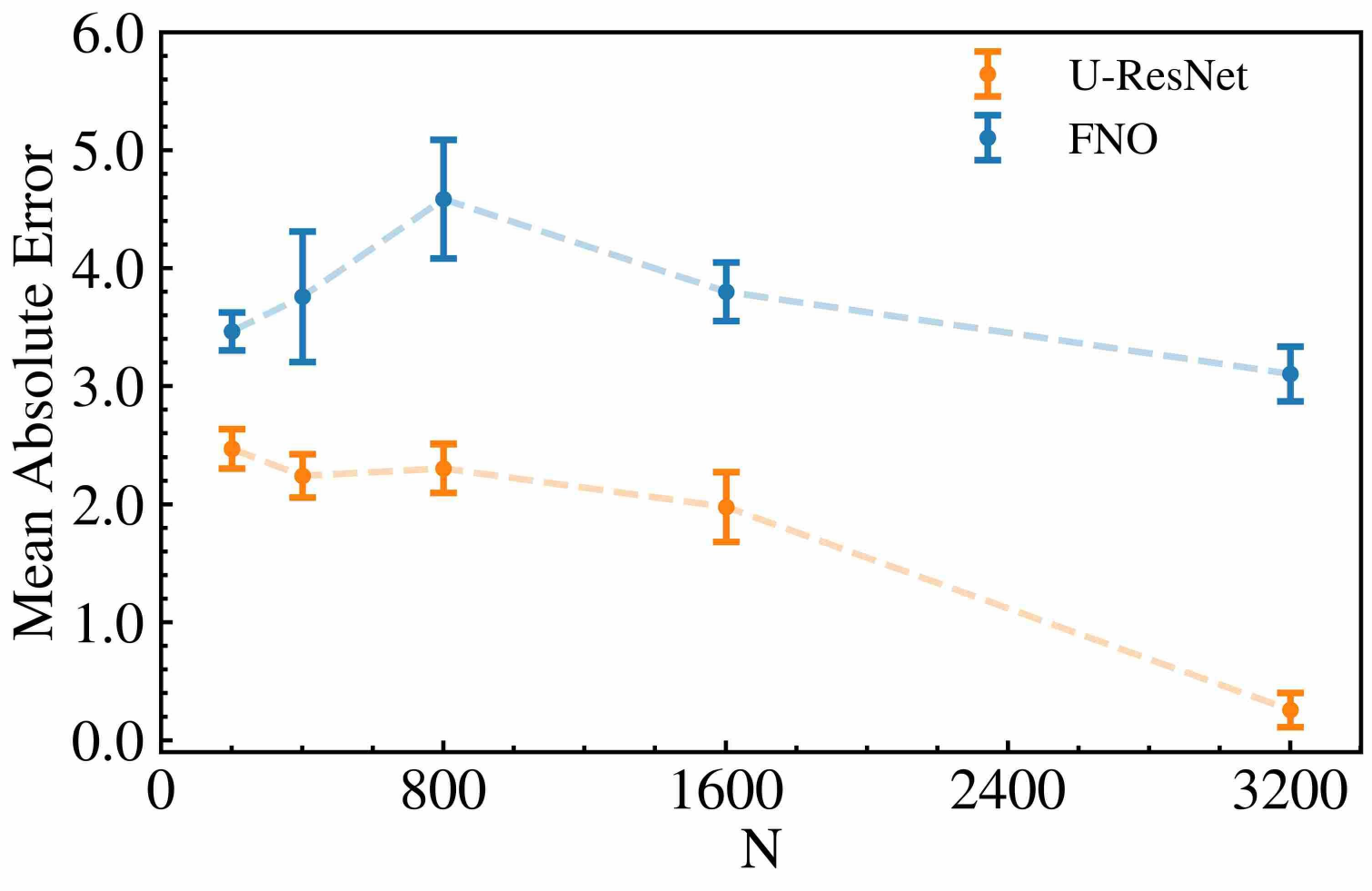}
            \put(-1,62){\small (a)}  
        \end{overpic}
    \end{subfigure}
    \hfill
    \begin{subfigure}[b]{0.49\textwidth}
        \begin{overpic}[width=\linewidth]{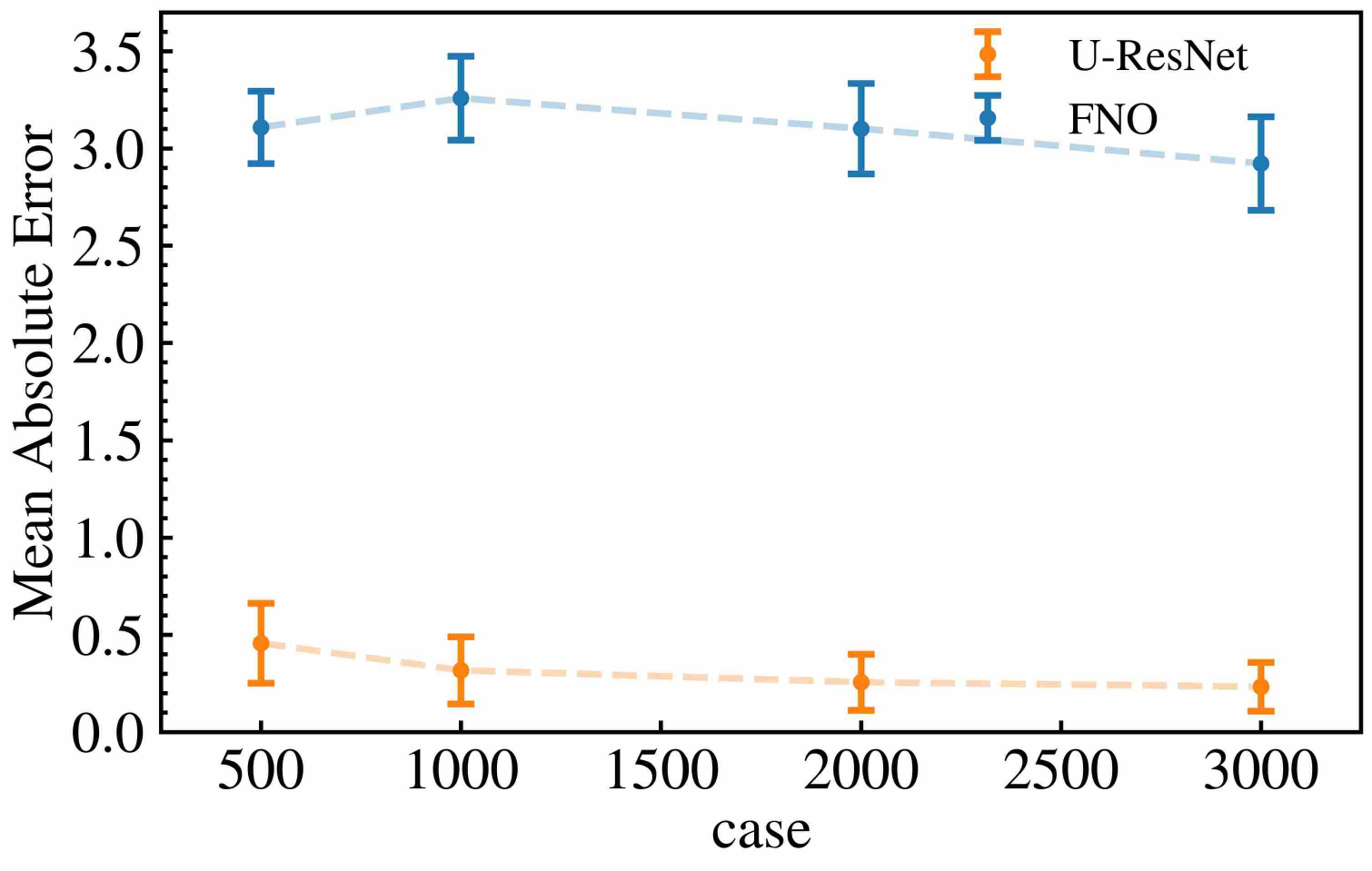}
            \put(-1,62){\small (b)}  
        \end{overpic}
    \end{subfigure}	
	\caption{Mean absolute error in vorticity prediction under varying (a) wall grid resolutions ($N_g=201$, 401, 801, 1601, 3201) and (b) training sample sizes ($N_s=500$, 1000, 2000, 3000). Here, the errors correspond to interpolation results.}\label{fig:23}
\end{figure}

Fig.~\ref{fig:23} evaluates the sensitivity of vorticity predictions to grid resolution and training dataset size for U-ResNet and FNO models. The results demonstrate distinct performance characteristics between the residual-based and spectral approaches when handling complex rotational flow structures. As illustrated in Fig.~\ref{fig:23}(a), U-ResNet exhibits pronounced grid sensitivity for vorticity predictions, with $MAE$ decreasing monotonically as resolution increases from $N_g=201$ to $N_g=3201$. This behavior mirrors the grid dependence observed in velocity predictions (Fig.~\ref{fig:20}(a)) and stems from vorticity's mathematical definition as the curl of velocity, making it particularly sensitive to spatial discretization. In contrast, FNO maintains relatively constant - albeit significantly higher error - across all grid resolutions, suggesting intrinsic limitations in capturing high-frequency vortical structures irrespective of grid refinement.

Fig.~\ref{fig:23}(b) demonstrates that U-ResNet achieves error convergence at approximately $N_s=2000$ training samples, with minimal improvement observed beyond this threshold. This convergence point represents the optimal trade-off between computational cost and prediction accuracy, requiring 87\% fewer samples than traditional Latin hypercube sampling approaches for comparable accuracy. FNO exhibits similar convergence behavior but with substantially higher asymptotic error, reflecting its fundamental challenges in representing sharp vorticity gradients typical of stenotic flows.

\subsection{Generalization results across Reynolds numbers} \label{generalization results}
To rigorously evaluate U-ResNet's generalization capability across varying flow regimes, we implemented a systematic cross-Reynolds number validation approach. The model was trained with 2000 samples for each of four baseline Reynolds numbers ($Re = 200, 400, 600, 800$), with inputs comprising x-coordinates from the 3201 grid points on each wall along with the corresponding Reynolds number. Validation and testing were then performed on 20 samples for each interpolated Reynolds number ($Re = 300, 500, 700$), totaling 60 test cases that challenge the model to predict hemodynamics in previously unseen flow regimes. Additionally, all test cases feature geometries that were not included in the training set.

\subsubsection{Pressure prediction generalization}
Table~\ref{tab:10} presents statistical performance metrics for pressure prediction using U-ResNet and FNO across all test cases. Notably, U-ResNet maintains high accuracy when generalizing to interpolated Reynolds numbers, achieving a mean absolute error (MAE) of $0.0128 \pm 0.0114$, which represents a 92\% error reduction compared to FNO's MAE of $0.1574 \pm 0.0853$. This performance is consistent with U-ResNet's accuracy on the training Reynolds numbers (Table~\ref{tab:5}), highlighting its ability to robustly parameterize the underlying flow physics rather than overfitting to specific training samples.

\begin{table}[tbp]
	\begin{center}
		\caption{Generalization performance of U-ResNet and FNO for pressure prediction at interpolated Reynolds numbers ($Re = 300, 500, 700$). Here, these errors correspond to interpolation results.}\label{tab:10}
		\begin{tabular*}{1\textwidth}{@{\extracolsep{\fill}} lcccc }
			\hline\hline
			\small    
			Model & MAE & NMAE($\%$) & RMSE & NRMSE($\%$) \\ \hline        
			U-ResNet & \textbf{0.0128 ± 0.0114}     & \textbf{1.9078 ± 0.8928}   & \textbf{0.0180 ± 0.0157} &  \textbf{2.6392 ± 1.1328}  \\ 
			FNO & 0.1574 ± 0.0853     & 24.1900 ± 6.3644   & 0.2072 ± 0.1052 &  32.0843 ± 7.4846  \\ \hline\hline            
		\end{tabular*}%
	\end{center}
\end{table}

\begin{figure}[htbp]
    \centering
    \begin{subfigure}[b]{0.49\textwidth}  % 调整宽度，使三张图并列
        \begin{overpic}[width=\linewidth]{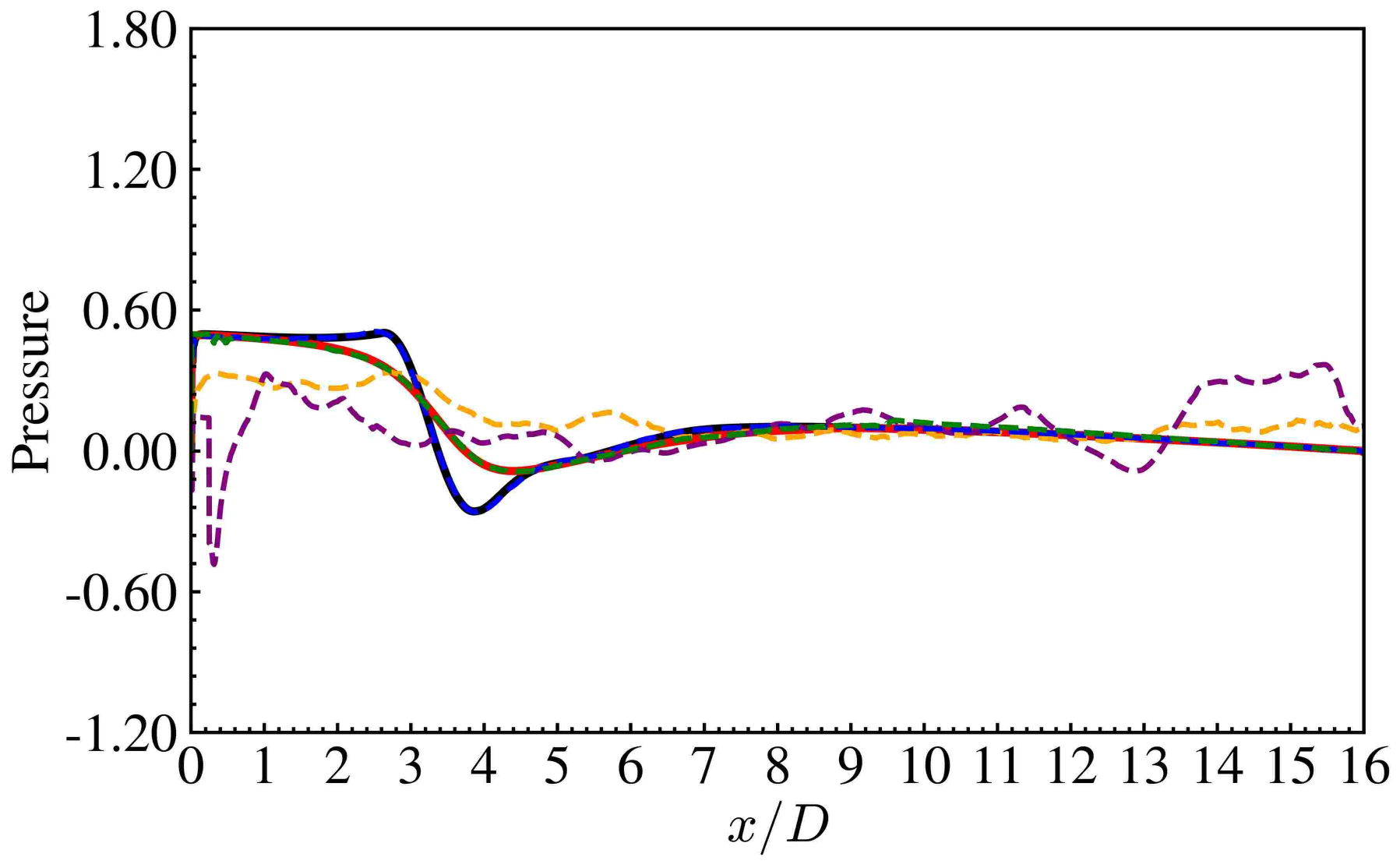}
            \put(-2,54){\small (a)}  
        \end{overpic}
    \end{subfigure}
    \hfill
    \begin{subfigure}[b]{0.49\textwidth}
        \begin{overpic}[width=\linewidth]{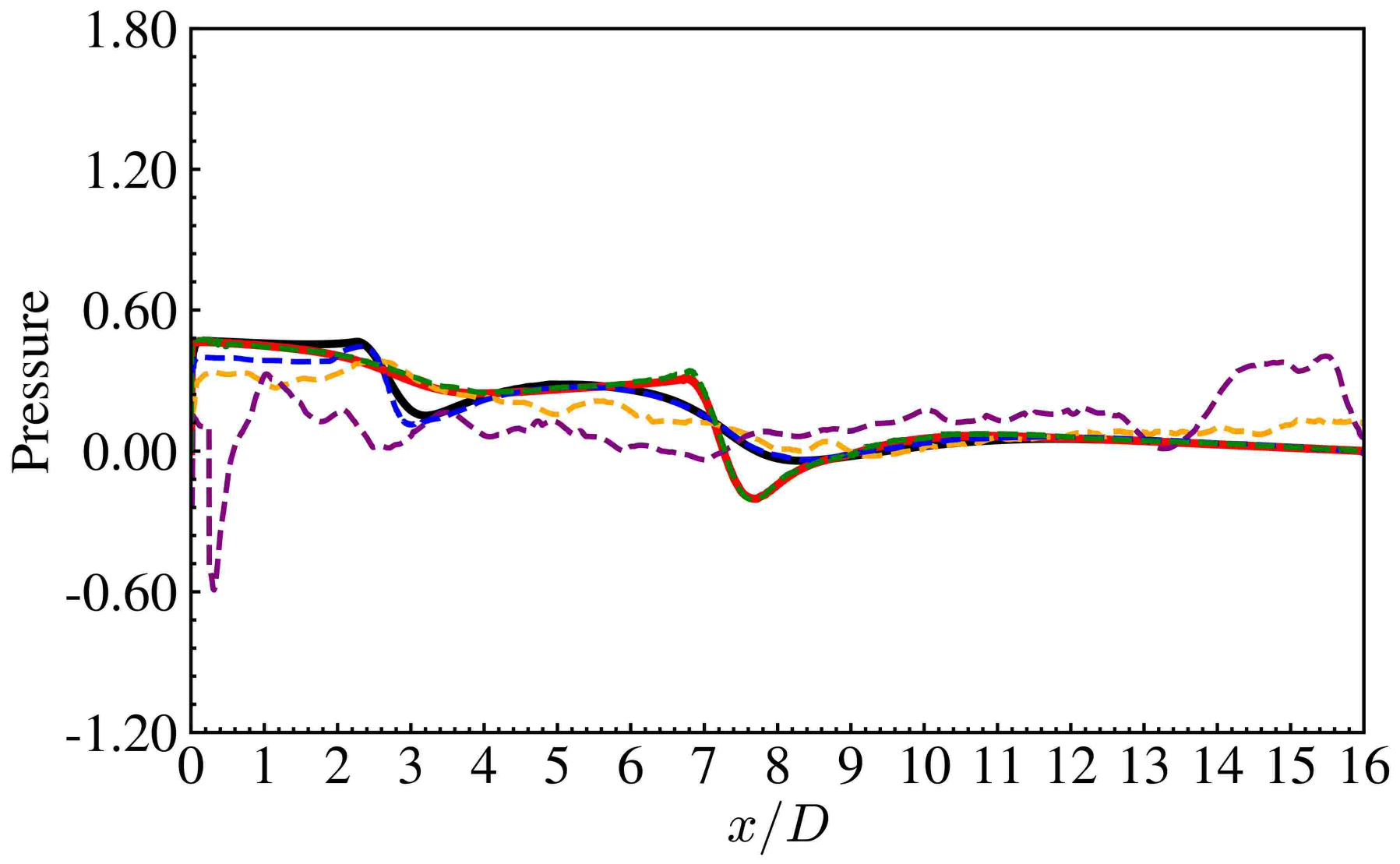}
            \put(-2,54){\small (b)}  
        \end{overpic}
    \end{subfigure}
    \vspace{0.1cm}
    \begin{subfigure}[b]{0.49\textwidth}
        \begin{overpic}[width=\linewidth]{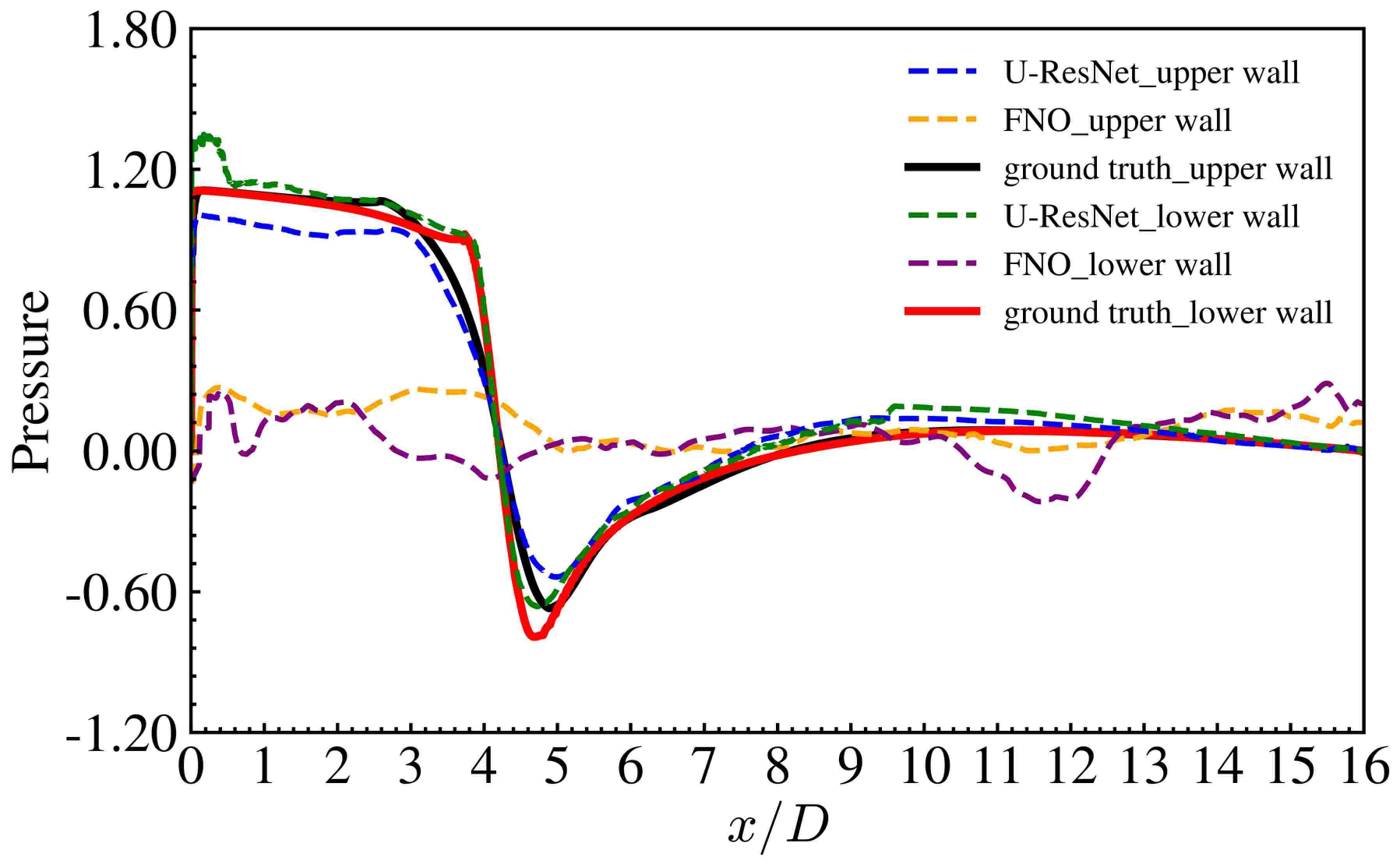}
            \put(-2,54){\small (c)}  
        \end{overpic}
    \end{subfigure}
	\caption{Pressure prediction generalization for U-ResNet and FNO: (a) Best performance case, (b) Moderate performance case, (c) Worst performance case.}\label{fig:24}
    \end{figure}

Fig.~\ref{fig:24} illustrates pressure predictions of U-ResNet and FNO for three representative cases selected from 60 test cases at interpolated Reynolds numbers ($Re=300$, 500, 700): (a) the best-performing case with $Re=500$ and unilateral stenosis, (b) a moderate-performing case with $Re=500$ and bilateral stenosis, and (c) the worst-performing case with $Re=300$ and bilateral stenosis. 
In both the best and moderate cases, U-ResNet predictions closely match CFD results, accurately capturing the magnitude and spatial distribution of pressure variations. In the worst case, although U-ResNet fails to accurately predict the minimum pressure drop and inlet pressure, it still reconstructs most regions well and successfully captures critical features such as pressure drop amplitudes that scale with Reynolds number, spatial gradients within the stenotic region, and post-stenotic pressure recovery.
In contrast, FNO fails to provide accurate predictions in all three cases, even in the best-performing scenario, demonstrating its inability to generalize pressure distributions across varying Reynolds numbers.

\begin{figure}[htbp]
    \centering
    \begin{subfigure}[b]{0.32\textwidth}  % 调整宽度，使三张图并列
        \begin{overpic}[width=\linewidth]{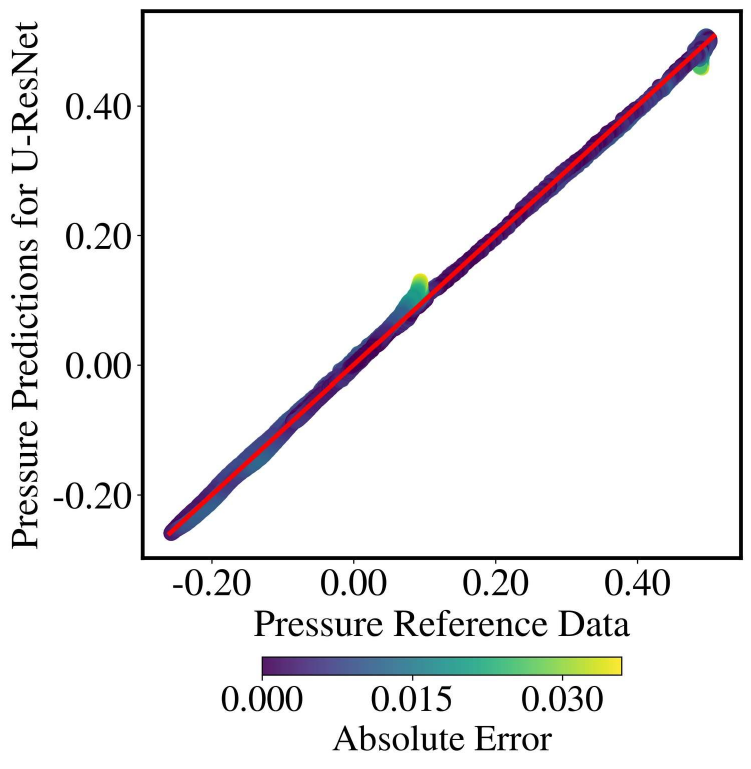}
            \put(-3,100){\small (a)}  
        \end{overpic}
    \end{subfigure}
    \hfill
    \begin{subfigure}[b]{0.32\textwidth}
        \begin{overpic}[width=\linewidth]{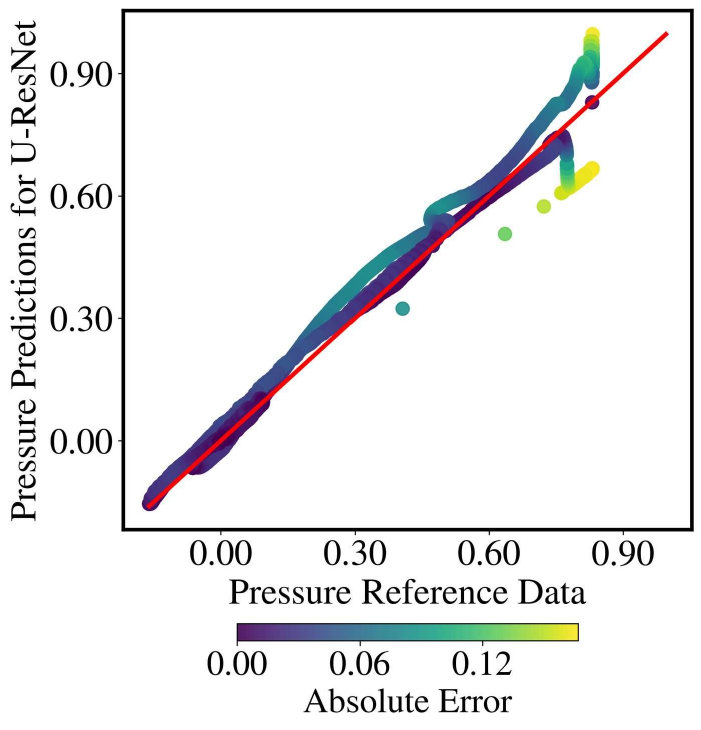}
            \put(-3,100){\small (b)}  
        \end{overpic}
    \end{subfigure}
    \hfill
    \begin{subfigure}[b]{0.32\textwidth}
        \begin{overpic}[width=\linewidth]{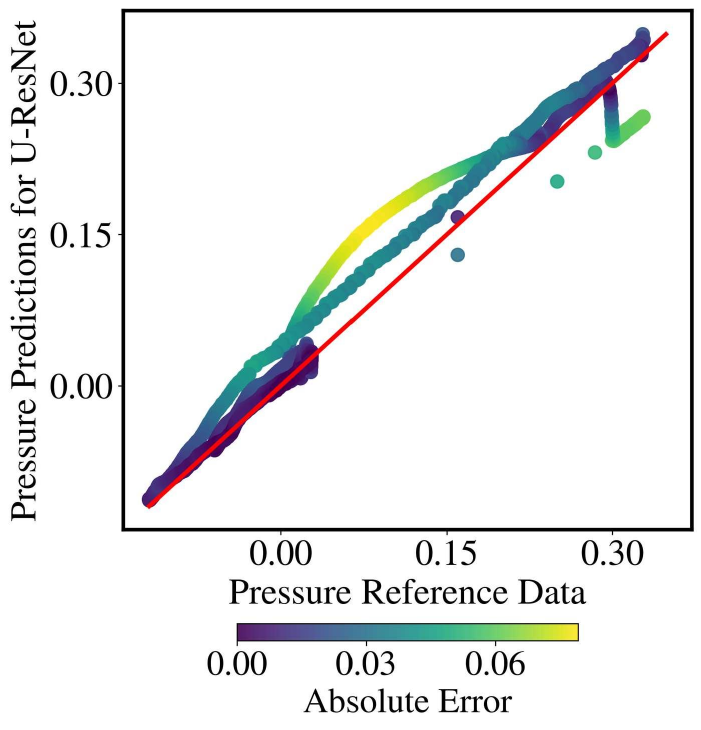}
            \put(-3,100){\small (c)}  
        \end{overpic}
    \end{subfigure}
	\caption{Pressure prediction of U-ResNet versus CFD data: (a) Best performance case, (b) Moderate performance case, (c) Worst performance case. Here, the errors correspond to interpolation results.}\label{fig:25}
    \end{figure}

To further assess prediction fidelity, Fig.~\ref{fig:25} compares predicted values of U-ResNet against CFD ground truth for three characteristic cases. Fig.~\ref{fig:25}(a) shows the highest-accuracy case (among the 60 test cases) where predictions nearly exactly match CFD data. Fig.~\ref{fig:25}(b) is a moderate-accuracy case (among the 60 test cases) with slight deviations in pressure gradient regions. Fig.~\ref{fig:25}(c) is the lowest-accuracy case (among the 60 test cases) showing more significant deviations. The majority of test cases ($>75\%$) achieve moderate to high accuracy, with $NMAE$ remaining below 2.5\% even for the most challenging geometries. This consistent performance across interpolated Reynolds numbers confirms U-ResNet's capacity to generalize across flow regimes, making it a robust tool for cardiovascular flow modeling where patient-specific Reynolds numbers may not precisely match training cases.

\subsubsection{WSS prediction generalization}
\begin{table}[tbp]
	\begin{center}
		\caption{Generalization performance of U-ResNet and FNO for wall shear stress prediction at interpolated Reynolds numbers ($Re = 300, 500, 700$). Here, these errors correspond to interpolation results.}\label{tab:11}
		\begin{tabular*}{1\textwidth}{@{\extracolsep{\fill}} lcccc }
			\hline\hline
			\small    
			Model & MAE & NMAE($\%$) & RMSE & NRMSE($\%$) \\ \hline        
			U-ResNet & \textbf{0.0009 ± 0.0010}     & \textbf{1.4534 ± 1.3185}   & \textbf{0.0012 ± 0.0012} &  \textbf{1.8278 ± 1.5324} \\
			FNO & 0.1472 ± 0.1701     & 420.8291 ± 570.4374   & 0.1967 ± 0.2312 &  564.8900 ± 773.2588 \\ \hline\hline            
		\end{tabular*}%
	\end{center}
\end{table}
To evaluate the generalization capability of U-ResNet and FNO in predicting wall shear stress (WSS) across interpolated Reynolds numbers, we conducted a comprehensive assessment using 60 test cases at previously unseen flow conditions ($Re = 300, 500, 700$). Table~\ref{tab:11} summarizes the statistical performance metrics for these tests. As shown, U-ResNet achieves excellent predictive accuracy, with a mean absolute error (MAE) of $0.0009 \pm 0.0010$, representing a 99\% error reduction compared to FNO ($0.1472 \pm 0.1701$), and a normalized mean absolute error (NMAE) of $1.4534 \pm 1.3185\%$, showing a 99.7\% reduction in error relative to FNO ($420.8291 \pm 570.4374\%$). These results are comparable to U-ResNet's performance on the training Reynolds numbers (Table~\ref{tab:6}), demonstrating its robust generalization capability. Furthermore, a root mean squared error (RMSE) of $0.0012 \pm 0.0012$ confirms the U-ResNet’s ability to avoid large outlier errors, even when extrapolating to previously unseen flow regimes.

\begin{figure}[htbp]
    \centering
    \begin{subfigure}[b]{0.49\textwidth}  % 调整宽度，使三张图并列
        \begin{overpic}[width=\linewidth]{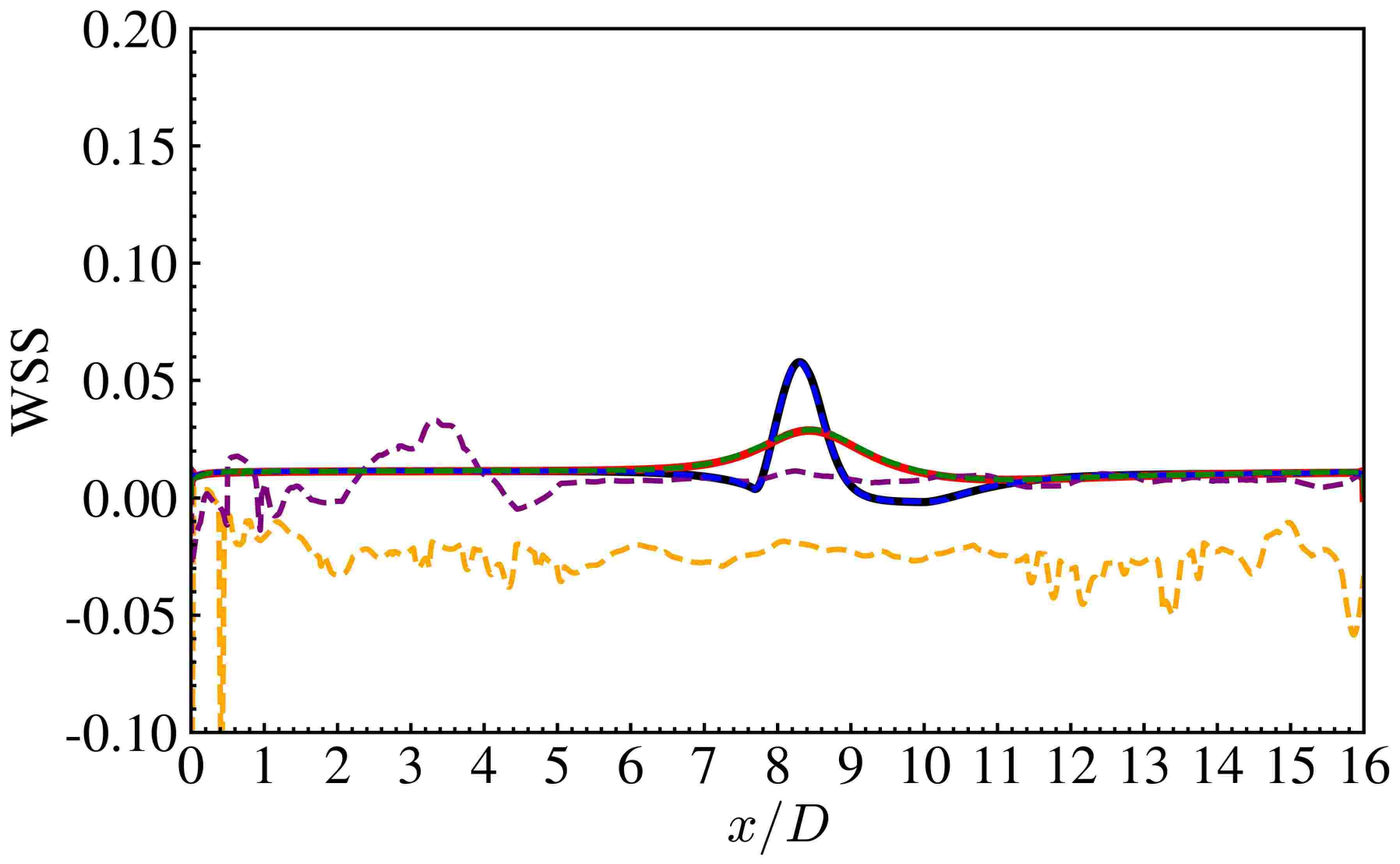}
            \put(-2,54){\small (a)}  
        \end{overpic}
    \end{subfigure}
    \hfill
    \begin{subfigure}[b]{0.49\textwidth}
        \begin{overpic}[width=\linewidth]{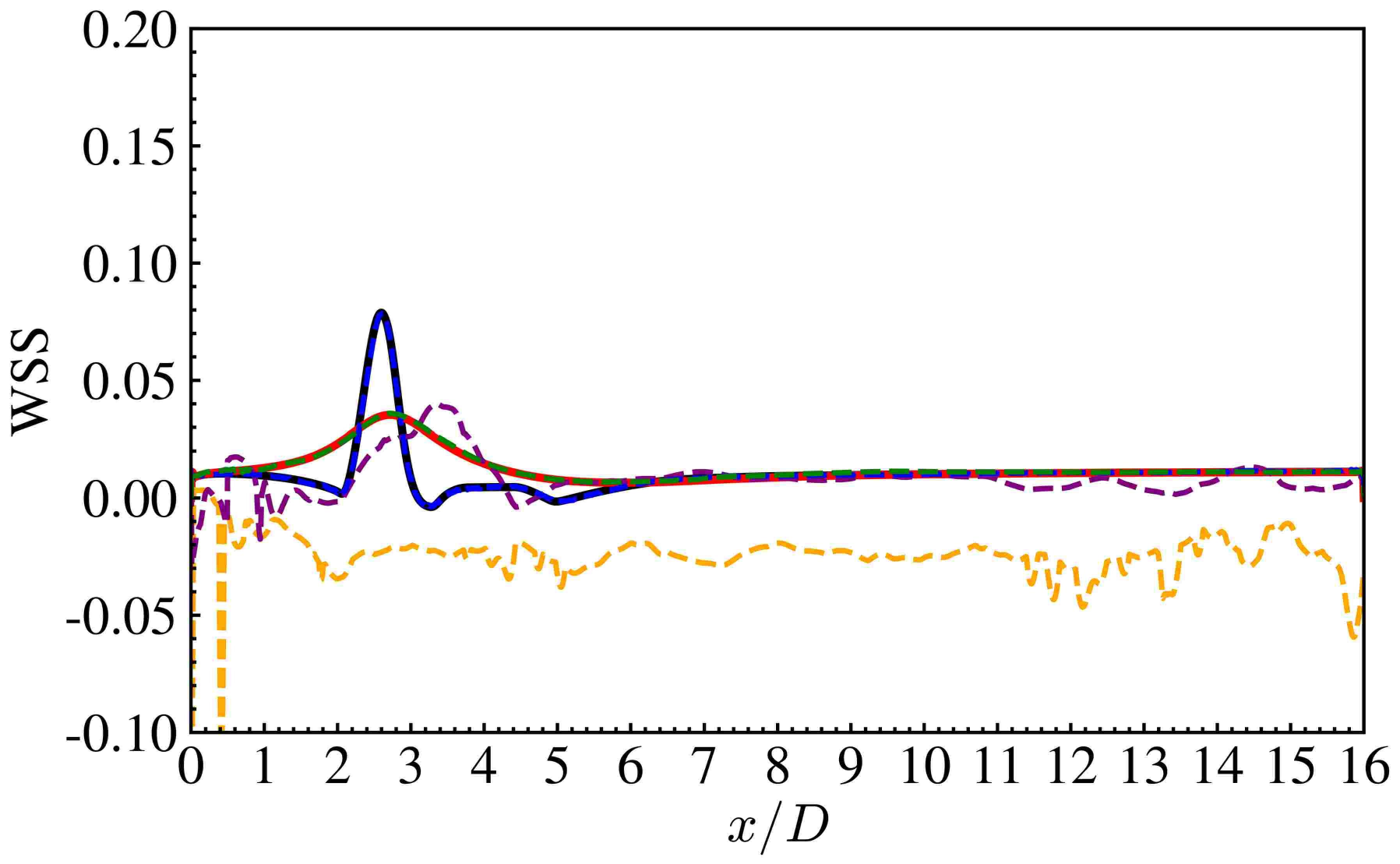}
            \put(-2,54){\small (b)}  
        \end{overpic}
    \end{subfigure}
    \vspace{0.1cm}
    \begin{subfigure}[b]{0.49\textwidth}
        \begin{overpic}[width=\linewidth]{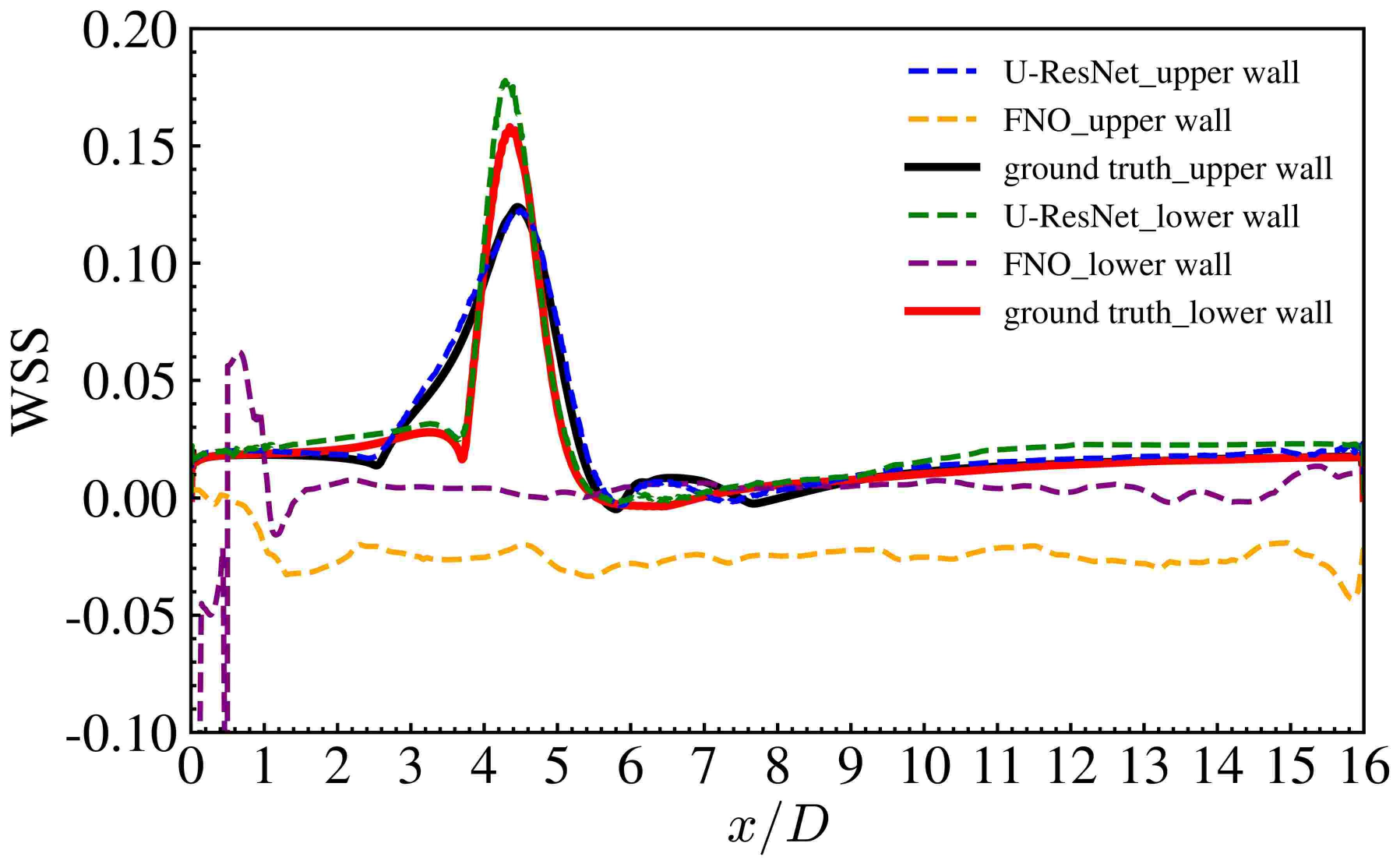}
            \put(-2,54){\small (c)}  
        \end{overpic}
    \end{subfigure}
	\caption{WSS prediction generalization for U-ResNet and FNO: (a) Best performance case, (b) Moderate performance case, (c) Worst performance case.}\label{fig:26}
    \end{figure}

Fig.~\ref{fig:26} presents WSS predictions from U-ResNet and FNO for three representative cases selected from 60 test cases at interpolated Reynolds numbers ($Re=300$, 500, 700): (a) the best-performing case with $Re=500$ and unilateral stenosis, (b) a moderate-performing case with $Re=500$ and unilateral stenosis, and (c) the worst-performing case with $Re=300$ and bilateral stenosis.
For both the best and moderate cases, U-ResNet demonstrates excellent agreement with CFD results, accurately capturing the magnitude and spatial distribution of WSS. In the worst case, while U-ResNet effectively reconstructs the overall profile, noticeable errors emerge near the WSS peaks. Nevertheless, these predictions highlight U-ResNet’s ability to generalize across varying stenotic geometries and flow regimes. The model successfully resolves key hemodynamic features, including peak WSS values that scale with Reynolds number, steep WSS gradients in stenotic regions, and post-stenotic recovery patterns indicative of flow separation.
Importantly, U-ResNet accurately predicts WSS peaks at the stenosis apex—a clinically significant feature linked to endothelial dysfunction and atherosclerotic progression. In contrast, FNO fails to capture meaningful WSS structures even in the best-performing case, indicating its limited generalization capability across different Reynolds numbers.

\begin{figure}[htbp]
    \centering
    \begin{subfigure}[b]{0.32\textwidth}  % 调整宽度，使三张图并列
        \begin{overpic}[width=\linewidth]{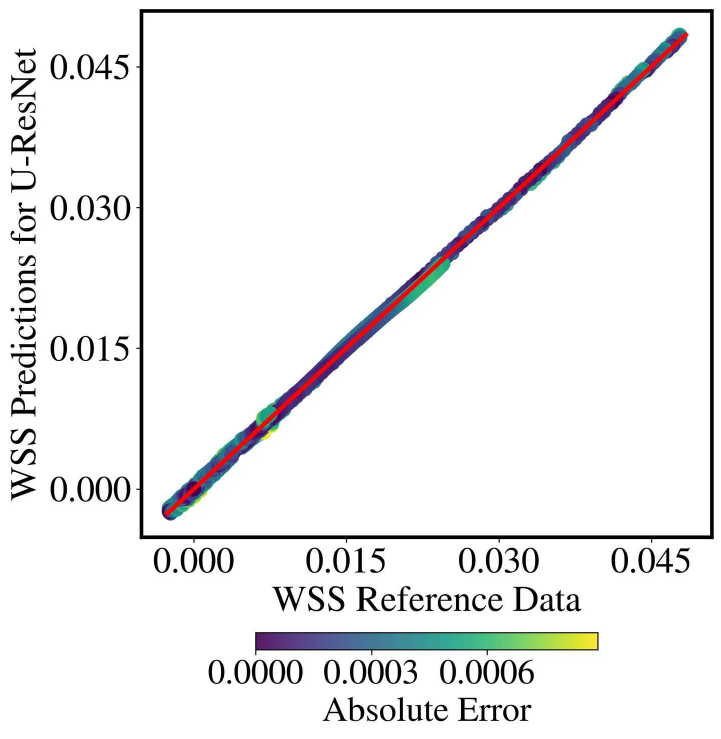}
            \put(-3,100){\small (a)}  
        \end{overpic}
    \end{subfigure}
    \hfill
    \begin{subfigure}[b]{0.32\textwidth}
        \begin{overpic}[width=\linewidth]{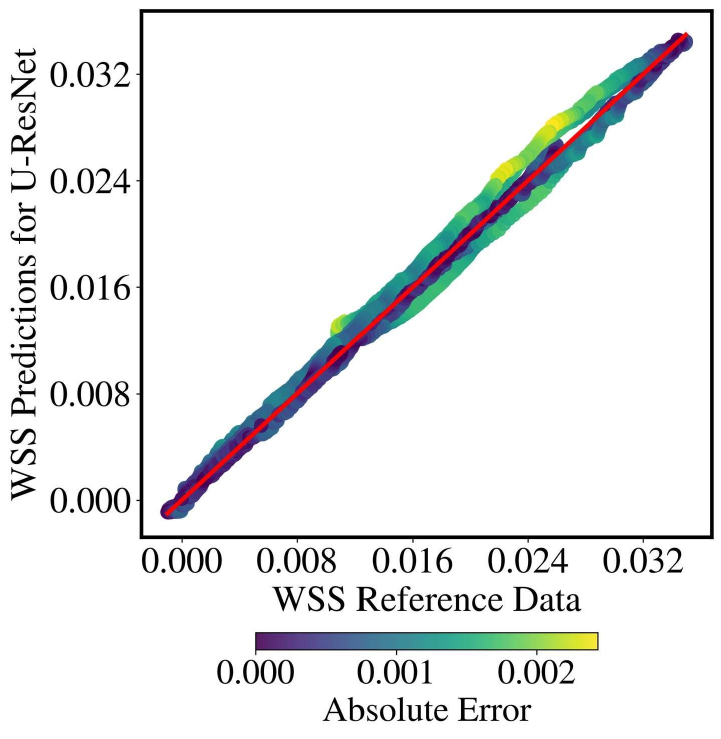}
            \put(-3,100){\small (b)}  
        \end{overpic}
    \end{subfigure}
    \hfill
    \begin{subfigure}[b]{0.32\textwidth}
        \begin{overpic}[width=\linewidth]{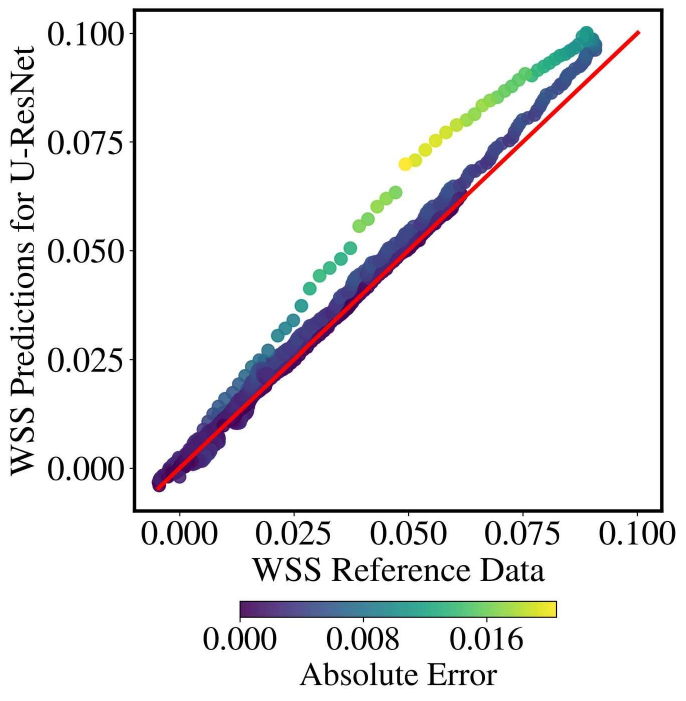}
            \put(-3,100){\small (c)}  
        \end{overpic}
    \end{subfigure}
	\caption{WSS prediction of U-ResNet versus CFD data: (a) Best performance case, (b) Moderate performance case, (c) Worst performance case. Here, the errors correspond to interpolation results.}\label{fig:27}
    \end{figure}

To provide a more granular assessment of WSS prediction fidelity, Fig.~\ref{fig:27} presents correlation plots comparing predicted values of U-ResNet against CFD ground truth for three characteristic cases. Fig.~\ref{fig:27}(a) present the highest-accuracy case (among the 60 test cases) where predictions closely align with the one-to-one correspondence line. Fig.~\ref{fig:27}(b) is a moderate-accuracy case (among the 60 test cases) showing slight deviations primarily in high-gradient regions. Fig.~\ref{fig:27}(c) shows the lowest-accuracy case (among the 60 test cases) exhibiting more substantial deviations, particularly for extreme WSS values. Quantitative analysis indicates that over 75\% of test cases achieve high to moderate accuracy, with $NMAE$ remaining below 1.5\% for the majority of geometries. This consistent performance across interpolated Reynolds numbers confirms U-ResNet's capacity to generalize across flow regimes, enabling reliable WSS prediction for patient-specific hemodynamic assessments where exact Reynolds number matching may not be feasible.
    
\subsubsection{Velocity field prediction generalization}
Tables~\ref{tab:12} and~\ref{tab:13} report the performance metrics of U-ResNet and FNO for velocity field prediction at interpolated Reynolds numbers ($Re = 300, 500, 700$). U-ResNet demonstrates strong generalization capabilities for both velocity components, maintaining high predictive accuracy across previously unseen flow regimes.

For the streamwise velocity component (Table~\ref{tab:12}), U-ResNet achieves a mean absolute error (MAE) of $0.0156 \pm 0.0087$, representing a 96\% reduction compared to FNO ($0.3937 \pm 0.2618$), and a normalized MAE (NMAE) of $0.8373 \pm 0.3535\%$, also reflecting a 96\% error reduction compared to FNO ($21.5131 \pm 13.8962\%$). The normalized root mean squared error (NRMSE) of $1.5277 \pm 0.6555\%$ further indicates consistent performance across diverse stenotic geometries. Notably, these results show a 30\% improvement in accuracy relative to the U-ResNet’s performance at the training Reynolds numbers (Table~\ref{tab:7}), suggesting enhanced interpolation capability between the training data points.

\begin{table}[tbp]
	\begin{center}
		\caption{Generalization performance of U-ResNet and FNO for streamwise velocity prediction at interpolated Reynolds numbers ($Re = 300, 500, 700$). Here, these errors correspond to interpolation results.}\label{tab:12}
		\begin{tabular*}{1\textwidth}{@{\extracolsep{\fill}} lcccc }
			\hline\hline
			\small    
			Model & MAE & NMAE($\%$) & RMSE & NRMSE($\%$) \\ \hline        
			U-ResNet & \textbf{0.0156 ± 0.0087 }    & \textbf{0.8373 ± 0.3535}   & \textbf{0.0285 ± 0.0160} &  \textbf{1.5277 ± 0.6555}\\ 
			FNO & 0.3937 ± 0.2618     & 21.5131 ± 13.8962   & 0.4571 ± 0.2694 &  25.0154 ± 14.2133\\ \hline\hline            
		\end{tabular*}%
	\end{center}
\end{table}
The vertical velocity component (Table~\ref{tab:13}) exhibits similarly strong predictive performance by U-ResNet, achieving a mean absolute error (MAE) of $0.0060 \pm 0.0024$, corresponding to a 73\% reduction in error compared to FNO ($0.0225 \pm 0.0125$). The normalized MAE (NMAE) is $2.8899 \pm 1.3132\%$, indicating an 86\% reduction compared to FNO ($12.0847 \pm 11.2746\%$). The relatively higher normalized errors compared to those of the streamwise velocity are expected, given the smaller magnitude of the vertical velocity in channel flows, where even minor absolute deviations can lead to larger normalized differences.
\begin{table}[tbp]
	\begin{center}
		\caption{Generalization performance of U-ResNet and FNO for vertical velocity prediction at interpolated Reynolds numbers ($Re = 300, 500, 700$). Here, these errors correspond to interpolation results.}\label{tab:13}
		\begin{tabular*}{1\textwidth}{@{\extracolsep{\fill}} lcccc }
			\hline\hline
			\small    
			Model & MAE & NMAE($\%$) & RMSE & NRMSE($\%$) \\ \hline        
			U-ResNet & \textbf{0.0060 ± 0.0024}     & \textbf{2.8899 ± 1.3132}  & \textbf{0.0127 ± 0.0054} &  \textbf{5.9273 ± 2.4172}  \\ 
			FNO & 0.0225 ± 0.0125     & 12.0847 ± 11.2746  & 0.0352 ± 0.0150 &  18.3453 ± 14.7579  \\ \hline\hline            
		\end{tabular*}%
	\end{center}
\end{table}

\begin{figure}[htbp]\centering
	\includegraphics[width=1.0\textwidth]{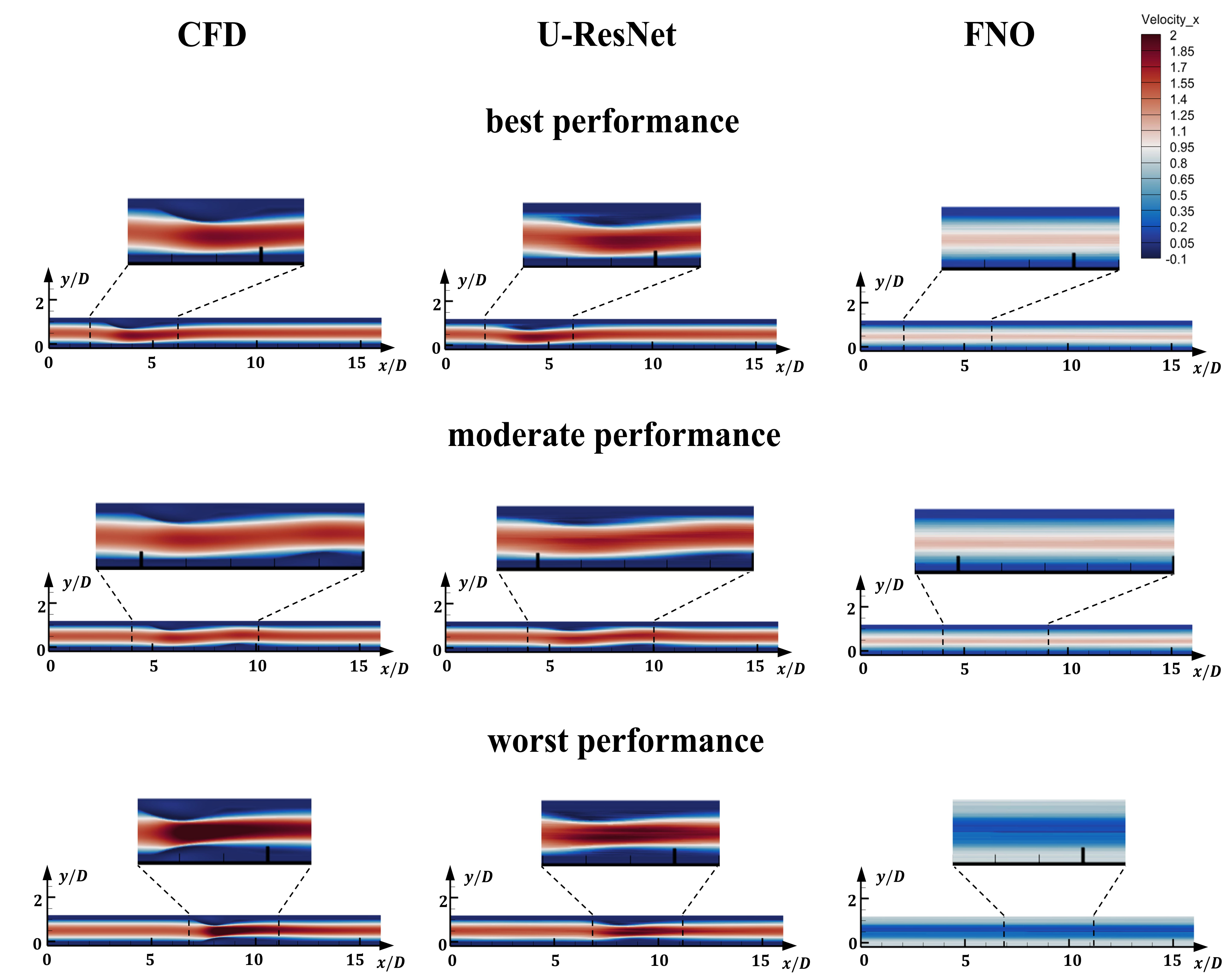}
	\caption{Streamwise velocity prediction generalization for U-ResNet and FNO with three representative (best, moderate, worst) stenotic flow cases.}\label{fig:28}
\end{figure}

Fig.~\ref{fig:28} presents streamwise velocity contours from U-ResNet and FNO for three representative cases selected from 60 test cases at interpolated Reynolds numbers ($Re=300$, 500, 700): (a) the best-performing case with $Re=500$ and unilateral stenosis, (b) a moderate-performing case with $Re=500$ and bilateral stenosis, and (c) the worst-performing case with $Re=300$ and bilateral stenosis.
For both the best and moderate cases, U-ResNet shows strong agreement with the CFD ground truth, accurately capturing the velocity field structures. In the worst case, although some errors arise near the stenosis region, U-ResNet still provides an overall accurate prediction of the flow dynamics.
These results highlight U-ResNet’s capability to generalize across varying stenotic geometries and Reynolds numbers. The model successfully resolves critical flow features, including peak velocity magnitudes that scale with $Re$ and the post-stenotic flow recovery associated with flow reattachment. In contrast, FNO fails to provide reasonable predictions in any of the cases, underscoring its poor generalization ability with respect to Reynolds number variations. Furthermore, the geometric configurations corresponding to each case in Fig.~\ref{fig:28} are detailed in Table~\ref{tab:geom3}.

\begin{table}[tbp]
	\begin{center}
		\caption{Geometric configurations for the best, moderate, and worst performance cases in streamwise velocity field generalization results.}\label{tab:geom3}
		\begin{tabular*}{1\textwidth}{@{\extracolsep{\fill}} lcccc }
			\hline\hline
			\small    
			Case & Peak height($D$) & Length($D$) & Tilt($D$) & Starting position($D$) \\ \hline
			Best: unilateral(up) & (0.301) & (2.573) & (-0.546) & (2.637)\\         
			Moderate: bilateral(up,down) & (0.174,0.208) & (1.683,2.429) & (-0.756,-0.641) & (4.951,7.874)\\
			Worst: bilateral(up,down) & (0.225,0.217)   & (2.183,4.963)  & (-0.359,-0.184) &  (6.920,7.451) \\ \hline\hline         
		\end{tabular*}%
	\end{center}
\end{table}

\begin{figure}[htbp]
    \centering
    \begin{subfigure}[b]{0.32\textwidth}  % 调整宽度，使三张图并列
        \begin{overpic}[width=\linewidth]{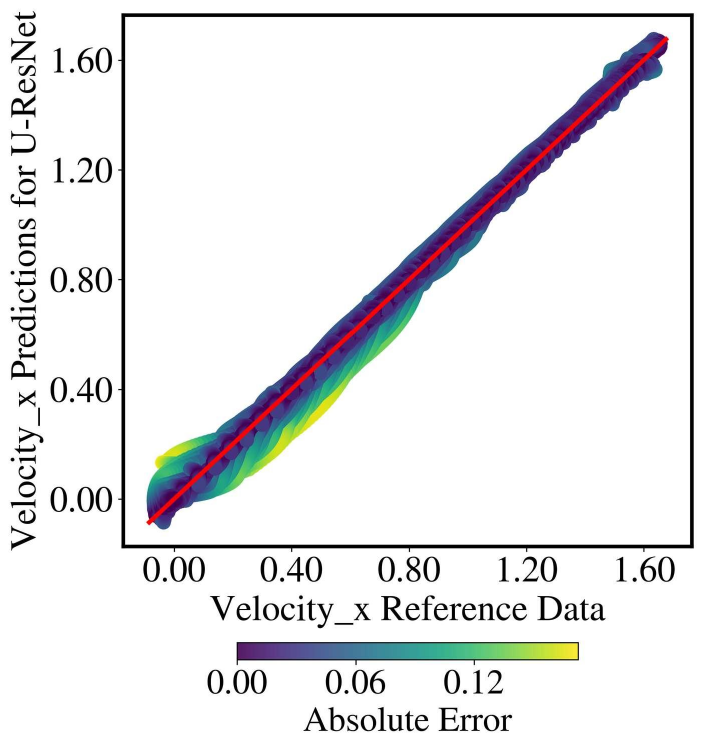}
            \put(-3,103){\small (a)}  
        \end{overpic}
    \end{subfigure}
    \hfill
    \begin{subfigure}[b]{0.32\textwidth}
        \begin{overpic}[width=\linewidth]{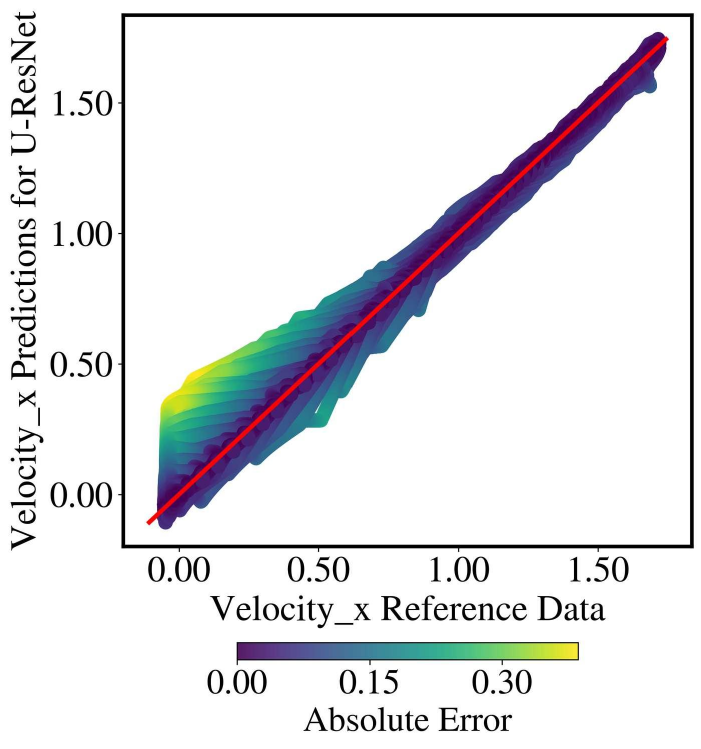}
            \put(-3,103){\small (b)}  
        \end{overpic}
    \end{subfigure}
    \hfill
    \begin{subfigure}[b]{0.32\textwidth}
        \begin{overpic}[width=\linewidth]{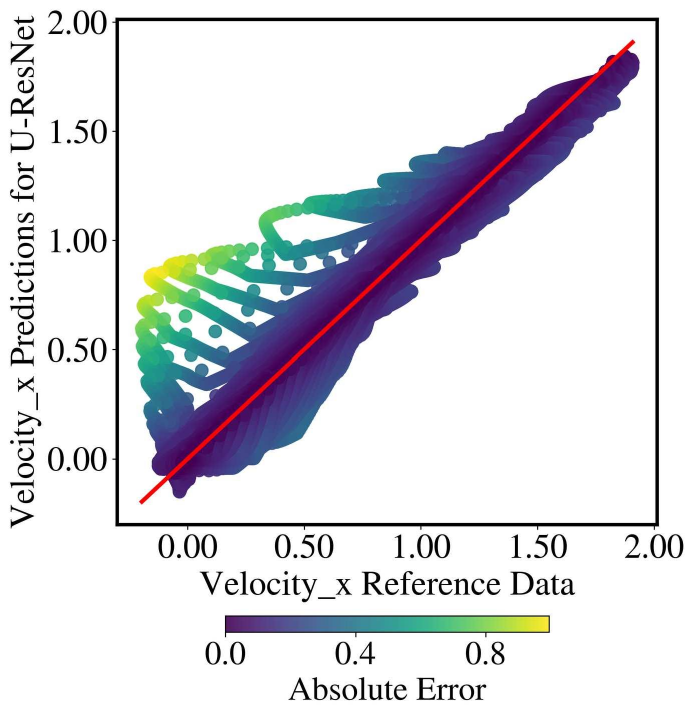}
            \put(-3,103){\small (c)}  
        \end{overpic}
    \end{subfigure}
	\caption{Streamwise velocity prediction of U-ResNet versus CFD data: (a) Best performance case, (b) Moderate performance case, (c) Worst performance case. Here, the errors correspond to interpolation results.}\label{fig:29}
    \end{figure}

The prediction accuracy of U-ResNet is further evaluated through quantitative error assessment in Fig.~\ref{fig:29}, which compares predicted values against CFD ground truth for three characteristic cases. Fig.~\ref{fig:29}(a) illustrates the highest-accuracy case (among the 60 test cases) with good correlation between predictions and CFD data. Fig.~\ref{fig:29}(b) depicts a moderate-accuracy case (among the 60 test cases) with slight deviations primarily in high-gradient regions. Fig.~\ref{fig:29}(c) shows the lowest-accuracy case (among the 60 test cases) with more substantial discrepancies.

The elevated error in velocity field prediction observed in Fig.~\ref{fig:29}(a), compared to the wall-based quantities in Figs.~\ref{fig:27}(a) (WSS) and ~\ref{fig:25}(a) (pressure), stems from fundamental differences in the spatial complexity and domain coverage of these hemodynamic parameters. Velocity fields require prediction across the entire two-dimensional domain, involving a dense grid of points with intricate flow features such as vortices, shear layers, and boundary layer transitions. This task demands resolution of high-frequency spatial variations and nonlinear interactions between flow structures, which are inherently more challenging for convolutional architectures with fixed receptive fields. Pressure and WSS are evaluated along the vessel wall (a 1-D manifold), where spatial variations are smoother and more gradual, reducing the model’s learning complexity . These parameters primarily reflect near-wall flow interactions, which are more predictable due to boundary layer regularity and reduced turbulent fluctuations. 

\subsubsection{Vorticity field prediction generalization}
\begin{table}[tbp]
	\begin{center}
		\caption{Generalization performance of U-ResNet and FNO for vorticity field prediction at interpolated Reynolds numbers ($Re = 300, 500, 700$). Here, these errors correspond to interpolation results.}\label{tab:14}
		\begin{tabular*}{1\textwidth}{@{\extracolsep{\fill}} lcccc }
			\hline\hline
			\small    
			Model & MAE & NMAE($\%$) & RMSE & NRMSE($\%$) \\ \hline        
			U-ResNet & \textbf{0.1518 ± 0.0747}     & \textbf{0.4712 ± 0.1526}   & \textbf{0.3771 ± 0.1797} &  \textbf{1.1465 ± 0.2694} \\ 
			FNO & 2.8108 ± 0.1798     & 9.3858 ± 2.3231   & 3.9804 ± 0.3256 &  13.2540 ± 3.2035 \\ \hline\hline            
		\end{tabular*}%
	\end{center}
\end{table}

Table~\ref{tab:14} summarizes the performance metrics of U-ResNet and FNO for vorticity field prediction at interpolated Reynolds numbers ($Re = 300, 500, 700$). U-ResNet maintains excellent predictive accuracy when generalizing to these previously unseen flow regimes, achieving a mean absolute error (MAE) of $0.1518 \pm 0.0747$, representing a 95\% reduction compared to FNO ($2.8108 \pm 0.1798$). The normalized MAE (NMAE) is $0.4712 \pm 0.1526\%$, also reflecting a 95\% reduction compared to FNO ($9.3858 \pm 2.3231\%$). These results correspond to a 40\% improvement in normalized error relative to the U-ResNet’s performance on the training Reynolds numbers (Table~\ref{tab:9}), highlighting U-ResNet’s robust capability to interpolate complex vortical structures across reference flow regimes.
The root mean squared error (RMSE) of $0.3771 \pm 0.1797$ and normalized RMSE (NRMSE) of $1.1465 \pm 0.2694\%$ further demonstrate the U-ResNet’s consistency in capturing peak vorticity magnitudes and spatial distributions across a variety of stenotic geometries. This sub-1.2\% normalized error is particularly noteworthy, as vorticity prediction involves sharp gradients and intricate rotational features that often pose significant challenges for data-driven approaches.

\begin{figure}[htbp]\centering
	\includegraphics[width=1.0\textwidth]{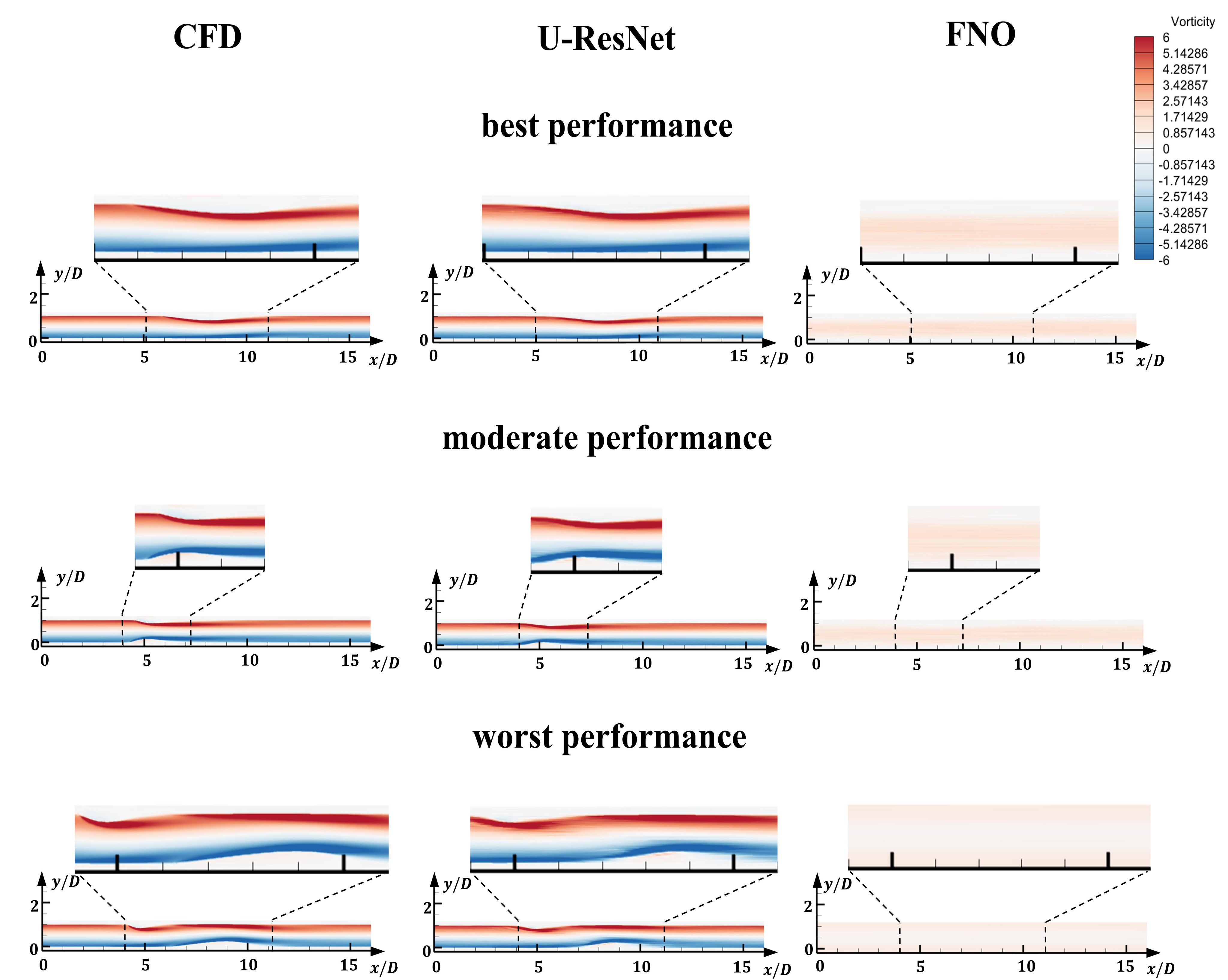}
	\caption{Vorticity prediction generalization for U-ResNet and FNO with three representative (best, moderate, worst) stenotic flow cases.}\label{fig:30}
\end{figure}

Fig.~\ref{fig:30} presents vorticity field predictions from U-ResNet and FNO for three representative cases selected from 60 test cases at interpolated Reynolds numbers ($Re=300$, 500, 700): (a) the best-performing case with $Re=700$ and unilateral stenosis, (b) a moderate-performing case with $Re=700$ and bilateral stenosis, and (c) the worst-performing case with $Re=300$ and bilateral stenosis.
As shown in Fig.~\ref{fig:30}, U-ResNet consistently provides accurate predictions across all three cases. Even in the worst-performing scenario, the model maintains a considerable level of accuracy in capturing the vorticity field.
These contour results demonstrate U-ResNet’s strong capability in capturing complex vortical structures, including shear-induced vorticity layers at stenosis boundaries, post-stenotic recirculation zones, secondary vortical formations between sequential stenoses, and the scaling behavior of vorticity magnitudes with increasing Reynolds number. Such qualitative agreement confirms the model’s robustness in resolving the underlying physics of vorticity transport across diverse flow regimes and stenotic geometries. In contrast, FNO fails to produce accurate predictions in all cases, primarily due to its inability to generalize across different Reynolds numbers. Moreover, Table~\ref{tab:geom4} presents the detailed geometric configurations for each case shown in Fig.~\ref{fig:30}.

\begin{table}[tbp]
	\begin{center}
		\caption{Geometric configurations for the best, moderate, and worst performance cases in vorticity field generalization results.}\label{tab:geom4}
		\begin{tabular*}{1\textwidth}{@{\extracolsep{\fill}} lcccc }
			\hline\hline
			\small    
			Case & Peak height($D$) & Length($D$) & Tilt($D$) & Starting position($D$) \\ \hline
			Best: unilateral(up) & (0.194)   & (3.153)  & (-0.384) & (5.982)\\         
			Moderate: bilateral(up,down) & (0.206,0.217) & (1.857,1.616) & (-0.412,-0.061) & (4.634,4.291) \\
			Worst: bilateral(up,down) & (0.187,0.232) & (1.613,3.895)  & (0.265,0.674) &  (4.245,6.391) \\ \hline\hline         
		\end{tabular*}%
	\end{center}
\end{table}

\begin{figure}[htbp]
    \centering
    \begin{subfigure}[b]{0.32\textwidth}  % 调整宽度，使三张图并列
        \begin{overpic}[width=\linewidth]{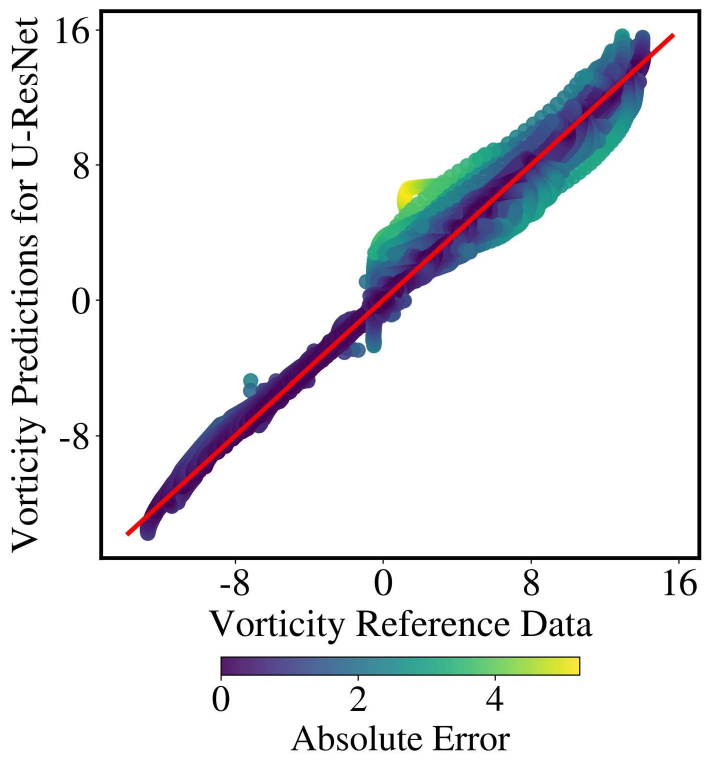}
            \put(-3,103){\small (a)}  
        \end{overpic}
    \end{subfigure}
    \hfill
    \begin{subfigure}[b]{0.32\textwidth}
        \begin{overpic}[width=\linewidth]{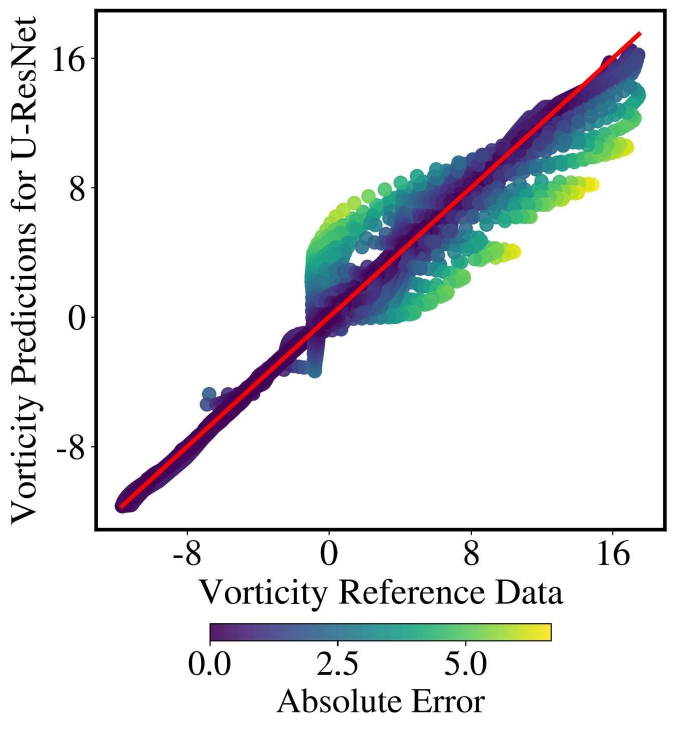}
            \put(-3,103){\small (b)}  
        \end{overpic}
    \end{subfigure}
    \hfill
    \begin{subfigure}[b]{0.32\textwidth}
        \begin{overpic}[width=\linewidth]{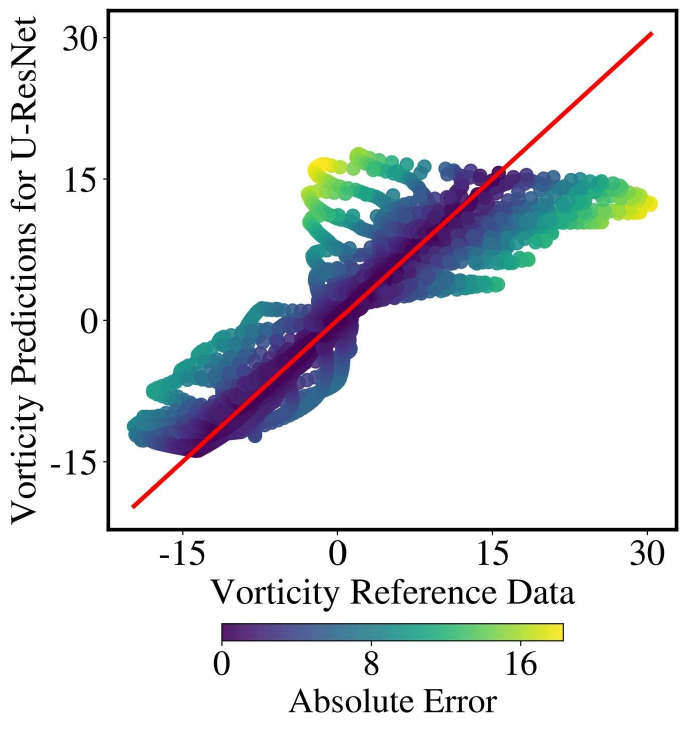}
            \put(-3,103){\small (c)}  
        \end{overpic}
    \end{subfigure}  
	\caption{Vorticity prediction of U-ResNet versus CFD data: (a) Best performance case, (b) Moderate performance case, (c) Worst performance case. Here, the errors correspond to interpolation results.}\label{fig:31}
    \end{figure}

Furthermore, Fig.~\ref{fig:31} provides a quantitative assessment of prediction fidelity through correlation plots comparing predicted values of U-ResNet against CFD ground truth for three characteristic cases. Fig.~\ref{fig:31}(a) illustrates the highest-accuracy case (among the 60 test cases) with excellent correlation between predictions and CFD data. Fig.~\ref{fig:31}(b) demonstrates a moderate-accuracy case (among the 60 test cases) with minor deviations primarily in high-gradient regions. Fig.~\ref{fig:31}(c) shows the lowest-accuracy case (among the 60 test cases) with more substantial discrepancies at intermediate vorticity values.

%======================================================
\section{Discussion}\label{discussion}
This study demonstrates that the U-ResNet significantly outperforms conventional convolutional networks and Fourier-based neural operators in predicting complex hemodynamics within stenotic channels. The superior accuracy of U-ResNet is quantitatively evident across multiple physical quantities, achieving normalized mean absolute errors ($NMAE$) of 1.10\% for pressure, 0.56\% for WSS, 1.06\% for streamwise velocity, 2.88\% for normal velocity, and 0.69\% for vorticity. These metrics represent a substantial improvement - often by an order of magnitude - over competing architectures when applied to geometrically complex stenotic flows at Reynolds numbers ranging from $200$ to $800$.

The exceptional performance of U-ResNet can be attributed to its architectural advantages in addressing the unique challenges of hemodynamic prediction. Residual connections facilitate gradient propagation throughout deep networks \citep{heDeepResidualLearning2015,heWhyResNetWorks2019}, which is particularly critical for resolving sharp hemodynamic gradients at stenotic boundaries. As demonstrated by \citet{freiAlgorithmdependentGeneralizationBounds2019}, these connections enable the network to learn identity mappings that preserve fine-scale features while simultaneously capturing multi-scale flow structures. This capability directly addresses the fundamental challenge in stenotic flow modeling: accurately resolving boundary layer dynamics while capturing global pressure distributions.

The comparative underperformance of FNO can be mechanistically explained through spectral bias - the inherent tendency of neural networks to prioritize low-frequency features \citep{rahamanSpectralBiasNeural2019}. In stenotic flows, where abrupt transitions in velocity and pressure occur near geometric constrictions, this bias is particularly detrimental. Our results in Fig.~\ref{fig:6} demonstrate that while FNO adequately captures the general pressure trend, it fails to resolve steep pressure gradients at the stenosis apex - a hemodynamically critical region where WSS peaks occur. Quantitatively, this limitation manifests as substantially higher errors for FNO in Tables~\ref{tab:7}–\ref{tab:9}, particularly for velocity ($NMAE = 6.48\%$) and vorticity ($NMAE = 8.93\%$) predictions.

Unlike spectrally-biased approaches, U-ResNet's multi-resolution design enables simultaneous capture of both global flow patterns and localized gradients. The architecture's superior performance in vorticity prediction ($NMAE = 0.69\%$) is particularly noteworthy, as vorticity inherently emphasizes high-frequency spatial variations that many neural networks struggle to resolve. This suggests that residual connections effectively mitigate the spectral bias problem by preserving high-frequency components throughout the network depth.

The present approach offers several distinctive advantages over existing deep learning models for hemodynamic prediction. First, our framework demonstrates robust generalization across Reynolds numbers ($Re = 300 -700$) despite being trained only on discrete values ($Re = 200, 400, 600, 800$). This cross-Reynolds generalization capability has rarely been achieved in previous studies and enables application across physiological flow regimes without retraining. Second, by employing non-dimensional formulations, our results maintain scalability across vessel sizes - a critical feature for translational applications. This universality distinguishes our approach from vessel-specific models that require retraining for different anatomical locations. The combined accuracy and generalization capability of this methodology position it as a valuable complement to traditional CFD, substantially enhancing the efficiency of rapid hemodynamic assessments in clinical settings.

The U-ResNet architecture demonstrates robust generalization capabilities across stenosis geometries that fall within the parametric distribution of the training dataset, which encompasses systematic variations in peak height ($P$), axial length ($L$), tilt ($T$), and position ($S$). As quantified in Section IV.B, the model maintains high predictive accuracy for unseen configurations, with normalized mean absolute errors ($NMAE$) consistently below 3.0\% for cases statistically aligned with the training distribution. For stenosis configurations exhibiting significant structural deviations from the training data - such as novel topologies (e.g., multifocal stenoses with complex interactions) or extreme geometric outliers - predictive performance may decline. To address such scenarios, a practical approach involves segmenting the vessel into anatomically distinct regions, applying the model to segments that match training characteristics, and potentially employing complementary physics-based solvers for highly aberrant segments. 

U-ResNet's training is typically a one-time investment. With a training dataset of 2000 samples, U-ResNet requires approximately 41.5 - 47.5 GPU·s per epoch on an NVIDIA A100 GPU, varying slightly based on the predicted physical quantity (pressure, WSS, velocity, or vorticity). Total training to convergence (typically 100 - 150 epochs) takes approximately 70–120 GPU minutes. For novel configurations outside the initial training distribution, the computational cost will scale linearly with dataset size. For example, doubling the training samples would roughly double the epoch time, while the number of epochs to convergence may vary. For moderately novel configurations, transfer learning (fine-tuning the pre-trained U-ResNet on a smaller dataset of new geometries) can significantly reduce training time compared to training from scratch, often requiring only 10–20\% of the original training data and epochs. By integrating parametric generalization with adaptive retraining strategies, U-ResNet maintains computational efficiency even when addressing evolving geometries (e.g., longitudinal studies of plaque progression for follow-up scans at 6-month intervals), positioning it as a scalable tool for both clinical and research applications.

The 2-D computational framework balances parametric flexibility and resolution fidelity, enabling systematic exploration of stenosis-flow interactions while avoiding the geometric and computational complexities of 3-D simulations. This approach provides a foundational understanding of hemodynamic mechanisms, which will inform future 3-D extensions for patient-specific applications. Despite its promising performance, translating this methodology to clinical applications requires addressing two fundamental limitations. First, the current study employs a 2-D representation of stenotic vessels, whereas physiological stenoses are inherently 3-D with complex morphologies. This simplification necessarily omits secondary flows and 3-D vortical structures that may significantly influence local hemodynamics, particularly in highly eccentric stenoses or at vessel bifurcations. Second, our models are trained and validated exclusively on synthetic CFD data without direct clinical validation. The absence of patient-specific measurements creates uncertainty regarding the framework's performance when confronted with physiological flow pulsatility, vessel compliance, and non-Newtonian blood rheology.

Future work will address these limitations through three principal avenues: (1) extending the framework to fully 3-D stenotic geometries to capture complex secondary flows; (2) incorporating physiologically realistic boundary conditions including flow pulsatility and vessel compliance; and (3) validating against clinical measurements such as Doppler ultrasound or phase-contrast MRI. By integrating clinical data with CFD simulations, we aim to develop a hybrid approach that leverages the complementary strengths of both data sources.

Furthermore, in clinical practice, repeated imaging of the same patient rarely produces identical stenosis geometries due to the dynamic progression of atherosclerotic disease. In the current study, synthetic and CFD-generated data with systematically varied stenosis geometries were utilized to ensure comprehensive coverage of training scenarios. This methodology enables exploration of the model’s generalization capability across diverse lesion morphologies. Direct application to clinical data necessitates the development of strategies to address inherent variability observed in longitudinal patient studies. To mitigate this limitation, future efforts will prioritize: (1) Incorporation of a broader range of clinically inspired and patient-specific geometries into training datasets; (2) Development of transfer learning \citep{daneker2024transfer} or domain adaptation techniques \citep{mutnuri2024using} to reconcile disparities between synthetic and real clinical data; (3) Collaboration with clinical partners to acquire and annotate longitudinal patient datasets, enabling validation and refinement of model robustness despite geometric variability.

The computational efficiency of neural surrogate models makes them particularly promising for uncertainty quantification and inverse design applications, where thousands of simulations may be required. Future implementations will explore these directions to enable robust risk stratification for individual patients and optimization of interventional approaches based on hemodynamic criteria.
%======================================================
\section{Conclusions}\label{conclusions}
This study establishes U-ResNet as an effective deep learning framework for predicting hemodynamic characteristics in 2-D asymmetric stenosis, a pathological condition prevalent in cardiovascular disease. Comprehensive quantitative evaluation demonstrates that U-ResNet significantly outperforms conventional neural architectures across all measured flow parameters, with normalized mean absolute errors ($NMAE$) consistently below 3\%.

U-ResNet's residual connections and multi-scale feature extraction enable exceptional fidelity in capturing steep hemodynamic gradients at stenotic boundaries. This capacity for detail preservation is particularly evident in WSS predictions, where U-ResNet achieves an $NMAE$ of 0.56\% - an order of magnitude improvement over U-Net (5.70\%) and FNO (3.99\%). The model demonstrates robust performance across interpolated Reynolds numbers ($Re = 300, 500, 700$) despite training only on discrete values ($Re = 200, 400, 600, 800$). This cross-regime generalization capability eliminates the need for regime-specific models, enabling seamless application across physiologically relevant flow conditions. With $5-8\times$ fewer parameters than FNO/UFNO architectures and 85-98\% reduced GPU memory requirements, U-ResNet delivers inference times of approximately 10 seconds per case - a $180\times$ acceleration compared to conventional CFD approaches (30 minutes). This computational advantage enables potential integration into clinical workflows where rapid hemodynamic assessment is essential. These advancements collectively position U-ResNet as a promising tool for both research and clinical applications in cardiovascular hemodynamics. The demonstrated capacity to accurately predict pressure distributions, wall shear stress patterns, and complex flow structures offers potential value for diagnostic assessment of stenotic severity, pre-interventional planning, and physiological flow analysis.

% \section*{Supplementary Material}
% \noindent Movie 1. Re=200 quasi-periodic oscillation of the collapsible tube.\\
% Movie 2. Re=200 quasi-periodic vortex shedding.
%================================================
\begin{acknowledgments}
This work was supported by the National Natural Science Foundation of China (NSFC, Grant No. 12172161). Additional support was provided by the Center for Computational Science and Engineering at the Southern University of Science and Technology and the Queensland University of Technology Early Career Research Ideas Scheme (ECRIS) 2024.
\end{acknowledgments}

%================================================
\section*{AUTHOR DECLARATIONS}
\subsection*{Conflict of Interest}
The authors have no conflicts to disclose.

%================================================
\section*{Data Availability}
The data that support the findings of this study are available from the corresponding author upon reasonable request.

%================================================
\appendix 
\section{Hyperparameter optimization for deep learning models}\label{appendix1}
This appendix provides a comprehensive overview of the hyperparameter configurations and optimization processes for the deep learning models evaluated in this study, including U-ResNet, FNO, U-Net, and UFNO. Training and testing loss metrics are reported for each model across four hemodynamic quantities: pressure, wall shear stress (WSS), velocity, and vorticity fields.

U-ResNet was trained using a consistent set of hyperparameters across all physical quantities to ensure uniformity in model architecture and training protocol (Table~\ref{tab:apd1}). The Adam optimizer~\cite{kingma2017adammethodstochasticoptimization} was employed with integrated weight decay regularization to mitigate overfitting. For FNO, an extensive grid search was conducted to identify optimal configurations of Fourier modes and channel widths for each hemodynamic quantity (Table~\ref{tab:apd2}). Configurations yielding minimal testing loss were selected for performance comparisons. Training epochs, learning rate, batch size, and weight decay were held consistent with U-ResNet to ensure equitable benchmarking.

%As detailed in Section~\ref{results}, U-ResNet is utilized to predict pressure, WSS, velocity, and vorticity fields using a consistent set of hyperparameters, as summarized in Table~\ref{tab:apd1}. The corresponding training and testing losses for each physical quantity are also reported in the same table. The weight decay parameter specified in the table is incorporated into the Adam optimizer~\cite{kingma2017adammethodstochasticoptimization}.

%For FNO, different sets of hyperparameters are employed for predicting pressure, WSS, velocity, and vorticity, along with their corresponding training and testing losses, as summarized in Table~\ref{tab:apd2}. For each physical quantity, we select the hyperparameter configuration that yields the lowest testing loss to serve as the representative FNO model for comparison with other models. Moreover, the number of training epochs, learning rate, batch size, and weight decay used in FNO remain consistent with those used in U-ResNet, as listed in Table~\ref{tab:apd1}.

%For U-Net, which is used to predict pressure and WSS in Section~\ref{results}, the same hyperparameters as those for U-ResNet are adopted. The detailed hyperparameter settings, along with the corresponding training and testing losses, are summarized in Table~\ref{tab:apd3}. UFNO, used for predicting pressure and WSS in Section~\ref{results}, shares the same training epochs, learning rate, batch size, and weight decay settings as U-ResNet. Its detailed hyperparameters, along with the corresponding training and testing losses, are summarized in Table~\ref{tab:apd4}.

U-Net and UFNO utilized identical training protocols (epochs, learning rate, batch size, weight decay) as U-ResNet for direct comparison (Tables~\ref{tab:apd3}--\ref{tab:apd4}). UFNO’s hybrid architecture combined Fourier neural operator layers with U-Net’s encoder-decoder structure, maintaining parameter counts comparable to other models.

\begin{table}[tbp]
	\begin{center}
		\caption{Optimized hyperparameters and loss metrics for U-ResNet across physical quantities.}\label{tab:apd1}
		\begin{tabular*}{1\textwidth}{@{\extracolsep{\fill}} lcccccc }
			\hline\hline
			\small    
			Physical quantity & Epochs & Learning rate & Batch size & Weight decay & Training loss & Testing loss\\ \hline    
			Pressure & 200 &   0.0005 &  1 &  0.0001 & 0.0329  & 0.0367 \\ 
			WSS & 200 &   0.0005   & 1  & 0.0001 &  0.0277  & 0.0307 \\ 
			Velocity field & 200 &   0.0005   & 1  & 0.0001 &  0.0502 & 0.0508\\
			Vorticity field & 200 &   0.0005   & 1  & 0.0001 & 0.1912  &  0.1947\\  \hline\hline             
		\end{tabular*}%
	\end{center}
\end{table}

\begin{table}[tbp]
	\begin{center}
		\caption{Hyperparameter sensitivity analysis for FNO across hemodynamic quantities.}\label{tab:apd2}
		\begin{tabular*}{1\textwidth}{@{\extracolsep{\fill}} lcccc }
			\hline\hline
			\small    
			Physical quantity & Modes & Width & Training loss & Testing loss\\ \hline      
            Pressure & 60 & 200  & \textbf{0.0575}  & \textbf{0.0583}\\
                     & 30 & 200  & 0.0660  & 0.0655\\
                     & 90 & 200  & 0.0587 & 0.0592 \\
                     & 60 & 150  &  0.0702 & 0.0699 \\
                     & 60 & 250  &  0.0626 & 0.0636\\ \hline
            WSS & 100 & 200  & 0.0682  & \textbf{0.0885}\\
                & 50 & 200  &  0.0692 & 0.0895\\
                & 150 & 200  & 0.0687 & 0.0912\\
                & 100 & 150  & \textbf{0.0658}  & 0.0893\\   
                & 100 & 250  & 0.0679  & 0.0889\\   \hline           
            Velocity field  & 100 & 200  & 0.0513  & \textbf{0.0528} \\
                            & 50 & 200  & 0.0588 & 0.0599\\
                            & 150 & 200  & \textbf{0.0511}  & 0.0529\\
                            & 100 & 150  &  0.0525 & 0.0537\\   
                            & 100 & 250  & 0.0588  & 0.0597\\   \hline            
            Vorticity field & 100 & 200   &  \textbf{0.2125} & \textbf{0.2184}\\ 
                            & 50 & 200  & 0.2604  & 0.2609\\
                            & 150 & 200  & 0.2672  & 0.2739\\
                            & 100 & 150  & 0.2284  & 0.2348\\   
                            & 100 & 250  & 0.2283  & 0.2321\\   \hline\hline
		\end{tabular*}%
	\end{center}
\end{table}

\begin{table}[tbp]
	\begin{center}
		\caption{Hyperparameters and loss metrics for U-Net.}\label{tab:apd3}
		\begin{tabular*}{1\textwidth}{@{\extracolsep{\fill}} lcccccc }
			\hline\hline
			\small    
			Physical quantity & Epochs & Learning rate & Batch size & Weight decay & Training loss & Testing loss\\ \hline    
			Pressure & 200 &   0.0005   & 1  & 0.0001 & 0.6262  &  0.6214 \\ 
			WSS & 200 &   0.0005   & 1  & 0.0001 & 0.3568  &  0.3537\\ \hline\hline 
		\end{tabular*}%
	\end{center}
\end{table}

\begin{table}[tbp]
	\begin{center}
		\caption{Hyperparameters and loss metrics for UFNO.}\label{tab:apd4}
		\begin{tabular*}{1\textwidth}{@{\extracolsep{\fill}} lcccc }
			\hline\hline
			\small    
			Physical quantity & Modes & Width & Training loss & Testing loss\\ \hline      
            Pressure & 48 & 128  &  0.0115  &  0.0143 \\
            WSS & 48 & 128 &  0.0091 & 0.0122 \\ \hline\hline
		\end{tabular*}%
	\end{center}
\end{table}

%\section{Additional Figures and Metrics}
%This appendix provides additional evaluation results, figures, or derivations that support the main findings presented in the main text.
%============================================
\bibliography{reference.bib}
\end{document}